\documentclass[twocolumn]{svjour3}  

\usepackage[utf8]{inputenc}
\usepackage{import}
\usepackage{tabu}
\usepackage{makecell}
\usepackage{hyperref}
\usepackage{xcolor}
\usepackage{amssymb}
\usepackage{graphicx}
\usepackage{tablefootnote}

\newcommand{\orcid}[1]{\href{https://orcid.org/#1}{\includegraphics[width=10pt]{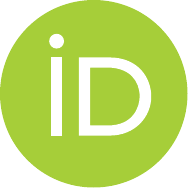}}}

\definecolor{red}{RGB}{0,0,0}

\begin{document}
\title{Scientific Paper Recommendation Systems: a Literature Review of recent Publications}
\titlerunning{Scientific Paper Recommendation Systems: a Literature Review of recent Publications}

\author{Christin Katharina Kreutz$^{1}$\orcid{0000-0002-5075-7699}
\and Ralf Schenkel$^{2}$\orcid{0000-0001-5379-5191}}

\institute{$^{1}$ Cologne University of Applied Sciences, Germany; work was done while at Trier University. \email{christin.kreutz@th-koeln.de}\\
$^{2}$ Trier University, Germany. \email{schenkel@uni-trier.de} 
}

\authorrunning{Kreutz and Schenkel}

\date{Received: date / Accepted: date}

\maketitle

\begin{abstract}
Scientific writing builds upon already published papers. Manual identification of publications to read, cite or consider as related papers relies on a researcher's ability to identify fitting keywords or initial papers from which a literature search can be started. The rapidly increasing amount of papers has called for automatic measures to find the desired \textit{relevant} publications, so-called paper recommendation systems. 

As the number of publications increases so does the amount of paper recommendation systems. Former literature reviews focused on discussing the general landscape of approaches throughout the years and highlight the main directions. We refrain from this perspective, instead we only consider a comparatively small time frame but analyse it fully.

In this literature review we discuss used methods, datasets, evaluations and open challenges encountered in all works first released between January 2019 and October 2021. The goal of this survey is to provide a comprehensive and complete overview of current paper recommendation systems.

\end{abstract}

\section{Introduction}

% reasons we need such systems
The rapidly increasing number of publications leads to a large quantity of possibly relevant papers~\cite{DBLP:conf/lwa/AlzoghbiA0L15} for more specific tasks such as finding related papers~\cite{DBLP:conf/jcdl/CollinsB19}, finding ones to read~\cite{DBLP:journals/corr/abs-2103-08819} or literature search in general to inspire new directions and understand the state-of-the-art approaches~\cite{KANG20212020BDP0008}.
Overall researchers typically spend a large amount of time on searching for relevant related work~\cite{DBLP:conf/webi/AmamiFSP17}. Keyword based search options are insufficient to find relevant papers~\cite{DBLP:journals/access/BaiWLYKX19,DBLP:journals/corr/abs-1304-5457,DBLP:journals/corr/abs-2103-08819}, they require some form of initial knowledge about a field. Oftentimes, users' information needs are not explicitly specified~\cite{DBLP:journals/tois/LiCPR19} which impedes this task further.

To close this gap, a plethora of paper recommendation systems have been proposed recently~\cite{DBLP:conf/aaai/GuoCZLDH20,DBLP:journals/scientometrics/HarunaIQKHMC20,DBLP:journals/access/SakibAABHHG21,DBLP:journals/concurrency/TangLQ21,DBLP:journals/www/YangLLLZZZZ19}. These systems should fulfil different functions:
% functions
for~ junior~ researchers~ systems~ should~ recommend a broad variety of papers, for senior ones the recommendations should align more with their already established interests~\cite{DBLP:journals/access/BaiWLYKX19} or help them discover relevant interdisciplinary research~\cite{DBLP:conf/jcdl/SugiyamaK11}. In general paper recommendation approaches positively affect researchers' professional lives as they enable finding relevant literature more easily and faster~\cite{DBLP:conf/recsys/LeKD19}.

% some types of recommendation systems
As there are many different approaches, their objectives and assumptions are also diverse.
% problem definition
A simple problem definition of a paper recommendation system could be the following: given one paper recommend a list of papers fitting the source paper~\cite{DBLP:conf/recsys/HassanSGMB19}. This definition would not fit all approaches as some specifically do not require any initial paper to be specified but instead observe a user as input~\cite{DBLP:conf/aaai/GuoCZLDH20}.
% different assumptions
Some systems recommend sets of publications fitting the queried terms only if \textcolor{red}{these papers are all} observed together~\cite{DBLP:conf/seke/LiuKCQ19,DBLP:journals/complexity/LiuKYQ20}, most of the approaches suggest a number of single publications as their result~\cite{DBLP:conf/aaai/GuoCZLDH20,DBLP:journals/scientometrics/HarunaIQKHMC20,DBLP:journals/access/SakibAABHHG21,DBLP:journals/www/YangLLLZZZZ19}\textcolor{red}{, such that any single one of these papers satisfies the information need of a user fully}.
Most approaches assume that all required data to run a system is present already~\cite{DBLP:conf/aaai/GuoCZLDH20,DBLP:journals/www/YangLLLZZZZ19} but some works~\cite{DBLP:journals/scientometrics/HarunaIQKHMC20,DBLP:journals/access/SakibAABHHG21} % ,DBLP:journals/corr/abs-1304-5457,DBLP:conf/pci/TsolakidisTSC16} 
explicitly crawl general publication information or even abstracts and keywords from the web.

In this literature review we observe papers recently published in the area of scientific paper recommendation between and including January 2019 and October 2021\footnote{The most recent surveys~\cite{DBLP:journals/access/BaiWLYKX19,DBLP:journals/ccsecis/LiZ19,DBLP:journals/tjs/ShahidAABZYC20} focusing on scientific paper recommendation appeared in 2019 such that this time frame is not yet covered.}. We strive to give comprehensive overviews on their utilised methods as well as their datasets, evaluation measures and open challenges of current approaches. 
Our contribution is four-fold:
\begin{itemize}
    \item We propose a current multidimensional character\-isation~ of~ current~ paper~ recommendation approaches.
    \item We compile a list of recently used datasets in evaluations of paper recommendation approaches.
    \item We compile a list of recently used evaluation measures for paper recommendation.
    \item We analyse existing open challenges and identify current novel problems in paper recommendation which could be specifically helpful for future approaches to address.
\end{itemize}

% structure of paper
In the following Section~\ref{sec:problemStatement} we describe the general problem statement for paper recommendation systems before we dive into the literature review in Section~\ref{sec:litReview}. Section~\ref{sec:datasets} gives insight into datasets used in current work. In the following Section~\ref{sec:evaluation} different definitions of relevance, relevance assessment as well as evaluation measures are analysed. Open challenges and objectives are discussed in detail in Section~\ref{sec:openChallenges}. Lastly Section~\ref{sec:conclusion} concludes this literature review.

\section{Problem Statement}
\label{sec:problemStatement}

Over~ the~ years~ different~ formulations~ for~ a~ problem statement of a paper recommendation system have emerged. In general they should specify the input for the recommendation system, the type of recommendation results, the point in time when the recommendation will be made and which specific goal an approach tries to achieve. Additionally, the target audience should be specified.

As \textit{input} we can either specify an initial paper~\cite{DBLP:conf/jcdl/CollinsB19}, keywords~\cite{DBLP:journals/www/YangLLLZZZZ19}, a user~\cite{DBLP:conf/aaai/GuoCZLDH20}, a user and a paper~\cite{DBLP:journals/kbs/AliQMAA20} or more~ complex~ information~ such~ as~ user-constructed knowledge graphs~\cite{DBLP:journals/corr/abs-2103-08819}. Users can be modelled as a combination~ of~ features~ of~ papers~ they~ interacted with~\cite{Bereczki1587420,Bulut2020}, e.g. their clicked~\cite{DBLP:journals/jodl/ChaudhuriSSS21} or authored publications~\cite{Bulut2018APR}. Papers can for example be represented by their textual content~\cite{DBLP:journals/access/SakibAABHHG21}.

\textit{As types of recommendation} we could either specify single (independent) papers~\cite{DBLP:conf/aaai/GuoCZLDH20} or a set of papers which is to be observed \textcolor{red}{completely to satisfy the information need}~\cite{DBLP:journals/complexity/LiuKYQ20}. A study by Beierle et al.~\cite{DBLP:journals/jodl/BeierleACB20} found that existing digital libraries recommend between three and ten single papers, in their case the optimal number of suggestions to display to users was five to six.

As for the \textit{point in time}, most work focuses on immediate recommendation of papers. Only a few approaches also consider delayed suggestion\footnote{\textcolor{red}{Non-immediate variants allow using methods which require more time to compute recommendations. Temporal patterns of user behaviour could be incorporated in the recommendation process to identify a fitting moment to present new recommendations to a user. The moment a recommendation is presented to a user influences their interest, as the delayed recommendation might no longer be relevant or does not fit the current task of a user.}} via newsletter for example~\cite{DBLP:journals/tois/LiCPR19}. 

In general, recommended papers should be relevant in one way or another to achieve certain \textit{goals}. %They could e.g. be related to an initial paper~\cite{DBLP:conf/jcdl/CollinsB19} or publications which should be read~\cite{DBLP:journals/corr/abs-2103-08819}. 
\textcolor{red}{The intended goal of authors of papers could e.g. either be to recommend papers which should be read~\cite{DBLP:journals/corr/abs-2103-08819} by a user or recommend papers which are simply somehow related to an initial paper~\cite{DBLP:conf/jcdl/CollinsB19}, by topic, citations or user interactions.}

Different \textit{target audiences}, for example junior or senior researcher, have different demands from paper recommendation systems~\cite{DBLP:journals/access/BaiWLYKX19}. Usually paper recommendation approaches target single users but there are also works which strive to recommend papers for sets of users~\cite{DBLP:journals/asc/WangWYXYY21,wang21}.

\section{Literature Review}
\label{sec:litReview}

In this chapter we first clearly define the scope of of our literature review (see Sect.~\ref{sec:litReview_scope}) before we conduct a meta analysis on the observed papers (see Sect.~\ref{sec:litReview_meta}). Afterwards our categorisation or lack thereof is discussed in depth (see Sect.~\ref{sec:litReview_categorisation}), before we give short overviews of all paper recommendation systems we found (see Sect.~\ref{sec:litReview_prs}) and some other relevant related work (see Sect.~\ref{sec:sec:litReview_others}).

\subsection{Scope}
\label{sec:litReview_scope}

To the best of our knowledge the literature reviews by Bai et al.~\cite{DBLP:journals/access/BaiWLYKX19}, Li and Zou~\cite{DBLP:journals/ccsecis/LiZ19} and Shahid et al.~\cite{DBLP:journals/tjs/ShahidAABZYC20} are the most recent ones targeting the domain of scientific paper recommendation systems. They were accepted for publication or published in 2019 so they only consider paper recommendation systems up until 2019 at most.
We want to bridge the gap between papers published after their surveys were finalised and current work so we only focus on the discussion of publications which appeared between January 2019 and October 2021 when this literature search was conducted. 

\begin{figure*}
    \centering
    \includegraphics[width=0.85\textwidth]{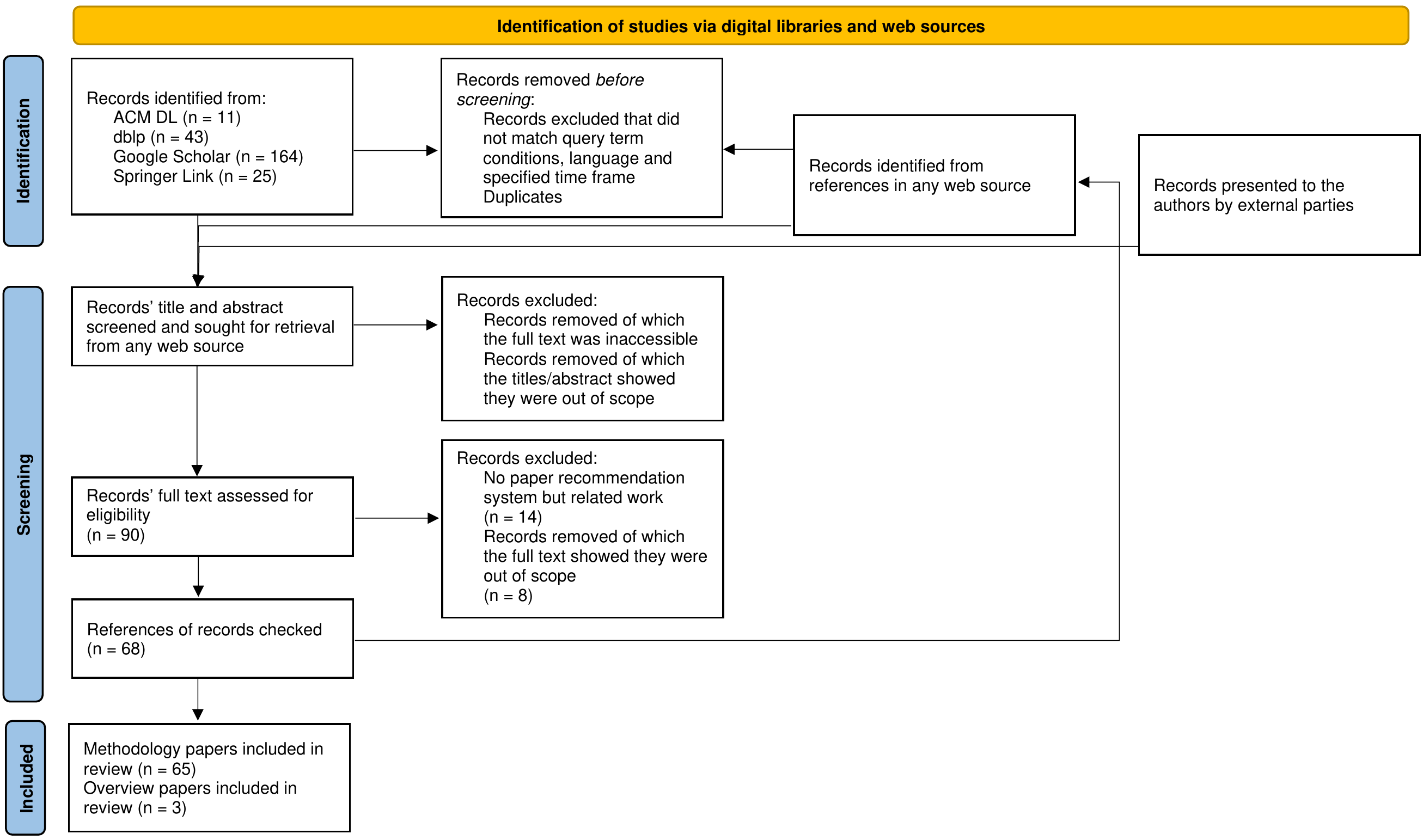}
    \caption{\textcolor{red}{PRISMA workflow of our literature review process.}}
    \label{fig:prisma}
\end{figure*}

We conducted our literature search on the following digital libraries:  ACM\footnote{\url{https://dl.acm.org/}}, dblp\footnote{\url{https://dblp.uni-trier.de/}}, GoogleScholar\footnote{\url{https://scholar.google.com/}} and Springer\footnote{\url{https://link.springer.com/}}. 
Titles of considered publications had to contain either \textit{paper}, \textit{article} or \textit{publication} as well as some form of \textit{recommend}. Papers had to be written in English to be observed. We judged relevance of retrieved publications by observing titles and abstracts if the title alone did not suffice to assess their topical relevance. In addition to these papers found by systematically searching digital libraries, we also considered their referenced publications if they were from the specified time period and of topical fit. 
For all papers their date of first publication determines their publication year \textcolor{red}{which decides if they lie in our time observed time frame or not}. E.g. for journal articles we consider the point in time when they were first published online instead of the date on which they were published in an issue, for conference articles we consider the date of the conference instead a later date when they were published online.
\textcolor{red}{Figure~\ref{fig:prisma} depicts the PRISMA~\cite{PRISMA} workflow for this study.}

% Abgrenzung
We refrain from including works in our study which do not identify as scientific paper recommendation systems such as Wikipedia article recommendation~\cite{wikipedia2,wikipedia3,wikipedia1} or general news article recommendation~\cite{news1,news2,news3}. Citation recommendation systems~\cite{DBLP:conf/ictai/Ng20,citrec2,DBLP:journals/kbs/ZhuLLSQN21} are also out of scope of this literature review. Even though citation and paper recommendation can be regarded as analogous~\cite{kanakia19}, we argue the differing functions of citations~\cite{garfield} and tasks of these recommendation systems~\cite{DBLP:journals/cit/MedicS20} should not be mixed with the problem of paper recommendation.
\textcolor{red}{Färber and Jatowt~\cite{DBLP:journals/jodl/FarberJ20} also support this view by stating that both are disjunctive, with paper recommendation pursuing the goal of providing papers to read and investigate while incorporating user interaction data and citation recommendation supporting users with finding citations for given text passages.}\footnote{\textcolor{red}{For a survey of current trends in citation recommendation refer to Färber and Jatowt~\cite{DBLP:journals/jodl/FarberJ20}.}}
%\footnote{\textcolor{red}{Medic and Snajder~\cite{DBLP:journals/cit/MedicS20} e.g. explicitly define the task of citation recommendation as getting an article or text snippet as input and recommending a list of papers to cite. This description excludes the multiplicity of paper recommendation systems taking more complex inputs such as user information and providing personalised recommendations.}
We also consciously refrain from discussing the plethora of more area-independent recommender systems which could be adopted to the domain of scientific paper recommendation.

Our literature research resulted in \textcolor{red}{82} relevant papers. \textcolor{red}{Of these, three were review articles.} We found \textcolor{red}{14} manuscripts which do not present paper recommendation systems but are relevant works for the area none\-theless, they are discussed in Section~\ref{sec:sec:litReview_others}. This left \textcolor{red}{65} publications describing paper recommendation systems for us to analyse in the following.

\subsection{Meta analysis}
\label{sec:litReview_meta}

For papers within our scope, we consider their publication year as stated in the citation information \textcolor{red}{for this meta analysis. This could affect the publication year of papers compared to the former definition of which papers are included in this survey. E.g. for journal articles we do not set the publication year as the point in time when they were first published online, instead for consistency (this data is present in the citation information of papers) for this analysis we use the year the issue was published in which the article is contained}. Of the \textcolor{red}{65} relevant system papers, 21 were published in 2019, 23 were published in 2020 and \textcolor{red}{21} were published in 2021. On average each paper has \textcolor{red}{4.0462} authors (std. dev.=\textcolor{red}{1.6955}) and \textcolor{red}{12.4154} pages (std. dev.=\textcolor{red}{9.2402}). \textcolor{red}{35 (53.85\%)} of the papers appeared as conference papers, 27 \textcolor{red}{(41.54\%)} papers were published in journals and there were two preprints \textcolor{red}{(3.08\%)} which have not yet been published otherwise. There has been one master's thesis \textcolor{red}{(1.54\%)} within scope. The most common venues for publications were the ones depicted in Table~\ref{tab:num_venue}.
Some papers~\cite{DBLP:conf/icadl/NishiokaHS19,DBLP:conf/ercimdl/NishiokaHS19,DBLP:journals/peerj-cs/NishiokaHS20,DBLP:journals/peerj-cs/ShahidAAAA21,shahid21a} described the same approach without modification or extension of the actual paper recommendation methodology e.g. by providing evaluations\footnote{\textcolor{red}{These papers could either be a demo paper and a later published full paper or the conference and journal version of the same approach, which is then slightly extended by more experiments. These paper clusters are no exact duplicates or fraudulent publications.}}. This left us with \textcolor{red}{62} different paper recommendation systems to discuss.

\begin{table}[t]
    \centering
    \begin{tabular}{l|l|l}
    
        Type & Venue & \#p \\ \hline
        Journal & IEEE Access & 5\\
       Journal & Scientometrics & 2\\ 
       Journal & PeerJ CS & 2\\\hline
       Conference & WWW & 2\\
       Conference & ChineseCSCW & 2\\
       Conference & CSCWD & 2\\
    \end{tabular}
    \caption{Top most common venues where relevant papers were published together with their type and number of papers (\#p). \textcolor{red}{Other venues had only one associated paper.}}
    \label{tab:num_venue}
\end{table}

\subsection{Categorisation}
\label{sec:litReview_categorisation}

\subsubsection{Former Categorisation}
\label{sec:litReview_former}

The~ already~ mentioned~ three~ most~ recent~\cite{DBLP:journals/access/BaiWLYKX19,DBLP:journals/ccsecis/LiZ19,DBLP:journals/tjs/ShahidAABZYC20} and one older but highly influential~\cite{DBLP:journals/jodl/BeelGLB16} literature reviews in scientific paper recommendation utilise different categorisations to group approaches.
Beel et al.~\cite{DBLP:journals/jodl/BeelGLB16} categorise observed papers by their underlying recommendation~ principle~ into~ stereotyping,~ content-based filtering, collaborative filtering, co-occurrence, graph-based, global relevance and hybrid models.
Bai et al.~\cite{DBLP:journals/access/BaiWLYKX19} only utilise the classes content-based filtering, collaborative filtering, graph-based methods, hybrid methods and other models. 
Li and Zou~\cite{DBLP:journals/ccsecis/LiZ19} use the categories content-based recommendation, hybrid recommendation, graph-based recommendation and recommendation based on deep learning.
Shahid et al.~\cite{DBLP:journals/tjs/ShahidAABZYC20} label approaches by the criterion they identify relevant papers with: content, metadata, collaborative filtering and citations.

The four predominant categories thus are content-based filtering, collaborative filtering, graph-based and hybrid systems. Most of these categories are defined precisely but graph-based approaches are not always characterised concisely:
\textit{Content-based filtering} (CBF) methods are said to be ones where user interest is inferred by observing their historic interactions with papers~\cite{DBLP:journals/access/BaiWLYKX19,DBLP:journals/jodl/BeelGLB16,DBLP:journals/ccsecis/LiZ19}. Recommendations are composed by observing features of papers and users~\cite{DBLP:journals/kbs/AliQMAA20}.
In \textit{collaborative filtering} (CF) systems the preferences of users similar to a current one are observed to identify likely relevant publications~\cite{DBLP:journals/access/BaiWLYKX19,DBLP:journals/jodl/BeelGLB16,DBLP:journals/ccsecis/LiZ19}. Current users' past interactions need to be similar to similar users' past interactions~\cite{DBLP:journals/access/BaiWLYKX19,DBLP:journals/jodl/BeelGLB16}.
\textit{Hybrid} approaches are ones which combine multiple types of recommendations~\cite{DBLP:journals/access/BaiWLYKX19,DBLP:journals/jodl/BeelGLB16,DBLP:journals/ccsecis/LiZ19}.

\textit{Graph-based} methods can be characterised in multiple ways. A very narrow definition only encompasses ones which observe the recommendation task as a link prediction problem or utilise random walk~\cite{DBLP:journals/kbs/AliQMAA20}. Another less strict definition identifies these systems as ones which construct networks of papers and authors and then~ apply~ some~ graph~ algorithm~ to~ estimate relevance~\cite{DBLP:journals/access/BaiWLYKX19}. Another definition specifies this class as one using graph metrics such as random walk with restart, bibliographic coupling or co-citation inverse document frequency~\cite{DBLP:conf/bigdataconf/TannerAH19}. Li and Zhou~\cite{DBLP:journals/ccsecis/LiZ19} abstain from clearly characterising this type of systems directly but give examples which hint that in their understanding of graph-based methods somewhere in the recommendation process, some type of graph information e.g. bibliographic coupling or co-citation strength, should be used. Beel et al.~\cite{DBLP:journals/jodl/BeelGLB16} as well as Bai et al.~\cite{DBLP:journals/access/BaiWLYKX19} follow a similar line, they characterise graph-based methods broadly as ones which build upon the existing connections in a scientific context to construct a graph network.

\begin{table}[t]
    \centering
    \scriptsize
    \begin{tabular}{l|l|l}
         Work & Label & c\\\hline
        \cite{DBLP:conf/flairs/AfsarCF21} &	knowledge-based	& $\times$\\
        \cite{ahmedi} &	hybrid & $\checkmark$	\\
        \cite{DBLP:conf/icmla/AlfarhoodC19} &	deep learning-based & $\checkmark$	\\
        \cite{DBLP:journals/kbs/AliQMAA20} &	unified model & $\times$	\\
        \cite{Bereczki1587420} &	graph-based	& $\checkmark$ \\
        \cite{Bulut2020} &	user-specific & $\times$ 	\\
        \cite{DBLP:journals/corr/abs-2107-07831} &	hybrid &$\checkmark$	\\
        \cite{DBLP:conf/aiccsa/DuGWHZG20} &	graph-based	& $\checkmark$\\
        \cite{DBLP:journals/tkde/DuTD21} &	active one-shot learning  & $\times$ \\
        \cite{DBLP:conf/aaai/GuoCZLDH20} &	collaborative filtering	& $\checkmark$\\
        \cite{DBLP:journals/scientometrics/HarunaIQKHMC20} &	hybrid & $\checkmark$	\\
        \cite{DBLP:conf/ifip12/HuMLH20} &	hybrid	& $\checkmark$\\
        \cite{DBLP:conf/ml4cs/JingY20} &	hybrid	& $\checkmark$\\
        \cite{kanakia19} &	hybrid	& $\checkmark$\\
        \cite{KANG20212020BDP0008} &	hybrid	& $\checkmark$\\
        \cite{li20} &	hybrid	& $\checkmark$\\
        \cite{DBLP:journals/dss/LiWNLL21} &	network-based & $\times$ 	\\
        \cite{lin21} &	content-based  & $\checkmark$ 	\\
        \cite{DBLP:journals/complexity/LiuKYQ20} &	graph-based	& $\checkmark$\\
        \cite{DBLP:conf/cscwd/LuHCP021} &	neuro-collaborative filtering & $\times$ 	\\
        \cite{DBLP:journals/access/MaW19a} &	meta-path based	& $\times$  	\\
        \cite{DBLP:journals/monet/MaZZ19} &	heterogeneous graph representation based & $\times$ 		\\
        \cite{DBLP:journals/ijiit/Manju} &	social network-based  & $\times$ 	\\
        \cite{MohamedHassan} &	hybrid & $\checkmark$	\\
        \cite{nair} &	content-based & $\checkmark$		\\
        \cite{DBLP:conf/icadl/NishiokaHS19,DBLP:conf/ercimdl/NishiokaHS19,DBLP:journals/peerj-cs/NishiokaHS20} &	content-based & $\checkmark$ 		\\
        \cite{DBLP:conf/iui/RahdariB19} &	hybrid	& $\checkmark$\\
        \cite{renuka} &	content-based & $\checkmark$ 		\\
        \cite{DBLP:journals/access/SakibAH20} &	collaborative filtering	& $\checkmark$\\
        \cite{DBLP:journals/access/SakibAABHHG21} &	hybrid	& $\checkmark$\\
        \cite{DBLP:journals/peerj-cs/ShahidAAAA21,shahid21a} & in-text citation frequencies-based & $\times$		\\
        \cite{DBLP:conf/ijcnn/ShiMZJLC20} &	hybrid	& $\checkmark$\\
        \cite{subathra} & content-based & $\checkmark$	\\
        \cite{DBLP:journals/concurrency/TangLQ21} &	hybrid	& $\checkmark$\\
        \cite{DBLP:conf/bigdataconf/TannerAH19} &	graph-based	& $\checkmark$\\
        \cite{DBLP:journals/access/WaheedIRMK19} &	hybrid	& $\checkmark$\\
        \cite{DBLP:conf/service/WangXTWX20} &	knowledge-aware path recurrent network & $\times$	\\
        \cite{DBLP:journals/corr/abs-2103-08819} &	graph-based	& $\checkmark$\\
        \cite{DBLP:journals/asc/WangWYXYY21} & hybrid & $\checkmark$\\
        \cite{wang21} &	hybrid & $\checkmark$	\\
        \cite{DBLP:journals/www/YangLLLZZZZ19} &	hybrid	& $\checkmark$\\
        \cite{DBLP:conf/ksem/YuHLZXLXY19} &	network & $\times$\\
        \cite{DBLP:journals/access/ZhaoKFMN20} & hybrid & $\checkmark$	\\
    \end{tabular}
    \caption{Indications as what type of paper recommendation system works describe themselves with indication if the description is a common used label (c).}
    \label{tab:self_labels}
\end{table}

%All of these classification schemes observe the basic recommendation principle of works.
When trying to classify approaches by their recommendation type, we encountered some problems:
\begin{enumerate}
    \item We have to refrain from only utilising the labels the works give themselves (see Table~\ref{tab:self_labels} for an overview of self-labels of works which do classify themselves). Works do not necessarily (clearly) state, which category they belong to~\cite{DBLP:conf/jcdl/CollinsB19,DBLP:conf/cscwd/LLP21,DBLP:conf/seke/LiuKCQ19}. Another problem with self-labelling is authors' individual definitions of categories while disregarding all possible ones (as e.g. seen with Afsar et al.~\cite{DBLP:conf/flairs/AfsarCF21} or Ali et al.~\cite{DBLP:journals/kbs/AliQMAA20}). Mis-definition or omitting of categories could lead to an incorrect classification.
    \item When considering the broadest definition of graph-based methods many recent paper recommendation systems tend to belong to the class of hybrid methods. Most of the approaches~\cite{DBLP:journals/kbs/AliQMAA20,KANG20212020BDP0008,DBLP:journals/tetc/KongMWLX21,DBLP:conf/cscwd/LLP21,DBLP:journals/dss/LiWNLL21,DBLP:journals/access/SakibAABHHG21,DBLP:conf/kdd/TangZYLZS08,DBLP:journals/www/YangLLLZZZZ19} utilise some type of graph structure information as part of the approach which would classify them as graph-based but as they also utilise historic user-interaction data or descriptions of paper features (see e.g. Li et al.~\cite{DBLP:journals/dss/LiWNLL21} who describe their approach as network-based while using a graph structure, textual components and user profiles) which would render them as either CF or CBF also.
\end{enumerate}

Thus we argue the former categories do not suffice to classify the particularities of current approaches in a meaningful way. So instead, we introduce more dimensions by which systems could be grouped.

\subsubsection{Current Categorisation}

Recent paper recommendation systems can be categorised in 20 different dimensions by \textcolor{red}{general information on the approach (G), already existing data directly taken from the papers used (D) and methods which might create or (re-)structure data, which are part of the approach (M)}: 

\begin{itemize}
    %\item Type: The approach has been published either as conference paper (c), journal paper (j), preprint (p) or something else, e.g. as a master's thesis (o).
    \item (G) Personalisation (person.): The approach produces personalised recommendations. \textcolor{red}{The recommended items depend on the person using the approach, if personalisation is not considered, the recommendation solely depends on the input keywords or paper. This dimension is related to the existence of user profiles.}
    \item (G) Input: The approach requires some form of input, either a paper (p), keywords (k), user (u) or something else, e.g. an advanced type of input (o). Hybrid forms are also possible. In some cases the input is not clearly specified throughout the paper so it is unknown (?).
    
    \item (D) Title: The approach utilises titles of papers.
    \item (D) Abstract (abs.): The approach utilises abstracts of papers.
    \item (D) Keyword (key.): The approach utilises keywords of papers. \textcolor{red}{These keywords are usually explicitly defined~ by~ the~ authors~ of~ papers,~ contrasting~ key phrases.}
    \item (D) Text: The approach utilises some type of text of papers which is not clearly specified as titles, abstracts or keywords. In the evaluation this approach might utilise specified text fragments of publications. 
    \item (D) Citation (cit.): The approach utilises citation information, e.g. numbers of citations or co-references.
    \item (D) Historic interaction (inter.): The approach uses some sort of historic user-interaction data, e.g. previously authored, cited or liked publications. An approach can only include historic user-interaction data if it also somehow contains user profiles.
    
    \item (M) User profile (user): The approach constructs some sort of user profile or utilises profile information. \textcolor{red}{Most approaches using personalisation also construct user profiles but some do not explicitly construct profiles but rather encode user information in the used structures.}
    \item (M) Popularity (popul.): The approach utilises some sort of popularity indication, e.g. CORE rank, numbers of citations\footnote{\textcolor{red}{The number of citations can be regarded both as an input data as well as a method to denote popularity.}} or number of likes. 
    
    \item (M) Key phrase (KP): The approach utilises key phrases. \textcolor{red}{Key phrases are not explicitly provided by authors of papers but are usually computed from the titles and abstracts of papers to provide a descriptive summary, contrasting keywords of papers.}
    \item (M) Embedding (emb.): The approach utilises some sort~ of~ text~ or~ graph~ embedding~ technique,~ e.g. BERT or Doc2Vec.
    \item (M) Topic model (TM): The approach utilises some sort of topic model, e.g. LDA.
    \item (M) Knowledge graph (KG): The approach utilises or builds some sort of knowledge graph. \textcolor{red}{This dimension surpasses the mere incorporation of a graph which describes a network of nodes and edges of different types. A knowledge graph is a sub-category of a graph.}
    \item (M) Graph: The approach actively builds or directly uses a graph structure, e.g. a knowledge graph or scientific heterogeneous network. Utilisation of a neural network is not considered in this dimension.
    \item (M) Meta-path (path): The approach utilises meta-paths. \textcolor{red}{They usually are composed from paths in a network.}
    \item (M) Random Walk (with Restart) (RW): The approach utilises Random Walk or Random Walk with Restart.
    \item (M) Advanced machine learning (AML): The approach utilises some sort of advanced machine learning component in its core such as a neural network. Utilisation of established embedding methods which themselves use neural networks (e.g. BERT) are not considered in this dimension. \textcolor{red}{We do not consider traditional and simple ML techniques such as k means in this dimension but rather mention methods explicitly defining a loss function, using multi-layer perceptrons or GCNs.}
    \item (M) Crawling (crawl.): The approach conducts some sort of web crawling step.
    \item (M) Cosine similarity (cosine): The approach utilis\-es cosine similarity at some point.
\end{itemize}

Of the observed paper recommendation systems, six were general systems or methods which were only applied on the domain of paper recommendation~\cite{ahmedi,DBLP:conf/icmla/AlfarhoodC19,DBLP:journals/corr/abs-2107-07831,DBLP:conf/seke/LiuKCQ19,DBLP:conf/ksem/YuHLZXLXY19,ZHANG2019616}. Two were targeting explicit set-based recommendation of publications where only all papers in the set together satisfy users' information needs~\cite{DBLP:conf/seke/LiuKCQ19,DBLP:journals/complexity/LiuKYQ20}, two recommend multiple papers~\cite{DBLP:conf/wise/HuaCLZZ20,nair} (e.g. on a path~\cite{DBLP:conf/wise/HuaCLZZ20}), all the other approaches focused on recommendation of $k$ single papers. Only two approaches focus on recommendation of papers to user groups instead of single users~\cite{DBLP:journals/asc/WangWYXYY21,wang21}.
Only one paper~\cite{DBLP:journals/tois/LiCPR19} supports subscription-based recommendation of papers, all other approaches solely regarded a scenario in which papers were suggested straight away.

Table~\ref{tab:dimensions_cat} classifies the observed approaches according to the afore discussed dimensions.

\begin{table*}[t]
    \centering
    \scriptsize
    \begin{tabu}{l|[1.5pt]l|l|[1.5pt]l|l|l|l|l|l|[1.5pt]l|l|l|l|l|l|l|l|l|l|l|l}
    &  \multicolumn{2}{c|[1.5pt]}{\textbf{General}} &  \multicolumn{6}{c|[1.5pt]}{\textbf{Data}} & \multicolumn{12}{c}{\textbf{Methods}} \\ \hline
         \rotatebox[origin=c]{90}{Work} &  \rotatebox[origin=c]{90}{Person.} & \rotatebox[origin=c]{90}{Input} &
         \rotatebox[origin=c]{90}{Title} & \rotatebox[origin=c]{90}{Abs.} & \rotatebox[origin=c]{90}{Key.} & \rotatebox[origin=c]{90}{Text} &
         \rotatebox[origin=c]{90}{Citat.} & \rotatebox[origin=c]{90}{Inter.} &
         \rotatebox[origin=c]{90}{User} &  \rotatebox[origin=c]{90}{Popul.} &  \rotatebox[origin=c]{90}{KP} & \rotatebox[origin=c]{90}{Emb.} & \rotatebox[origin=c]{90}{TM} & \rotatebox[origin=c]{90}{KG} & \rotatebox[origin=c]{90}{Graph}& \rotatebox[origin=c]{90}{Path}& \rotatebox[origin=c]{90}{RW} & \rotatebox[origin=c]{90}{AML} & \rotatebox[origin=c]{90}{Crawl.} & \rotatebox[origin=c]{90}{Cosine}\\\hline
        \cite{DBLP:conf/flairs/AfsarCF21} & 			$\bullet$&	u	&	$\bullet$&		&		&		&		&	$\bullet$&		$\bullet$&		&			&		&		&	$\bullet$&	$\bullet$&		&		&	$\bullet$&		&		\\
        \cite{DBLP:journals/elektrik/AhmadA20} & 				&	p	&	$\bullet$&	$\bullet$&		&		&	$\bullet$&		&			&		&		$\bullet$&		&		&		&	$\bullet$	&		&		&		&	$\bullet$&	$\bullet$\\
        \cite{ahmedi} & 			$\bullet$&	u	&		&		&		&	$\bullet$&		&	$\bullet$&		$\bullet$&		&			&		&	$\bullet$&		&		&		&		&	$\bullet$&		&		\\
        \cite{DBLP:conf/icmla/AlfarhoodC19} & 			$\bullet$&	u	&		&		&		&	$\bullet$&		&	$\bullet$&		$\bullet$&		&			&		&		&		&		&		&		&	$\bullet$&		&		\\
        \cite{DBLP:journals/kbs/AliQMAA20} & 			$\bullet$&	pu	&		&		&		&	$\bullet$&	$\bullet$&	$\bullet$&		$\bullet$&		&			&	$\bullet$&	$\bullet$&		&	$\bullet$&		&		&	$\bullet$&		&	$\bullet$\\
        \cite{Bereczki1587420} & 			$\bullet$&	u	&		&		&		&	$\bullet$&		&	$\bullet$&			$\bullet$	&	&			&	$\bullet$&		&		&	$\bullet$&		&		&	$\bullet$&		&		\\
        \cite{Bulut2018APR} & 			$\bullet$&	u	&	$\bullet$&		&	$\bullet$&		&	$\bullet$&	$\bullet$&		$\bullet$&	$\bullet$&			&		&		&		&		&		&		&		&		&		\\
        \cite{Bulut2020} & 			$\bullet$&	u	&	$\bullet$&	$\bullet$&	$\bullet$&		&		&		&		$\bullet$&		&			&	$\bullet$&		&		&		&		&		&		&		&	$\bullet$\\
        \cite{DBLP:conf/tencon/ChaudhuriSS19} & 				&	k	&		&		&	$\bullet$&	$\bullet$&	$\bullet$&		&			&	$\bullet$&			&		&		&		&		&		&		&		&		&		\\
        \cite{DBLP:journals/corr/abs-2107-07831} & 			$\bullet$&	k	&		&		&	$\bullet$&	$\bullet$&	$\bullet$&		&		$\bullet$&	$\bullet$&			&		&		&		&		&		&		&		&		&		\\
        \cite{DBLP:journals/jodl/ChaudhuriSSS21} & 			$\bullet$&	ku	&	$\bullet$&		&		&		&		&	$\bullet$&		$\bullet$&		&			&	$\bullet$&	$\bullet$&		&		&		&		&	$\bullet$&		&		\\
        \cite{chenban} & 			&	u	&		&		&		&	$\bullet$&		&	$\bullet$&		$\bullet$&		&			&		&	$\bullet$&		&		&		&		&		&		&		\\
        \cite{DBLP:conf/jcdl/CollinsB19} & 				&	p	&	$\bullet$&	$\bullet$&		&		&		&		&			&		&		$\bullet$&	$\bullet$&		&		&		&		&		&		&		&	$\bullet$\\
        \cite{DBLP:conf/aiccsa/DuGWHZG20} & 				&	p	&		&		&		&	$\bullet$&	$\bullet$&		&			&		&			&	$\bullet$&		&		&	$\bullet$&		&	$\bullet$	&	$\bullet$&		&	$\bullet$\\
        \cite{DBLP:journals/tkde/DuTD21} & 			$\bullet$&	pu	&		&		&		&	$\bullet$&		&	$\bullet$&		$\bullet$&		&			&	$\bullet$&		&		&		&		&		&	$\bullet$&		&	$\bullet$\\
        \cite{DBLP:conf/aaai/GuoCZLDH20} & 			$\bullet$&	u	&	$\bullet$&	$\bullet$&		&		&		&	$\bullet$&		$\bullet$&		&			&	$\bullet$&		&		&		&		&		&	$\bullet$&		&		\\
        \cite{DBLP:journals/scientometrics/HabibA19} & 				&	p	&		&		&		&	$\bullet$&	$\bullet$&		&			&		&			&		&		&		&		&		&		&		&	$\bullet$&		\\
        \cite{DBLP:journals/scientometrics/HarunaIQKHMC20} & 				&	p	&	$\bullet$&	$\bullet$&		&		&	$\bullet$&		&			&		&			&		&		&		&		&		&		&		&	$\bullet$&	$\bullet$\\
        \cite{DBLP:conf/ifip12/HuMLH20} & 			$\bullet$&	u	&		&		&		&		&	$\bullet$&	$\bullet$&		$\bullet$&	$\bullet$&			&	$\bullet$&		&		&	$\bullet$&		&		&	$\bullet$&		&	$\bullet$\\
        \cite{DBLP:conf/wise/HuaCLZZ20} & 				&	p	&		&		&		&	$\bullet$&		&		&			&		&			&		&		&		&	$\bullet$&	$\bullet$&	$\bullet$&	$\bullet$&		&	$\bullet$\\
        \cite{DBLP:conf/ml4cs/JingY20} & 				&	p	&		&		&		&	$\bullet$&	$\bullet$&		&			&		&			&	$\bullet$&		&		&	$\bullet$&		&	$\bullet$&		&		&		\\
        \cite{kanakia19} & 				&	p	&	$\bullet$&	$\bullet$&	$\bullet$&		&	$\bullet$&		&			&		&			&	$\bullet$&		&		&	$\bullet$&		&		&		&		&		\\
        \cite{KANG20212020BDP0008} & 		&	p	&		&		&		&	$\bullet$&	$\bullet$&		&			&		&		$\bullet$&		&		&		&	$\bullet$&		&		&		&	$\bullet$&	$\bullet$\\
        \cite{DBLP:journals/tetc/KongMWLX21} & 				&	p	&		&		&		&	$\bullet$&	$\bullet$&		&			&		&			&	$\bullet$&		&		&	$\bullet$&		&	$\bullet$&	$\bullet$&		&		$\bullet$\\
        \cite{DBLP:conf/cscwd/LLP21} & 			$\bullet$&	u	&		&		&		&		&		&	$\bullet$&		$\bullet$&		&			&		&		&		&	$\bullet$&		&		&	$\bullet$&		&		\\
        \cite{DBLP:journals/tois/LiCPR19} & 		$\bullet$&	u	&	$\bullet$&	$\bullet$&	$\bullet$&		&	$\bullet$&	$\bullet$&		$\bullet$&	$\bullet$	&	&	$\bullet$&		&	$\bullet$&	$\bullet$&		&		&	$\bullet$&		&	$\bullet$\\
        \cite{li20} &			$\bullet$&	k	&	$\bullet$&	$\bullet$&	$\bullet$&		&		&	$\bullet$&		$\bullet$&	&			&	$\bullet$&		&		&	$\bullet$	&		&		&		&		&	$\bullet$\\
        \cite{DBLP:journals/dss/LiWNLL21} & 			$\bullet$&	u	&		&		&		&	$\bullet$&	$\bullet$&	$\bullet$&		$\bullet$&		&			&		&		&		&	$\bullet$&	$\bullet$&	$\bullet$&	$\bullet$&		&		\\
        \cite{lin21} & 				&	k	&		&		&	$\bullet$&		&	$\bullet$&		&			&	$\bullet$&			&		&		&		&		&		&		&		&		&	$\bullet$\\
        \cite{DBLP:conf/seke/LiuKCQ19} & 				&	k	&		&		&	$\bullet$&		&	$\bullet$&		&			&		&			&		&		&		&	$\bullet$&		&		&		&		&	$\bullet$\\
        \cite{DBLP:journals/complexity/LiuKYQ20} & 				&	k	&		&		&	$\bullet$&		&	$\bullet$&		&			&		&			&		&		&		&	$\bullet$&		&		&		&		&		\\
        \cite{DBLP:conf/cscwd/LuHCP021} & 			$\bullet$&	u	&	$\bullet$&	$\bullet$&		&		&		&	$\bullet$&		$\bullet$&		&			&	$\bullet$&		&		&		&		&		&	$\bullet$&		&		\\
        \cite{DBLP:journals/access/MaW19a} & 			$\bullet$&	u	&		&		&		&		&		&	$\bullet$&		$\bullet$&		&			&		&		&		&	$\bullet$&	$\bullet$&	$\bullet$&	$\bullet$&		&		\\
        \cite{DBLP:journals/monet/MaZZ19} & 			$\bullet$&	u	&	$\bullet$&	$\bullet$&	$\bullet$&		&		&	$\bullet$&		$\bullet$&		&			&	$\bullet$&		&		&	$\bullet$&	$\bullet$&		&	$\bullet$&		&	$\bullet$\\
        \cite{DBLP:journals/ijiit/Manju} & 			$\bullet$&	pu	&	$\bullet$&		&	$\bullet$&		&	$\bullet$&	$\bullet$&		$\bullet$&	$\bullet$&			&		&		&		&	$\bullet$&		&	$\bullet$&		&		&		\\
        \cite{MohamedHassan} & 			$\bullet$&	u	&	$\bullet$&	$\bullet$&		&		&		&	$\bullet$&		$\bullet$&		&			&		&		&		&		&		&		&	$\bullet$&		&		\\
        \cite{nair} & 			$\bullet$&	p	&	$\bullet$&		&		&		&		&		&			&		&			&	$\bullet$&		&		&		&		&		&	$\bullet$&		&		\\
        \cite{DBLP:conf/icadl/NishiokaHS19,DBLP:conf/ercimdl/NishiokaHS19,DBLP:journals/peerj-cs/NishiokaHS20} & 			$\bullet$&	u	&		&		&		&	$\bullet$&		&		&		$\bullet$&		&			&		&		&		&		&		&		&		&		&	$\bullet$\\
        \cite{DBLP:conf/iui/RahdariB19} & 			$\bullet$&	u	&	$\bullet$&	$\bullet$&	$\bullet$&		&		&	$\bullet$&		$\bullet$&	$\bullet$&			&		&		&		&		&		&		&		&		&	$\bullet$\\
        \cite{renuka} & 				&	p	&		&		&	$\bullet$&	$\bullet$&		&		&			&		&		$\bullet$&		&		&		&		&		&		&		&		&	$\bullet$\\
        \cite{DBLP:journals/access/SakibAH20} & 				&	p	&		&		&		&		&	$\bullet$&		&			&		&			&		&		&		&	$\bullet$	&		&		&		&		&		\\
        \cite{DBLP:journals/access/SakibAABHHG21} & 				&	p	&	$\bullet$&	$\bullet$&	$\bullet$&		&	$\bullet$&		&			&		&			&		&		&		&	$\bullet$&		&		&		&	$\bullet$&	$\bullet$\\
        \cite{DBLP:journals/peerj-cs/ShahidAAAA21,shahid21a} & 				&	p	&		&		&		&	$\bullet$&	$\bullet$&		&			&		&			&		&		&		&		&		&		&		&	$\bullet$&		\\
        \cite{Sharma} & 				&	p	&		&		&		&	$\bullet$&		&		&			&		&			&		&		&	$\bullet$&	$\bullet$	&	&		&		&		&	$\bullet$\\
        \cite{DBLP:conf/ijcnn/ShiMZJLC20} & 	&	p	&	$\bullet$&	$\bullet$&		&		&	$\bullet$&		&			&	$\bullet$&			&		&		&		&	$\bullet$&		&		&	$\bullet$&		&		\\
        \cite{subathra} & 				&	k	&		&		&		&	$\bullet$&		&		&			&		&			&		&	$\bullet$&		&		&		&		&		&	$\bullet$&	$\bullet$\\
        \cite{DBLP:journals/concurrency/TangLQ21} & 			$\bullet$&	u	&	$\bullet$&	$\bullet$&	$\bullet$&		&	$\bullet$&	$\bullet$&		$\bullet$&		&			&	$\bullet$&	&	$\bullet$&	$\bullet$&		&		&	$\bullet$&		&		\\
        \cite{DBLP:conf/bigdataconf/TannerAH19} & 				&	p	&		&		&		&	$\bullet$&	$\bullet$&		&			&		&			&		&		&		&	$\bullet$&		&	$\bullet$&		&		&		\\
        \cite{tao} & 				&	p	&		&		&		&	$\bullet$&	$\bullet$&		&			&		&			&	$\bullet$&	$\bullet$&		&	$\bullet$&		&	$\bullet$&		&		&	$\bullet$\\
        \cite{DBLP:journals/access/WaheedIRMK19} & 				&	p	&		&		&		&		&	$\bullet$&		&			&	$\bullet$&			&		&		&		&	$\bullet$&		&		&		&		&		\\
        \cite{DBLP:conf/service/WangXTWX20} & 			$\bullet$&	u	&		&		&	$\bullet$&	$\bullet$&	$\bullet$&	$\bullet$&		$\bullet$&		&			&	$\bullet$&		&	$\bullet$&	$\bullet$	&	&		&	$\bullet$&		&	$\bullet$\\
        \cite{DBLP:journals/corr/abs-2103-08819} & 			$\bullet$&	ko	&	$\bullet$&	$\bullet$&	$\bullet$&		&		&		&			&		&			&	$\bullet$&		&	$\bullet$&	$\bullet$&		&		&		&	$\bullet$&	$\bullet$\\
        \cite{DBLP:journals/asc/WangWYXYY21} & 			$\bullet$&	u	&		&		&		&	$\bullet$&		&	$\bullet$&		$\bullet$&		&			&		&		&		&		&		&		&	$\bullet$&	$\bullet$&	$\bullet$\\
        \cite{wang21} & 			$\bullet$&	u	&		&		&	$\bullet$&		&		&	$\bullet$&		$\bullet$&		&			&		&		&		&		&		&		&	$\bullet$&		&	$\bullet$\\
        \cite{DBLP:conf/sigir/XieSB21} & 			$\bullet$&	ku	&		&		&	$\bullet$&	$\bullet$&	$\bullet$&	$\bullet$&		$\bullet$&	$\bullet$&			&	$\bullet$&		&		&		&		&		&	$\bullet$&		&		\\
        \cite{xie21} & 				&	pk	&	$\bullet$&	$\bullet$&	$\bullet$&		&	$\bullet$&		&			&		&			&		&	$\bullet$&		&		&		&		&	$\bullet$&		& \\
        \cite{DBLP:journals/www/YangLLLZZZZ19} & 				&	k	&	$\bullet$&	$\bullet$&		&		&	$\bullet$&		&			&	$\bullet$&			&		&	$\bullet$&		&	$\bullet$&		&	$\bullet$&		&		&	$\bullet$\\
        \cite{DBLP:conf/ksem/YuHLZXLXY19} & 			$\bullet$&	u	&		&		&		&		&		&	$\bullet$&		$\bullet$&		&			&	$\bullet$&		&		&		&		&		&	$\bullet$&		&		\\
        \cite{zarvel} & 	$\bullet$ & o & & & $\bullet$ & & $\bullet$ & $\bullet$ & $\bullet$ & $\bullet$ &  & $\bullet$ & & & & & & &\\

        \cite{ZHANG2019616} &				&	?	&	$\bullet$&	$\bullet$&		&		&	$\bullet$&		&			&		&			&		&		&		&	$\bullet$&		&	$\bullet$&		&		&	$\bullet$\\
        \cite{zhang20} & 				&	?	&	$\bullet$&	$\bullet$&		&		&	$\bullet$&		&			&		&			&		&		&		&	$\bullet$&		&		$\bullet$	&		&		&	$\bullet$\\
        \cite{DBLP:journals/access/ZhaoKFMN20} & 			$\bullet$&	u	&	$\bullet$&	$\bullet$&		&		&		&	$\bullet$&		$\bullet$&		&			&	$\bullet$&		&		&		&		&		&	$\bullet$&		&		\\
    \end{tabu}
    \caption{Indications whether works utilise the specific data or methods. Papers describing the same approach without extension of the methodology (e.g. only describing more details or an evaluation) are regarded in combination with each other.}
    \label{tab:dimensions_cat}
\end{table*}

\subsection{Comparison of Paper Recommendation Systems in different Categories}

In this Section, we describe the scientific directions associated with the categories we presented in the previous section as the 65 relevant publications. We focus only on the methodological categories and describe how they are incorporated in the respective approaches.

\subsubsection{User Profile}

32~ approaches~ construct~ explicit~ user~ profiles.~ They utilise different components to describe users. We differentiate~ between~ profiles~ derived~ from~ user~ interac\-tions and ones derived from papers.

Most user profiles are constructed from \textit{users' actual interactions}:
unspecified historical interaction~\cite{DBLP:journals/tkde/DuTD21,DBLP:conf/aaai/GuoCZLDH20,DBLP:journals/tois/LiCPR19,DBLP:journals/dss/LiWNLL21,DBLP:journals/monet/MaZZ19,DBLP:conf/ksem/YuHLZXLXY19}, the mean of the representation of interacted with papers~\cite{Bereczki1587420}, time decayed interaction behaviour~\cite{DBLP:conf/cscwd/LuHCP021},~ liked~ papers~\cite{MohamedHassan,DBLP:journals/access/ZhaoKFMN20},~ bookmarked~ papers~\cite{DBLP:conf/iui/RahdariB19,zarvel}, read papers~\cite{wang21,DBLP:conf/service/WangXTWX20}, rated papers~\cite{ahmedi,DBLP:conf/icmla/AlfarhoodC19,DBLP:journals/asc/WangWYXYY21}, clicked on papers~\cite{DBLP:journals/corr/abs-2107-07831,DBLP:journals/jodl/ChaudhuriSSS21,DBLP:conf/cscwd/LLP21}, categories of clicked papers~\cite{DBLP:conf/flairs/AfsarCF21}, features of clicked papers~\cite{DBLP:journals/concurrency/TangLQ21}, tweets~\cite{DBLP:conf/icadl/NishiokaHS19,DBLP:conf/ercimdl/NishiokaHS19,DBLP:journals/peerj-cs/NishiokaHS20}, social interactions~\cite{DBLP:journals/ijiit/Manju} and explicitly defined topics of interest tags~\cite{zarvel}.

Some approaches derived user profiles from \textit{users' written papers}:
authored papers~\cite{DBLP:journals/kbs/AliQMAA20,Bulut2020,Bulut2018APR,li20,DBLP:journals/access/MaW19a,DBLP:conf/icadl/NishiokaHS19,DBLP:conf/ercimdl/NishiokaHS19,DBLP:journals/peerj-cs/NishiokaHS20,xie21}, a partitioning of authored papers~\cite{chenban}, research fields~ of~ authored~ papers~\cite{DBLP:conf/ifip12/HuMLH20}~ and~ referenced~ papers~\cite{xie21}.

\subsubsection{Popularity}

We found 13 papers using some type of popularity measure. Those can be defined on authors, venues or papers.

For \textit{author-based popularity} measures we found unspecified ones~\cite{DBLP:journals/ijiit/Manju} such as authority~\cite{xie21} as well as ones regarding the citations an author received: citation count of papers~\cite{Bulut2018APR,DBLP:conf/ijcnn/ShiMZJLC20,DBLP:journals/access/WaheedIRMK19,zarvel}, change in citation count~\cite{DBLP:conf/tencon/ChaudhuriSS19,DBLP:journals/jodl/ChaudhuriSSS21}, annual citation count~\cite{DBLP:journals/jodl/ChaudhuriSSS21}, number of citations related to papers~\cite{lin21}, h-index~\cite{DBLP:journals/jodl/ChaudhuriSSS21}. We found two definitions of author's popularity using the graph structure of scholarly networks, namely the number of co-authors~\cite{DBLP:conf/ifip12/HuMLH20} and a person's centrality~\cite{DBLP:journals/access/WaheedIRMK19}.

For \textit{venue-based popularity} measures, we found an unspecific reputation notion~\cite{xie21} as well as incorporation of the impact factor~\cite{DBLP:journals/jodl/ChaudhuriSSS21,DBLP:journals/www/YangLLLZZZZ19}.

For \textit{paper-based popularity} measures we encountered some citation-based definitions such as vitality~\cite{DBLP:journals/www/YangLLLZZZZ19}, citation count of papers~\cite{Bulut2018APR} and theirs centrality~\cite{DBLP:conf/ijcnn/ShiMZJLC20} in the citation network. Additionally, some approaches incorporated less formal interactions: number of downloads~\cite{DBLP:journals/tois/LiCPR19}, social media mentions~\cite{zarvel} and normalised number of bookmarks~\cite{DBLP:conf/iui/RahdariB19}.

\subsubsection{Key Phrase}

Only four papers use key phrases in some shape or form: Ahmad and Afzal.~\cite{DBLP:journals/elektrik/AhmadA20} construct key terms from preprocessed titles and abstracts using tf-idf to represent papers.
Collins and Beel~\cite{DBLP:conf/jcdl/CollinsB19} use the Distiller Framework~\cite{distiller} to extract uni-, bi- and tri-gram key phrase candidates from tokenised, part-of-speech tagged and stemmed titles and abstracts. Key phrase candidates were weighted and the top 20 represent candidate papers.
Kang et al.~\cite{KANG20212020BDP0008} extract key phrases from CiteSeer to describe the diversity of recommended papers.
%do not further describe how they crawled key phrases to represent paper content, on which similarity is computed.
Renuka et al.~\cite{renuka} apply rapid automatic keyword extraction.

In summary, different length key phrases usually get constructed from titles and abstracts with automatic methods such as tf-idf or the Distiller Framework to represent the most important content of publications.

\subsubsection{Embedding}

We found a lot of approaches utilising some form of embedding based on existing document representation methods. We distinguish by embedding of papers, users and papers and sophisticated embedding from the proposed approaches.

Among the most common methods was their application on \textit{papers}:
in an unspecified representation~\cite{DBLP:journals/tkde/DuTD21,zarvel},~ Word2Vec~\cite{Bereczki1587420,DBLP:conf/aaai/GuoCZLDH20,DBLP:conf/ml4cs/JingY20,kanakia19,li20,DBLP:journals/concurrency/TangLQ21,DBLP:conf/service/WangXTWX20},~ Word2Vec of LDA top words~\cite{DBLP:journals/corr/abs-2107-07831,tao}, 
Doc2vec~\cite{Bulut2020,DBLP:conf/jcdl/CollinsB19,DBLP:journals/tetc/KongMWLX21,DBLP:conf/cscwd/LuHCP021,DBLP:journals/access/MaW19a,tao}, Doc2Vec of word pairs~\cite{DBLP:journals/corr/abs-2103-08819}, BERT~\cite{DBLP:journals/access/ZhaoKFMN20} and SBERT~\cite{Bereczki1587420,DBLP:journals/kbs/AliQMAA20}.
Most times these approaches do not mention which part of the paper to use as input but some specifically mention the following parts: titles~\cite{DBLP:conf/aaai/GuoCZLDH20}, titles and abstracts~\cite{DBLP:conf/jcdl/CollinsB19,kanakia19}, titles, abstracts and bodies~\cite{DBLP:journals/tetc/KongMWLX21}, keywords and paper~\cite{zarvel}.

Few approaches observed \textit{user profiles and papers}, here  Word2Vec~\cite{Bulut2020} and NPLM~\cite{DBLP:conf/aiccsa/DuGWHZG20} embeddings were used.

Several approaches embed the information in their own~ model~ embedding:~ a~ heterogeneous~ information network~\cite{DBLP:journals/kbs/AliQMAA20}, a two-layer NN~\cite{DBLP:conf/aaai/GuoCZLDH20}, a scientific social reference network~\cite{DBLP:conf/ifip12/HuMLH20}, the TransE model~\cite{DBLP:journals/tois/LiCPR19}, node embeddings~\cite{DBLP:journals/access/MaW19a}, paper, author and venue embedding~\cite{xie21}, user and item embedding~\cite{DBLP:conf/ksem/YuHLZXLXY19}, a GRU and association rule mining model~\cite{nair}, a GCN embedding of users~\cite{DBLP:journals/concurrency/TangLQ21} and an LSTM model~\cite{DBLP:conf/service/WangXTWX20}.

\subsubsection{Topic Model}

Eight approaches use some topic modelling component. Most of them use LDA to represent papers' content~\cite{ahmedi,DBLP:journals/kbs/AliQMAA20,DBLP:journals/corr/abs-2107-07831,chenban,tao,DBLP:journals/www/YangLLLZZZZ19}. Only two of them do not follow this method: Subathra and Kumar~\cite{subathra} use LDA on papers to find their top $n$ words, then they use LDA again on these words' Wikipedia articles. Xie et al.~\cite{DBLP:conf/sigir/XieSB21} use a hierarchical LDA adoption on papers, which introduces a discipline classification.

\subsubsection{Knowledge Graph}
\label{class:kg}

Only six of the observed papers incorporate knowledge graphs. Only one uses a predefined one, the Watson for Genomics knowledge graph~\cite{Sharma}.
Most of the approaches build their own knowledge graphs, only one \textit{asks users to construct} the graphs: Wang et al.~\cite{DBLP:journals/corr/abs-2103-08819} build two knowledge graphs, one in-domain and one cross-domain graph. The graphs are user-constructed and include representative papers for the different concepts.

All other approaches \textit{do not rely on users} building the knowledge graph:
Afsar et al.~\cite{DBLP:conf/flairs/AfsarCF21} utilise an expert-built knowledge base as a source for their categorisation of papers, which are then recommended to users. 
Li et al.~\cite{DBLP:journals/tois/LiCPR19} employ a knowledge graph-based embedding of authors, keywords and venues.
Tang et al.~\cite{DBLP:journals/concurrency/TangLQ21} link words with high tf-idf weights from papers to LOD and then merge this knowledge graph with the user-paper graph.
Wang et al.~\cite{DBLP:conf/service/WangXTWX20} construct a knowledge graph consisting of users and papers.

\subsubsection{Graph}

In terms of graphs, we found 33 approaches explicitly mentioning the graph structure they were utilising. We can describe which graph structure is used and which algorithms or methods are applied on the graphs. 

Of the observed approaches, most specify some form of (heterogeneous) \textit{graph structure}. Only a few of them are unspecific and mention an undefined heterogeneous graph~\cite{DBLP:journals/access/MaW19a,DBLP:journals/monet/MaZZ19,DBLP:journals/ijiit/Manju} or a multi-layer~\cite{DBLP:journals/tetc/KongMWLX21} graph. Most works clearly define the type of graph they are using: author-paper-venue-label-topic graph~\cite{DBLP:journals/kbs/AliQMAA20}, author-paper-venue-keyword graph~\cite{DBLP:journals/tois/LiCPR19,DBLP:journals/dss/LiWNLL21}, paper-author graph~\cite{Bereczki1587420,DBLP:conf/aiccsa/DuGWHZG20,li20,DBLP:journals/concurrency/TangLQ21},~~ paper-topic~~ graph~\cite{DBLP:conf/aiccsa/DuGWHZG20},~~ author-paper-venue graph~\cite{DBLP:conf/wise/HuaCLZZ20,ZHANG2019616,zhang20},~ author~ graph~\cite{DBLP:conf/ifip12/HuMLH20},~ paper-paper graph~\cite{DBLP:conf/wise/HuaCLZZ20,DBLP:conf/cscwd/LLP21},~ citation~ graph~\cite{DBLP:journals/elektrik/AhmadA20,DBLP:conf/ml4cs/JingY20,kanakia19,KANG20212020BDP0008,DBLP:journals/access/SakibAABHHG21,DBLP:journals/access/SakibAH20,DBLP:journals/access/WaheedIRMK19,DBLP:conf/bigdataconf/TannerAH19,DBLP:journals/www/YangLLLZZZZ19} or undirected citation graph~\cite{DBLP:conf/seke/LiuKCQ19,DBLP:journals/complexity/LiuKYQ20}.
Some approaches specifically mention usage of co-citations~\cite{DBLP:journals/elektrik/AhmadA20,kanakia19}, bibliographic coupling or both~\cite{DBLP:journals/access/SakibAABHHG21,DBLP:journals/access/SakibAH20,DBLP:conf/ijcnn/ShiMZJLC20,DBLP:journals/access/WaheedIRMK19}.

As for \textit{algorithms or methods used on these graphs}, we encountered usage of centrality measures in different graph types~\cite{DBLP:conf/ifip12/HuMLH20,DBLP:conf/ijcnn/ShiMZJLC20,DBLP:journals/access/WaheedIRMK19}, some use knowledge graphs (see Section~\ref{class:kg}), some using meta-paths (see Section~\ref{class:mp}), some using random walks e.g. in form of PageRank or hubs and authorities (see Section~\ref{class:rw}), construction of Steiner trees~\cite{DBLP:journals/complexity/LiuKYQ20}, usage of the graph as input for a GCN~\cite{DBLP:journals/concurrency/TangLQ21}, BFS~\cite{DBLP:conf/service/WangXTWX20}, clustering~\cite{DBLP:journals/www/YangLLLZZZZ19} or calculation of a closeness degree~\cite{DBLP:journals/www/YangLLLZZZZ19}.

\subsubsection{Meta-Path}
\label{class:mp}

We found only four approaches incorporating meta-paths.
Hua et al.~\cite{DBLP:conf/wise/HuaCLZZ20} construct author-paper-author and author-paper-venue-paper-author paths by applying beam search. Papers on the most similar paths are recommended to users.
Li et al.~\cite{DBLP:journals/dss/LiWNLL21} construct meta-paths of a max length between users and papers and use random walk on these paths.
Ma et al.~\cite{DBLP:journals/access/MaW19a,DBLP:journals/monet/MaZZ19} use meta-paths to measure the proximity between nodes in a graph.

\subsubsection{Random Walk (with Restart)}
\label{class:rw}

We found twelve approaches using some form of random walk in their methodology. We differentiate between ones using random walk, random walk with restart and algorithms using a random walk component.

Some methods use \textit{random walk} on heterogeneous graphs~\cite{DBLP:conf/aiccsa/DuGWHZG20,DBLP:journals/ijiit/Manju} and weighted multi-layer graphs~\cite{DBLP:journals/tetc/KongMWLX21}.
A few approaches use random walk to identify~\cite{DBLP:conf/wise/HuaCLZZ20,DBLP:journals/dss/LiWNLL21} or determine the proximity between~\cite{DBLP:journals/monet/MaZZ19} meta-paths.

Three approaches explicitly utilise \textit{random walk with restart}. They determine similarity between papers~\cite{DBLP:conf/bigdataconf/TannerAH19}, identify papers to recommend~\cite{DBLP:conf/ml4cs/JingY20} or find most relevant papers in clusters~\cite{DBLP:journals/www/YangLLLZZZZ19}.

Some~ approaches~ use~ algorithms~ which~ \textit{incorpo\-rate a random walk component}: PageRank~\cite{tao} and the identifications of hubs and authorities~\cite{zhang20} with PageRank~\cite{ZHANG2019616}.

\subsubsection{Advanced Machine Learning}

29 approaches utilised some form of advanced machine learning. We encountered different methods being used and some papers specifically presenting novel machine learning models. All of these papers surpass mere usage of a topic model or typical pre-trained embedding method.

We found a multitude of \textit{machine learning methods} being used, from multi armed bandits~\cite{DBLP:conf/flairs/AfsarCF21}, LSTM~\cite{DBLP:journals/corr/abs-2107-07831,DBLP:conf/aaai/GuoCZLDH20,DBLP:conf/service/WangXTWX20}, multi-layer perceptrons~\cite{DBLP:conf/cscwd/LuHCP021,DBLP:conf/ijcnn/ShiMZJLC20,DBLP:journals/concurrency/TangLQ21}, (bi-)GRU~\cite{DBLP:conf/aaai/GuoCZLDH20,MohamedHassan,nair,DBLP:journals/access/ZhaoKFMN20}, matrix factorisation~\cite{DBLP:conf/icmla/AlfarhoodC19,DBLP:conf/cscwd/LuHCP021,MohamedHassan,DBLP:journals/asc/WangWYXYY21,wang21}, gradient ascent or descent~\cite{DBLP:conf/ifip12/HuMLH20,DBLP:journals/dss/LiWNLL21,DBLP:journals/access/MaW19a,xie21}, some form of simple neural network~\cite{DBLP:journals/tois/LiCPR19,DBLP:conf/aaai/GuoCZLDH20,DBLP:journals/tkde/DuTD21}, some form of graph neural network~\cite{DBLP:journals/concurrency/TangLQ21,DBLP:conf/cscwd/LLP21,Bereczki1587420}, autoencoder~\cite{DBLP:conf/icmla/AlfarhoodC19}, neural collaborative filtering~\cite{DBLP:conf/cscwd/LuHCP021}, learning methods~\cite{DBLP:journals/access/ZhaoKFMN20,DBLP:journals/tkde/DuTD21} to DTW~\cite{DBLP:journals/tetc/KongMWLX21}.
Three approaches ranked the papers to recommend~\cite{DBLP:journals/tois/LiCPR19,DBLP:journals/dss/LiWNLL21,DBLP:conf/ksem/YuHLZXLXY19} with e.g. Bayesian Personalized Ranking. Two of the observed papers proposed topic modelling approaches~\cite{DBLP:conf/sigir/XieSB21,ahmedi}. 

Several papers proposed \textit{models}: a bipartite network embedding~\cite{DBLP:journals/kbs/AliQMAA20}, heterogeneous graph embeddings~\cite{DBLP:conf/aiccsa/DuGWHZG20,DBLP:journals/access/MaW19a,DBLP:conf/wise/HuaCLZZ20,DBLP:journals/tetc/KongMWLX21}, a scientific social reference network~\cite{DBLP:conf/ifip12/HuMLH20}, a paper-author-venue embedding~\cite{xie21} and a relation prediction model~\cite{DBLP:journals/monet/MaZZ19}.

\subsubsection{Crawling}

We found nine papers incorporating a crawling step as part of their approach. PDFs are oftentimes collected from CiteSeer~\cite{DBLP:journals/scientometrics/HabibA19,KANG20212020BDP0008} or CiteSeerX~\cite{DBLP:journals/elektrik/AhmadA20,shahid21a,DBLP:journals/peerj-cs/ShahidAAAA21}, in some cases~\cite{DBLP:journals/scientometrics/HarunaIQKHMC20,DBLP:journals/access/SakibAABHHG21,DBLP:journals/asc/WangWYXYY21} the sources are not explicitly mentioned. Fewer used data sources are Wikipedia for articles explaining the top words from papers~\cite{subathra} or papers from ACM, IEEE and EI~\cite{DBLP:journals/corr/abs-2103-08819}. 
Some approaches explicitly mention the extraction of citation information~\cite{DBLP:journals/elektrik/AhmadA20,DBLP:journals/scientometrics/HabibA19,DBLP:journals/scientometrics/HarunaIQKHMC20,KANG20212020BDP0008,DBLP:journals/access/SakibAABHHG21,shahid21a,DBLP:journals/peerj-cs/ShahidAAAA21} e.g. to identify co-citations.

\subsubsection{Cosine Similarity}

Some form of cosine similarity was encountered in most (31) paper recommendation approaches. It is often applied between papers, between users, between users and papers and in other forms.

For application \textit{between papers} we encountered the possibility of using unspecified embeddings: \textit{unspecified word or vector representations} of papers~\cite{DBLP:journals/tkde/DuTD21,DBLP:journals/tetc/KongMWLX21,DBLP:journals/asc/WangWYXYY21,tao}, papers' key terms or top words~\cite{DBLP:journals/elektrik/AhmadA20,subathra} and key phrases~\cite{KANG20212020BDP0008}. We found some approaches using \textit{vector space model} variants: unspecified~\cite{lin21}, tf vectors~\cite{DBLP:journals/scientometrics/HarunaIQKHMC20,DBLP:journals/access/SakibAABHHG21}, tf-idf vectors~\cite{DBLP:conf/wise/HuaCLZZ20,Sharma,wang21}, dimensionality reduced tf-idf vectors~\cite{renuka} and lastly, tf-idf and entity embeddings~\cite{DBLP:journals/tois/LiCPR19}.
Some approaches incorporated more advanced embedding techniques: SBERT embeddings~\cite{DBLP:journals/kbs/AliQMAA20}, Doc2Vec embeddings~\cite{DBLP:conf/jcdl/CollinsB19}, Doc2Vec embeddings with incorporation of their emotional score~\cite{DBLP:journals/corr/abs-2103-08819} and NPLM representations~\cite{DBLP:conf/aiccsa/DuGWHZG20}.

Cosine similarity was used \textit{between preferences or profiles of users and papers} in the following ways: unspecified representations~\cite{DBLP:journals/access/MaW19a,DBLP:conf/service/WangXTWX20,DBLP:conf/sigir/XieSB21,DBLP:conf/iui/RahdariB19}, Boolean representation of users and keywords~\cite{DBLP:conf/seke/LiuKCQ19}, tf-idf vectors~\cite{Bulut2020,DBLP:conf/icadl/NishiokaHS19,DBLP:journals/peerj-cs/NishiokaHS20,DBLP:conf/ercimdl/NishiokaHS19},~ cf-idf~ vectors~\cite{DBLP:conf/icadl/NishiokaHS19,DBLP:journals/peerj-cs/NishiokaHS20,DBLP:conf/ercimdl/NishiokaHS19}~ and~ hcf-idf vectors~\cite{DBLP:conf/icadl/NishiokaHS19,DBLP:journals/peerj-cs/NishiokaHS20,DBLP:conf/ercimdl/NishiokaHS19}.

For \textit{between users} application of cosine similarity, we found unspecified representations~\cite{DBLP:conf/ifip12/HuMLH20} and time-decayed Word2Vec embeddings of users' papers' keyword~\cite{li20}.

\textit{Other} applications include the usage between input keywords and paper clusters~\cite{DBLP:journals/www/YangLLLZZZZ19} and between nodes in a graph represented by their neighbouring nodes~\cite{ZHANG2019616,zhang20}.

\subsection{Paper Recommendation Systems}
\label{sec:litReview_prs}

The \textcolor{red}{65} relevant works identified in our literature search are described in this section. We deliberately refrain from trying to structure the section by classifying papers by an arbitrary dimension and instead point to Table~\ref{tab:dimensions_cat} to identify those dimensions in which a reader is interested to navigate the following short descriptions.
The works are ordered by the surname of the first author and ascending publication year. An exception to this rule are papers presenting extensions of previous approaches with different first authors. These papers are ordered to their preceding approaches.

Afsar et al.~\cite{DBLP:conf/flairs/AfsarCF21} propose KERS, a multi armed bandit approach for patients to help with medical treatment decision making. It consists of two phases: first an exploration phase identifies categories users are implicitly interested in. This is supported by an expert-built knowledge base. Afterwards an exploitation phase takes place where articles from these categories are recommended until a user's focus changes and another exploitation phase is initiated. The authors strive to minimise the exploration efforts while maximising users' satisfaction.

Ahmedi et al.~\cite{ahmedi} propose a personalised approach which can also be applied to more general recommendation scenarios which include user profiles. They utilise Collaborative~ Topic~ Regression~ to~ mine~ association rules from historic user interaction data.

Alfarhood and Cheng~\cite{DBLP:conf/icmla/AlfarhoodC19} introduce Collaborative Attentive Autoencoder, a deep learning-based model for general recommendation targeting the data sparsity problem. They apply probabilistic matrix factorisation while also utilising textual information to train a model which identifies latent factors in users and papers.

Ali et al.~\cite{DBLP:journals/kbs/AliQMAA20}~ construct~ PR-HNE,~ a~ personalised probabilistic paper recommendation model based on a joint representation of authors and publications. They utili\-se graph information such as citations as well as co-author\-ships, venue information and topical relevance to suggest papers. They apply SBERT and LDA to represent author embeddings and topic embeddings respectively.

Bereczki~\cite{Bereczki1587420} models users and papers in a bipartite graph. Papers are represented by their contents' Word2Vec or BERT embeddings, users' vectors consist of representations of papers they interacted with. These vectors are then aggregated with simple graph convolution.

Bulut et al.~\cite{Bulut2018APR} focus on current user interest in their approach which utilises k-Means and KNN. Users' profiles are constructed from their authored papers. Recommended papers are the highest cited ones from the cluster most similar to a user.
In a subsequent work they extended their research group to again work in the same domain. Bulut et al.~\cite{Bulut2020} again focus on users' features. They represent users as the sum of features of their papers. These representations are then compared with all papers' vector representations to find the most similar ones. Papers can be represented by TF-IDF, Word2Vec or Doc2Vec vectors.

%Chaudhari et al.~\cite{chaudhari} suggest new articles to users based on their authored publications. 

Chaudhuri et al.~\cite{DBLP:conf/tencon/ChaudhuriSS19} use indirect features derived from direct features of papers in addition to direct ones in their paper recommendation approach: keyword diversification, text complexity and citation analysis.
In an extended group Chaudhuri et al.~\cite{DBLP:journals/jodl/ChaudhuriSSS21} later propose usage of more indirect features such as quality in paper recommendation. Users' profiles are composed of their clicked papers.
Subsequently they again worked on an approach in the same area but in a slightly smaller group. Chaudhuri et al.~\cite{DBLP:journals/corr/abs-2107-07831} propose the general Hybrid Topic Model and apply it on paper recommendation. It learns users' preferences and intentions by combining LDA and Word2Vec. They compute user's interest from probability distributions of words of clicked papers and dominant topics in publications.

Chen and Ban~\cite{chenban} introduce CPM, a recommendation model based on topically clustered user interests mined from their published papers. They derive user need models from these clusters by using LDA and pattern equivalence class mining. Candidate papers are then ranked against the user need models to identify the best-fitting suggestions.

Collins and Beel~\cite{DBLP:conf/jcdl/CollinsB19} propose the usage of their paper recommendation system Mr. DLib as a recommender as-a-service. They compare representing papers via Doc2Vec with a key phrase-based recommender and TF-IDF vectors.

Du et al.~\cite{DBLP:conf/aiccsa/DuGWHZG20} introduce HNPR, a heterogeneous network method using two different graphs. The approach incorporates citation information, co-author relations and research areas of publications. They apply random walk on the networks to generate vector representations of papers.

Du et al.~\cite{DBLP:journals/tkde/DuTD21} propose Polar++, a personalised active~ one-shot~ learning-based~ paper~ recommendation system where new users are presented articles to vote on before they obtain recommendations. The model trains a neural network by incorporating a matching score between a query article and the recommended articles as well as a personalisation score dependant on the user.

Guo et al.~\cite{DBLP:conf/aaai/GuoCZLDH20} recommend publications based on papers initially liked by a user. They learn semantics between titles and abstracts of papers on word- and sentence-level, e.g. with Word2Vec and LSTMs to represent user preferences.

Habib and Afzal~\cite{DBLP:journals/scientometrics/HabibA19} crawl full texts of papers from CiteSeer. They then apply bibliographic coupling between input papers and a clusters of candidate papers to identify the most relevant recommendations.
In a subsequent work Afzal again used a similar technique. Ahmad and Afzal~\cite{DBLP:journals/elektrik/AhmadA20} crawled papers from CiteSeerX. Cosine similarity of TF-IDF representations of key terms from titles and abstracts is combined with co-citation strength of paper pairs. This combined score then ranks the most relevant papers the highest.

Haruna et al.~\cite{DBLP:journals/scientometrics/HarunaIQKHMC20} incorporate paper-citation relations combined with contents of titles and abstracts of papers to recommend the most fitting publications for an input query corresponding to a paper.

Hu et al.~\cite{DBLP:conf/ifip12/HuMLH20} present ADRCR, a paper recommendation~ approach~ incorporating~ author-author~ and author-paper citation relationships as well as authors' and papers' authoritativeness. A network is built which uses citation information as weights. Matrix decomposition helps learning the model.

Hua et al.~\cite{DBLP:conf/wise/HuaCLZZ20} propose PAPR which recommends relevant paper sets as an ordered path. They strive to overcome recommendation merely based on similarity by observing topics in papers changing over time. They combine similarities of TF-IDF paper representations with random-walk on different scientific networks.

Jing and Yu~\cite{DBLP:conf/ml4cs/JingY20} build a three-layer graph model which they traverse with random-walk with restart in an algorithm named PAFRWR. The graph model consists of one layer with citations between papers' textual content represented via Word2Vec vectors, another layer modelling co-authorships between authors and the third layer encodes relationships between papers and topics contained in them.

Kanakia et al.~\cite{kanakia19} build their approach upon the MAG dataset and strive to overcome the common problems of scalability and cold-start. They combine TF-IDF and Word2Vec representations of the content with co-citations of papers to compute recommendations. Speedup is achieved by comparing papers to clusters of papers instead of all other single papers.

Kang et al.~\cite{KANG20212020BDP0008} crawl full texts of papers from CiteSeer and construct citation graphs to determine candidate papers. Then they compute a combination of section-based citation and key phrase similarity to rank recommendations.

Kong et al.~\cite{DBLP:journals/tetc/KongMWLX21} present VOPRec, a model combining textual components in form of Doc2vec and Paper2Vec paper representations with citation network information in form of Struc2vec. Those networks of papers connect the most similar publications based on text and structure. Random walk on these graphs contributes to the goal of learning vector representations.

L et al.~\cite{DBLP:conf/cscwd/LLP21} base their recommendation on lately accessed papers of users as they assume future accessed papers are similar to recently seen ones. They utilise a sliding window to generate sequences of papers, on those they construct a GNN to aggregate neighbouring papers to identify users' interests. 

Li et al.~\cite{DBLP:journals/tois/LiCPR19}~ introduce~ a~ subscription-based~ approach which learns a mapping between users' browsing history and their clicks in the recommendation mails. They learn a re-ranking of paper recommendations by using its metadata, recency, word representations and entity representations by knowledge graphs as input for a neural network. Their defined target audience are new users.

Li et al.~\cite{li20} present HNTA a paper recommendation method utilising heterogeneous networks and changing user interests. Paper similarities are calculated with Word2Vec representations of words recommended for each paper. Changing user interest is modelled with help of an exponential time decay function on word vectors.

Li et al.~\cite{DBLP:journals/dss/LiWNLL21} utilise user profiles with a history of preferences to construct heterogeneous networks where they apply random walks on meta-paths to learn personalised weights. They strive to discover user preference patterns and model preferences of users as their recently cited papers.

Lin et al.~\cite{lin21} utilise authors' citations and years they have been publishing papers in their recommendation approach. All candidate publications are matched against user-entered keywords, the two factors of authors of these candidate publications are combined to identify the overall top recommendations.

Liu et al.~\cite{DBLP:conf/seke/LiuKCQ19} explicitly do not require all recommended publications to fit the query of a user perfectly. Instead they state the set of recommended papers fulfils the information need only in the complete form. Here they treat paper recommendation as a link prediction problem incorporating publishing time, keywords and author influence.
In a subsequent work, part of the previous research group again observes the same problem. In this work Liu et al.~\cite{DBLP:journals/complexity/LiuKYQ20} propose an approach utilising numbers of citations (author popularity) and relationships between publications in an undirected citation graph. They compute Steiner trees to identify the sets of papers to recommend.

Lu et al.~\cite{DBLP:conf/cscwd/LuHCP021} propose TGMF-FMLP, a paper recommendation approach focusing on the changing preferences of users and novelty of papers. They combine category attributes (such as paper type, publisher or journal), a time-decay function, Doc2Vec representations of the papers' content and a specialised matrix factorisation to compute recommendations.

Ma et al.~\cite{DBLP:journals/monet/MaZZ19} introduce HIPRec, a paper recommendation approach on heterogeneous networks of authors, papers, venues and topics specialised on new publications. They use the most interesting meta-paths to construct significant meta-paths. With these paths and features from these paths they train a model to identify new papers fitting users. 
Together with another researcher Ma further pursued this research direction. Ma and Wang~\cite{DBLP:journals/access/MaW19a} propose HGRec, a heterogeneous graph representation learning-based model working on the same network. They use meta-path-based features and Doc2Vec paper embeddings to learn the node embeddings in the network.

Manju et al.~\cite{DBLP:journals/ijiit/Manju} attempt to solve the cold-start problem with their paper recommendation approach coding social interactions as well as topical relevance into a heterogeneous graph. They incorporate believe propagation into the network and compute recommendations by applying random walk.

Mohamed Hassan et al.~\cite{MohamedHassan} adopt an existing tag prediction model which relies on a hierarchical attention network to capture semantics of papers. Matrix factorisation then identifies the publications to recommend.

Nair et al.~\cite{nair} propose C-SAR, a paper recommendation approach using a neural network. They input GloVe embeddings of paper titles into their Gated Recurrent Union model to compute probabilities of similarities of papers. The resulting adjacency matrix is input to an association rule mining a priori algorithm which generates the set of recommendations.

Nishioka et al.~\cite{DBLP:conf/icadl/NishiokaHS19,DBLP:conf/ercimdl/NishiokaHS19} state serendipity of recommendations as their main objective. They incorporate users' tweets to construct profiles in hopes to model recent interests and developments which did not yet manifest in users' papers. They strive to diversity the list of recommended papers.
In more recent work Nishioka et al.~\cite{DBLP:journals/peerj-cs/NishiokaHS20} explained their evaluation more in depth.

Rahdari and Brusilovsky~\cite{DBLP:conf/iui/RahdariB19} observe paper recommendation~ for~ participants~ of~ scientific~ conferences. Users' profiles are composed of their past publications. Users control the impact of features such as publication similarity, popularity of papers and its authors to influence the ordering of their suggestions.

Renuka et al.~\cite{renuka} propose a paper recommendation approach utilising TF-IDF representations of automatically extracted keywords and key phrases. They then either use cosine similarity between vectors or a clustering method to identify the most similar papers for an input paper.

Sakib et al.~\cite{DBLP:journals/access/SakibAH20} present a paper recommendation approach utilising second-level citation information and citation context. They strive to not rely on user profiles in the paper recommendation process. Instead they measure similarity of candidate papers to an input paper based on co-occurred or co-occurring papers.
In a follow-up work with a bigger research group Sakib et al.~\cite{DBLP:journals/access/SakibAABHHG21} combine contents of titles, keywords and abstracts with their previously mentioned collaborative filtering approach. They again utilise second-level citation relationships between papers to find correlated publications.

Shahid et al.~\cite{shahid21a} utilise in-text citation frequencies and assume a reference is more important to a referencing paper the more often it occurs in the text. They crawl papers from CiteSeerX to retrieve the top 500 citing papers.
In a follow-up work with a partially different research group Shahid et al.~\cite{DBLP:journals/peerj-cs/ShahidAAAA21} evaluate the previously presented approach with a user study.

Sharma et al.~\cite{Sharma} propose IBM PARSe, a paper recommendation system for the medical domain to reduce the number of papers to review for keeping an existing knowledge graph up-to-date. Classifiers identify new papers from target domains, named entity recognition finds relevant medical concepts before papers' TF-IDF vectors are compared to ones in the knowledge graph. New publications most similar to already relevant ones with matching entities are recommended to be included in the knowledge base.

Subathra and Kumar~\cite{subathra} constructed an paper recommendation system which applies LDA on Wikipedia articles twice. Top related words are computed using pointwise mutual information before papers are recommended for these top words. 

Tang et al.~\cite{DBLP:journals/concurrency/TangLQ21} introduce CGPrec, a content-based and knowledge graph-based paper recommendation system. They focus on users' sparse interaction history with papers and strive to predict papers on which users are likely to click. They utilise Word2Vec and a Double Convolutional Neural Network to emulate users' preferences directly from paper content as well as indirectly by using knowledge graphs.

Tanner et al.~\cite{DBLP:conf/bigdataconf/TannerAH19} consider relevance and strength of citation relations to weigh the citation network. They fetch citation information from the parsed full texts of papers. On the weighted citation networks they run either weighted co-citation inverse document frequency, weighted bibliographic coupling or random walk with restart to identify the highest scoring papers. 

Tao et al.~\cite{tao} use embeddings and topic modelling to compute paper recommendations. They combine LDA and Word2Vec to obtain topic embeddings. Then they calculate most similar topics for all papers using Doc2Vec vector representations and afterwards identify the most similar papers. With PageRank on the citation network they re-rank these candidate papers.

Waheed et al.~\cite{DBLP:journals/access/WaheedIRMK19} propose CNRN, a recommendation approach using a multilevel citation and authorship network to identify recommendation candidates. From these candidate papers ones to recommend are chosen by combining centrality measures and authors' popularity.
Highly correlated but unrelated Shi et al.~\cite{DBLP:conf/ijcnn/ShiMZJLC20} present AMHG, an approach utilising a multilayer perceptron. They also construct a multilevel citation network as described before with added author relations. Here they additionally utilise vector representations of publications and recency.

Wang et al.~\cite{DBLP:conf/service/WangXTWX20} introduce a knowledge-aware path recurrent network model. An LSTM mines path information from the knowledge graphs incorporating papers and users. Users are represented by their downloaded, collected and browsed papers, papers are represented by TF-IDF representations of their keywords. 

Wang et al.~\cite{DBLP:journals/corr/abs-2103-08819} require users to construct knowledge graphs to specify the domain(s) and enter keywords for which recommended papers are suggested. From the keywords they compute initially selected papers. They apply Doc2Vec and emotion-weighted similarity between papers to identify recommendations.

Wang et al.~\cite{DBLP:journals/asc/WangWYXYY21} regard paper recommendation targeting a group of people instead of single users and introduce GPRAH\_ER. They employ a two-step process which first individually predicts papers for users in the group before recommended papers are aggregated. Here users in the group are not considered equal, different importance and reliability weights are assigned such that important persons' preferences are more decisive of the recommended papers.
Together with a different research group two authors again pursued this definition of the paper recommendation problem. Wang et al.~\cite{wang21} recommend papers for groups of users in an approach called GPMF\_ER. As with the previous approach they compute TF-IDF vectors of keywords of papers to calculate most similar publications for each user. Probabilistic matrix factorisation is used to integrate these similarities in a model such that predictive ratings of all users and papers can be obtained. In the aggregation phase the number of papers read by a user is determined to replace the importance component.

Xie et al.~\cite{xie21} propose JTIE, an approach incorporating contents, authors and venues of papers to learn paper embeddings. Further, directed citation relations are included into the model. Based on users' authored and referenced papers personalised recommendations are computed.
They consider explainability of recommendations.~ 
In~ a~ subsequent~ work~ part~ of~ the~ researchers again work on this topic. Xie et al.~\cite{DBLP:conf/sigir/XieSB21} specify on recommendation of papers from different areas for user-provided keywords or papers. They use hierarchical LDA to model evolving concepts of papers and citations as evidence of correlation in their approach.

Yang et al.~\cite{DBLP:journals/www/YangLLLZZZZ19} incorporate the age of papers and impact factors of venues as weights in their citation network-based approach named PubTeller. Papers are clustered by topic, the most popular ones from the clusters most similar to the query terms are recommendation candidates. In this approach, LDA and TF-IDF are used to represent publications.

Yu et al.~\cite{DBLP:conf/ksem/YuHLZXLXY19} propose ICMN, a general collaborative memory network approach.% focusing on attractiveness of items (papers) and potential user-item relationships (citations). 
User and item embeddings are composed by incorporating papers' neighbourhoods and users' implicit preferences.

\textcolor{red}{Zavrel et al.~\cite{zarvel} present the scientific literature recommendation~ platform~ Zeta~ Alpha,~ which~ bases their recommended papers on examples tagged in user-defined categories. The approach includes these user-defined tags as well as paper content embeddings, social media mentions and citation information in their ensemble learning approach to recommend publications.}

Zhang et al.~\cite{ZHANG2019616} propose W-Rank, a general approach weighting edges in a heterogeneous author, paper and venue graph by incorporating citation relevance and author contribution. They apply their method on paper recommendation. Network- (via citations) and semantic-based (via AWD) similarity between papers is combined for weighting edges between papers, harmonic counting defines weights of edges between authors and papers. A HITS-inspired algorithm computes the final authority scores.
In a subsequent work in a slightly smaller group they focus on a specialised approach~ for~ paper~ recommendation.~ Here~ Zhang~ et al.~\cite{zhang20} strive to emulate a human expert recommending papers. They construct a heterogeneous network with authors, papers, venues and citations. Citation weights are determined by semantic- and network-level similarity~ of~ papers.~ Lastly,~ recommendation~ candidates are re-ranked while combining the weighted heterogeneous network and recency of papers.

Zhao et al.~\cite{DBLP:journals/access/ZhaoKFMN20} present a personalised approach focusing on diversity of results which consists of three parts. First LFM extracts latent factor vectors of papers and users from the users' interactions history with papers. Then BERT vectors are constructed for each word of the papers, with those vectors as input and the latent factor vectors as label a BiGRU model is trained.
Lastly, diversity and a user's rating weights determine the ranking of recommended publications for the specific user.

\subsection{Other relevant Work}
\label{sec:sec:litReview_others}

We now briefly discuss some papers which did not pre\-sent novel paper recommendation approaches but are relevant in the scope of this literature review nonetheless.

\subsubsection{Surrounding Paper Recommendation}
Here we present two works which could be classified as ones to use on top of or in combination with existing paper recommendation systems:
Lee et al.~\cite{lee} introduce LIMEADE, a general approach for opaque recommendation systems which can for example be applied on any paper recommendation system. They produce explanations for recommendations as a list of weighted interpretable features such as influential paper terms. 

Beierle~ et~ al.~\cite{DBLP:journals/jodl/BeierleACB20}~ use~ the~ recommendation-as-a-service provider Mr. DLib to analyse choice overload in user evaluations. They report several click-based measures and discuss effects of different study parameters on engagement of users.

\subsubsection{(R)Evaluations}

The following four works can be grouped as ones which provide (r)evaluations of already existing approaches. Their results could be useful for the construction of novel systems:
Ostendorff~\cite{DBLP:journals/corr/abs-2008-00202} suggests considering the context of paper similarity in background, methodology and findings sections instead of undifferentiated textual similarity for scientific paper recommendation.

Mohamed Hassan et al.~\cite{DBLP:conf/recsys/HassanSGMB19} compare different text embedding methods such as BERT, ELMo, USE and InferSent to express semantics of papers. They perform paper recommendation and re-ranking of recommendation candidates based on cosine similarity of titles. 

Le et al.~\cite{DBLP:conf/recsys/LeKD19} evaluate the already existing paper recommendation system Mendeley Suggest, which provides recommendations with different collaborative or content-based approaches. They observe different usage behaviours and state utilisation of paper recommendation systems does positively effect users' professional lives.

\textcolor{red}{Barolli et al.~\cite{barolli} compare similarities of paper pairs utilising n-grams, tf-idf and a transformer based on BERT. They model cosine similarities of these pairs into a paper connection graph and argue for the combination of content-based and graph based methods in the context of COVID-19 paper recommendation systems.}

\subsubsection{Living Labs}

Living labs help researchers conduct meaningful evaluations by providing an environment, in which recommendations produced by experimental systems are shown to real users in realistic scenarios~\cite{DBLP:conf/ecir/BeelCKDK19}. We found three relevant works for the area of scientific paper recommendation:
Beel et al.~\cite{DBLP:conf/ecir/BeelCKDK19} proposed a living lab for scholarly recommendation built on top of Mr. DLib, their recommender-as-a-service system. They log users' actions such as clicks, downloads and purchases for related recommended papers. Additionally, they plan to extend their living lab to also incorporate research grant or research collaborator recommendation.

Gingstad et al.~\cite{DBLP:conf/cikm/GingstadJB20} propose ArXivDigest, an online living lab for explainable and personalised paper recommendations from arXiv. Users can either be suggested papers while browsing their website or via email as a subscription-type service. Different approaches can be hooked into ArXivDigest, the recommendations generated by them can then be evaluated by users. A simple text-based baseline compares user-input topics with articles. Target values of evaluations are users' clicked and saved papers.

Schaer et al.~\cite{DBLP:conf/clef/SchaerBCWST21} held the Living Labs for Academic Search (LiLAS) where they hosted two shared tasks: dataset recommendation for scientific papers and ad-hoc multi-lingual retrieval of most relevant publications regarding specific queries.
To overcome the gap between real-world and lab-based evaluations they allowed integrating participants' systems into real-world academic search systems, namely LIVIO and GESIS Search.

\subsubsection{Multilingual/Cross-lingual Recommendation} 

The previous survey by Li and Zhou~\cite{DBLP:journals/ccsecis/LiZ19} identifies cross-language paper recommendation as a future research direction. The following two works could be useful for this aspect:
Keller and Munz~\cite{DBLP:conf/clef/KellerM21} present their results of participating on the CLEF LiLAS challenge where they tackled recommendation of multilingual papers based on queries. They utilised a pre-computed ranking approach, Solr and pseudo-relevance feedback to extend queries and identify fitting papers.

Safaryan et al.~\cite{DBLP:conf/aist/SafaryanFYKN20} compare different already existing techniques for cross-language recommendation of publications. They compare word by word translation, linear projection from a Russian to an English vector representation, VecMap alignment and MUSE word embeddings.

\subsubsection{Related Recommendation Systems}

Some recommendation approaches are slightly out of scope of pure paper recommendation systems but could still provide inspiration or relevant results:
Ng~\cite{Ng2020-gf} proposes CBRec, a children's book recommendation system utilising matrix factorisation. His goal is to encourage good reading habits of children. The approach combines readability levels of users and books with TF-IDF representations of books to find ones which are similar to ones which a child may have already liked.

Patra et al.~\cite{PATRA2020103399} recommend publications relevant for datasets to increase reusability. Those papers could describe the dataset, use it or be related literature. The authors represent datasets and articles as vectors and use cosine similarity to identify the best fitting papers. Re-ranking them with usage of Word2Vec embeddings results in the final recommendation.

\section{Datasets}
\label{sec:datasets}

\begin{table}[t]
\scriptsize
        \resizebox{.47\textwidth}{!}{
    \centering

    \begin{tabu}{l|l|l}
        Name & A? & Used by\\ \hline
        
         DBLP + Citations v1~\cite{DBLP:conf/kdd/TangZYLZS08} & $\checkmark$ & %Yang et al.~
         \cite{DBLP:journals/www/YangLLLZZZZ19}\\
        DBLP + Citations v8~\cite{DBLP:conf/kdd/TangZYLZS08} & $\times$ & %Ma and Wang~
        \cite{DBLP:journals/access/MaW19a}, %Ma et al.~
        \cite{DBLP:journals/monet/MaZZ19} \\
        DBLP + Citations v11 & $\checkmark$ & %Ali et al.~
        \cite{DBLP:journals/kbs/AliQMAA20} \\

        dblp + IEEE + ACM + Pubmed & $\times$ & %Bulut et al.~
        \cite{Bulut2018APR}\\
        DBLP paths & $\times$ & %Hua et al.~
        \cite{DBLP:conf/wise/HuaCLZZ20}\\
        DBLP-Citation-network f. AMiner & $\times$ & %Jing and Yu~
        \cite{DBLP:conf/ml4cs/JingY20}\\
        dblp & $\times$ & %Li et al.~
        \cite{DBLP:journals/dss/LiWNLL21}\\
        DBLP-REC & $\times$ & %Shi et al.~
        \cite{DBLP:conf/ijcnn/ShiMZJLC20} \\
        dblp + AMiner KG & $\times$ & %Wang et al.~
        \cite{DBLP:conf/service/WangXTWX20}\\
        dblp + AMiner + venue & $\times$ & %Xie et al.~
        \cite{xie21}\\
        %ACM + IEEE + dblp & $\times$ & Bulut et al.~\cite{Bulut2020}\\

        \hline
        
        %Sugiyama Kan 10~\cite{DBLP:conf/jcdl/SugiyamaK10} & $\checkmark$ &  & \makecell[tl]{}\\
        SPRD\_Senior & $\checkmark$ & %Chen and Ban~
        \cite{chenban}\\

        SPRD~\cite{DBLP:conf/jcdl/SugiyamaK13} & $\checkmark$ & \makecell[tl]{%Sugiyama and Kan~\cite{DBLP:journals/jodl/SugiyamaK15}, 
        %Haruna et al.~
        \cite{DBLP:journals/scientometrics/HarunaIQKHMC20}, %Sakib et al.~
        \cite{DBLP:journals/access/SakibAH20}, %Sakib et al.~
        \cite{DBLP:journals/access/SakibAABHHG21}} \\
        
        \hline

        Citeulike-a~\cite{DBLP:conf/ijcai/WangCL13} & $\checkmark$ & \makecell[tl]{%Ahmedi et al.~
        \cite{ahmedi}, %Alfarhood and Cheng~
        \cite{DBLP:conf/icmla/AlfarhoodC19}, %Guo et al.~
        \cite{DBLP:conf/aaai/GuoCZLDH20}, %L et al.~
        \cite{DBLP:conf/cscwd/LLP21}, %Mohammed Hassan et al.~
        \cite{MohamedHassan}, \\%Tang et al.~
        \cite{DBLP:journals/concurrency/TangLQ21}, %Yu et al.~
        \cite{DBLP:conf/ksem/YuHLZXLXY19}, %Zhao et al.~
        \cite{DBLP:journals/access/ZhaoKFMN20}} \\
        Citeulike-t~\cite{DBLP:conf/ijcai/WangCL13} & $\checkmark$ & %Alfarhood and Cheng~
        \cite{DBLP:conf/icmla/AlfarhoodC19}\\
        Citeulike\_huge & $\times$ & %Lu et al.~
        \cite{DBLP:conf/cscwd/LuHCP021}\\
        Citeulike\_medium & $\times$ & %Wang et al.~
        \cite{DBLP:journals/asc/WangWYXYY21}\\
        Citeulike\_tiny & $\times$ & %Wang et al.~
        \cite{wang21}\\
        \hline
       
        ACM paths & $\times$ & %Hua et al.~
        \cite{DBLP:conf/wise/HuaCLZZ20}\\

        ACM citation network V8 & $\times$ & \makecell[tl]{%Nishioka et al.~
        \cite{DBLP:conf/icadl/NishiokaHS19}, %Nishioka et al.~
        \cite{DBLP:conf/ercimdl/NishiokaHS19}, %Nishioka et al.~
        \cite{DBLP:journals/peerj-cs/NishiokaHS20}}\\

        \hline
        
        Scopus\_tiny & $\times$ & %Chaudhuri et al.~
        \cite{DBLP:journals/corr/abs-2107-07831,DBLP:journals/jodl/ChaudhuriSSS21}\\
        ScienceDirect+Scopus & $\times$ & %Li et al.~
        \cite{DBLP:journals/tois/LiCPR19}\\
        Scopus & $\times$ & %Xie et al.~
        \cite{DBLP:conf/sigir/XieSB21}\\

        \hline
        
        AMiner & $\times$ & %Li et al.~
        \cite{DBLP:journals/dss/LiWNLL21}\\
        AMiner + Wanfang & $\times$ & %Du et al.~
        \cite{DBLP:conf/aiccsa/DuGWHZG20}\\
        AMiner\_tiny & $\times$ & %Du et al.~
        \cite{DBLP:journals/tkde/DuTD21}\\
        AMiner\_huge &  $\times$ & %Waheed et al.~
        \cite{DBLP:journals/access/WaheedIRMK19}\\
        ACM C-D & $\times$ & %Xie et al.~
        \cite{DBLP:conf/sigir/XieSB21}\\ 

        \hline
        
        AAN\_original~\cite{radev} & $\checkmark$ & %Nair et al.~
        \cite{nair}\\
        AAN\_modified & $\times$ & %Ali et al.~
        \cite{DBLP:journals/kbs/AliQMAA20}, %L et al.~
        \cite{DBLP:conf/cscwd/LLP21}\\
        AAN\_tiny & $\times$ & %Tanner et al.~
        \cite{DBLP:conf/bigdataconf/TannerAH19}\\

        \hline
        Sowiport &$\times$ & %Collins and Beel~
        \cite{DBLP:conf/jcdl/CollinsB19}\\
        RARD\_tiny & $\times$ & %Du et al.~
        \cite{DBLP:journals/tkde/DuTD21}\\

        \hline
        
        CiteSeer & $\times$ & %Kang et al.~
        \cite{KANG20212020BDP0008}\\
        CiteSeer\_tiny  & $\times$ & %Shahid et al.~
        \cite{shahid21a}\\
        CiteSeer\_medium & $\times$ & %Shahid et al.~
        \cite{DBLP:journals/tjs/ShahidAABZYC20}\\

        \hline

        Patents\_tiny & $\times$ & %Du et al.~
        \cite{DBLP:journals/tkde/DuTD21}\\
        Patents & $\times$ & %Xie et al.~
        \cite{xie21}\\
        ACM H-I & $\times$ & %Xie et al.~
        \cite{DBLP:conf/sigir/XieSB21}\\

        \hline
        
        Hep-TH graph & $\times$ & %Liu et al.~
        \cite{DBLP:journals/complexity/LiuKYQ20}\\
        arXiv Hep-TH & $\times$ & %Zhang et al.~
        \cite{ZHANG2019616}\\
        \hline
        
        MSA & $\times$ & %Yang et al.~
        \cite{DBLP:journals/www/YangLLLZZZZ19}\\
        MAG 2017 & $\times$ & %Zhang et al.~
        \cite{ZHANG2019616}\\
        MAG 2018 & $\times$ & %Kanakia et al.~
        \cite{kanakia19}\\
        
        \hline
        
        BBC & $\checkmark$ & %Afsar et al.~
        \cite{DBLP:conf/flairs/AfsarCF21}\\
        PRSDataset & $\checkmark$ & %Guo et al.~
        \cite{DBLP:conf/aaai/GuoCZLDH20}, %L et al.~
        \cite{DBLP:conf/cscwd/LLP21}\\
        
        Physical Review A & $\times$ & %Kong et al.~
        \cite{DBLP:journals/tetc/KongMWLX21}\\

        ACL selection network & $\times$ & %Tao et al.~
        \cite{tao}\\

        prostate cancer & $\times$ & %Afsar et al.~
        \cite{DBLP:conf/flairs/AfsarCF21}\\
       
        Peltarion  & $\times$ & %Bereczki~
        \cite{Bereczki1587420}\\
        Jabref & $\times$ & %Collins and Beel~
        \cite{DBLP:conf/jcdl/CollinsB19}\\
        
        DM & $\times$ & %Hu et al.~
        \cite{DBLP:conf/ifip12/HuMLH20}\\
        
        Graphs & $\times$ & %L et al.~
        \cite{DBLP:conf/cscwd/LLP21}\\
        SCHOLAT & $\times$ & %Li et al.~
        \cite{li20}\\
        IEEE Xplore & $\times$ & %Lin et al.~
        \cite{lin21}\\

        %EC-TEL 2018 & $\times$ & Rahdari and Brusilovsky~\cite{DBLP:conf/iui/RahdariB19}\\

        KGs & $\times$ & %Wang et al.~
        \cite{DBLP:journals/corr/abs-2103-08819}\\

        Wanfang & $\times$ & %Kang et al.~
        \cite{KANG20212020BDP0008}\\

        Watson\texttrademark for Genomics & $\times$ & %Sharma et al.~
        \cite{Sharma}\\
        
        Wikipedia & $\times$ & %Subathra and Kumar~
        \cite{subathra}\\
        
        LibraryThing & $\times$ & %Zhao et al.        ~
        \cite{DBLP:journals/access/ZhaoKFMN20}\\

    \end{tabu}
    }
    \caption{Overview of datasets utilised in most recent related work with (unofficial) names, public availability of the possibly modified dataset which was used (A?), and a list of papers it was used in. Datasets are grouped by their underlying data source if possible.}
    %and a short description (papers P, authors A, venues V, users U.}
    \label{tab:datasets}
\end{table}

\begin{table*}[t]
\scriptsize
    \centering
    \begin{tabu}{l|l|l}
    
        Name & Used by & Description\\\hline
        
        \hline
     DBLP + Citations v8~\cite{DBLP:conf/kdd/TangZYLZS08} & \cite{DBLP:journals/access/MaW19a,DBLP:journals/monet/MaZZ19} & \makecell[tl]{2,133 \textit{p} from 20
\textit{v} from 2000 to 2016, 39,530 \textit{a}, 15,708 \textit{p} topics}\\
        dblp + IEEE + ACM + Pubmed & \cite{Bulut2018APR} & \makecell[tl]{sources: dblp, IEEE, ACM, Pubmed. 3,394,616 \textit{p} (titles), \textit{a}, publication years, \\keywords, \textit{r}}\\        
        DBLP paths & \cite{DBLP:conf/wise/HuaCLZZ20} & \makecell[tl]{1,782,700 \textit{p} (titles, abstracts, keywords), 2,052,414 \textit{a}, 18,936 \textit{v}, 100,000 \textit{t}, 9,590,600 \textit{i}}\\
        DBLP-Citation-network f. AMiner & \cite{DBLP:conf/ml4cs/JingY20} & \makecell[tl]{63,469 \textit{p} from 2013 to 2019, 152,586 \textit{a}}\\
        dblp & \cite{DBLP:journals/dss/LiWNLL21} & \makecell[tl]{2,126,267 \textit{p}, 8686 \textit{v}, 1,221,259 \textit{a}, 256,214 \textit{t}, 3765 \textit{u} relations}\\

        DBLP-REC & \cite{DBLP:conf/ijcnn/ShiMZJLC20} & \makecell[tl]{DBLP-Citation-network v11 + ScienceDirect + IEEE, 3,590,853 \textit{p}, 3,276,803 \textit{a}, \\35,254,530 \textit{c}}\\
        dblp + AMiner KG& \cite{DBLP:conf/service/WangXTWX20} & \makecell[tl]{KG with 223,431 \textit{a}, 337,561 \textit{p}, 5578 \textit{v}, 1179 keyword nodes, 16,328,642 \textit{c}}\\
        dblp + AMiner + venue & \cite{xie21} & \makecell[tl]{3,056,388 \textit{p} (titles, abstracts, keywords), 1,752,401 \textit{a}, 354,693 keywords, 11,397 \textit{v}, \textit{c},\\ discipline labels}\\ \hline
        Citeulike\_huge & \cite{DBLP:conf/cscwd/LuHCP021} & \makecell[tl]{210,137 \textit{p}, 3,039 \textit{u}, 284,960 \textit{u-p i} from Nov 2004 to Dec 2007}\\
        Citeulike\_medium & \cite{DBLP:journals/asc/WangWYXYY21} & \makecell[tl]{2,065 users, 718 groups, 85,542 \textit{p}}\\
        Citeulike\_tiny & \cite{wang21} & \makecell[tl]{1,659 users, 718 groups, 82,376 \textit{p}, 198,744 \textit{i}}\\
        \hline
       
        \hline
        ACM paths & \cite{DBLP:conf/wise/HuaCLZZ20} & \makecell[tl]{2,385,057 \textit{p} (titles, abstracts, keywords), 2,004,398 \textit{a}, 269,467 \textit{v}, 61,618 \textit{t}, 12,048,682 \textit{i}}\\

        ACM citation network V8 & \cite{DBLP:conf/icadl/NishiokaHS19,DBLP:conf/ercimdl/NishiokaHS19,DBLP:journals/peerj-cs/NishiokaHS20} & \makecell[tl]{1,669,237 \textit{p} (titles, abstracts), \textit{v}, \textit{a}}\\

        \hline
        
        Scopus\_tiny & \cite{DBLP:journals/corr/abs-2107-07831,DBLP:journals/jodl/ChaudhuriSSS21} & \makecell[tl]{2,000 \textit{p}}\\
        ScienceDirect + Scopus &\cite{DBLP:journals/tois/LiCPR19} & \makecell[tl]{\textit{u}'s browsed \textit{p} prior to first email from ScienceDirect, \textit{p} metadata from Scopus, 4,392 \\recommendation sessions (emails with clicks on \textit{p}, \textit{u}' browsing history)}\\
        Scopus & \cite{DBLP:conf/sigir/XieSB21} & \makecell[tl]{528,224 \textit{p}, \textit{a}, \textit{r}, discipline tags}\\
        Scopus + venue & \cite{xie21} & \makecell[tl]{1,304,907 \textit{p} (titles, abstracts, keywords),  482,602 \textit{a}, 127,630 keywords, 7653 \textit{v}, \textit{c}, \\discipline labels}\\
        \hline
        
        AMiner & \cite{DBLP:journals/dss/LiWNLL21} & \makecell[tl]{2,070,699 \textit{p}, 263,250 \textit{v}, 1,557,147 \textit{a}, 735,059 \textit{t}, 9398 \textit{u} relations}\\
        AMiner + Wanfang & \cite{DBLP:conf/aiccsa/DuGWHZG20} & \makecell[tl]{4 mio \textit{p}. 3 sets: data from 2018 and 2019 (221,076 \textit{p}, 503,945 \textit{a}), mathematical\\ analysis (98,702 \textit{p}, 117,183 \textit{a}), image processing (49,098 \textit{p}, 107,290 \textit{a})}\\
        AMiner\_tiny & \cite{DBLP:journals/tkde/DuTD21} & \makecell[tl]{188 input \textit{p}, 10 candidate \textit{p} for each input}\\
        AMiner\_huge & \cite{DBLP:journals/access/WaheedIRMK19} & \makecell[tl]{2,092,356 \textit{p}, 1,712,433 \textit{a}, 8,024,869 \textit{c}, 4,258,615 co-autorships}\\
        ACM C-D & \cite{DBLP:conf/sigir/XieSB21} & \makecell[tl]{43,380 \textit{p} from AMiner, \textit{a}, ACM CSS tags}\\

        \hline
        AAN\_modified & \cite{DBLP:journals/kbs/AliQMAA20,DBLP:conf/cscwd/LLP21} & \makecell[tl]{21,455 \textit{p} from 312 \textit{v} from NLP, 17,342 \textit{a}, 113,367 \textit{c}}\\
        AAN\_tiny & \cite{DBLP:conf/bigdataconf/TannerAH19} & \makecell[tl]{2082 \textit{p} (ids, titles, publication year), 8194 \textit{c}, avg. 7.87 \textit{c} per \textit{p}, \textit{a}, \textit{v}}\\

        \hline
        
        Sowiport & \cite{DBLP:conf/jcdl/CollinsB19} & \makecell[tl]{\textit{u} \textit{i} data from Mar 2017 to Oct 2018, 0.1\% click-through rate}\\
        RARD\_tiny & \cite{DBLP:journals/tkde/DuTD21} & \makecell[tl]{800 input \textit{p} from Related-Article Recommendation Dataset from Sowiport~\cite{rard}}\\        
        
        \hline
        
        CiteSeer & \cite{KANG20212020BDP0008} & \makecell[tl]{1,100 \textit{p}, 10 sets of relevant \textit{p}}\\
        CiteSeer\_tiny & \cite{shahid21a} & \makecell[tl]{400 \textit{c}-pairs, 1,230 \textit{c} contexts}\\
        CiteSeer\_medium & \cite{DBLP:journals/tjs/ShahidAABZYC20} & \makecell[tl]{10 \textit{p}, 226 \textit{c}-pairs}\\

        \hline
        Patents\_tiny & \cite{DBLP:journals/tkde/DuTD21} & \makecell[tl]{67 input patents, 20 candidate patents for each input}\\
        
        Patents & \cite{xie21} & \makecell[tl]{182,260 patents, 73,974 \textit{a}}\\
        
        ACM H-I & \cite{DBLP:conf/sigir/XieSB21} & \makecell[tl]{70,090 patents with ownership from 2017, \textit{r}, ACM CSS tags}\\

        \hline
        Hep-TH graph & \cite{DBLP:journals/complexity/LiuKYQ20} & \makecell[tl]{graph with 8,721 \textit{p} (keywords)}\\

        arXiv Hep-TH & \cite{ZHANG2019616} & \makecell[tl]{$\sim$29,000 \textit{p}, 350,000 \textit{c}, 14,909 \textit{a}, 428 journals}\\
        
        \hline
        
        MSA & \cite{DBLP:journals/www/YangLLLZZZZ19} & \makecell[tl]{101,205 \textit{p}, 190,146 \textit{c} in 300 conferences}\\
        MAG 2017 & \cite{ZHANG2019616} & \makecell[tl]{based on data until 2017, area: intrusion detection in cyber security, 6428 \textit{p}, 94,887 \textit{c},\\ 18,890 \textit{a}, 6428 journals}\\
        MAG 2018 & \cite{kanakia19} & \makecell[tl]{based on MAG Azure database from Oct 2018, 206,676,892 \textit{p}}\\
        
        \hline
        
        Physical Review A & \cite{DBLP:journals/tetc/KongMWLX21} & \makecell[tl]{393 \textit{p} from 2007 to 2009 with 2,664 \textit{c} from  American Physical Society}\\%~\tablefootnote{\url{https://journals.aps.org/datasets}}}\\

        ACL selection network & \cite{tao} & \makecell[tl]{18,718 \textit{p} (titles, summaries) from ACL proceedings}\\

        prostate cancer &\cite{DBLP:conf/flairs/AfsarCF21} & \makecell[tl]{500 \textit{p} tagged with 5 categories}\\
       
        Peltarion  & \cite{Bereczki1587420} & \makecell[tl]{290 \textit{p}, \textit{u} \textit{i} from Dec 2018 to May 2021 of \textit{u} of Peltarion Knowledge Center who have \\read $\ge$ 5 \textit{p}}\\

        Jabref & \cite{DBLP:conf/jcdl/CollinsB19} & \makecell[tl]{\textit{u} \textit{i} data from Mar 2017 to Oct 2018, 0.22\% click-through rate}\\

        DM & \cite{DBLP:conf/ifip12/HuMLH20} & \makecell[tl]{8,301 \textit{p} from journals: DMKD, TKDE + conferences: KDD, ICDM, SDM}\\
        
        Graphs & \cite{DBLP:conf/cscwd/LLP21} & \makecell[tl]{Cora (1 graph, 2.7k nodes), TU-IMDB (1.5k graphs, ~13 nodes each), TU-MUTAG \\(188 molecules, 18 nodes)}\\
        SCHOLAT & \cite{li20} & \makecell[tl]{34,518 \textit{p} (titles, abstracts, keywords), \textit{a}}\\
        IEEE Xplore & \cite{lin21} & \makecell[tl]{3 \textit{p} (keywords), \textit{r}, \textit{a} appeared in IEEE between 2010 and 2017}\\

        %EC-TEL 2018 & \cite{DBLP:conf/iui/RahdariB19} & \makecell[tl]{\textit{p} from EC-TEL 2018, click data of \textit{u} from that conference}\\
        KGs & \cite{DBLP:journals/corr/abs-2103-08819} & \makecell[tl]{knowledge graphs,  600 \textit{p} from information retrieval + machine learning}\\

        Wanfang & \cite{KANG20212020BDP0008} & \makecell[tl]{500 \textit{p}, 5 sets of relevant \textit{p}}\\

        Watson\texttrademark for Genomics & \cite{Sharma} & \makecell[tl]{15,320 \textit{p} from top 10 percentile genomics journals from Jun 2016}\\
        
        Wikipedia & \cite{subathra} & \makecell[tl]{1000 \textit{p} from Wikipedia, 20 topics}\\
        
        LibraryThing & \cite{DBLP:journals/access/ZhaoKFMN20} & \makecell[tl]{120,150 books (titles, abstracts), \textit{u}, 185,210 favourites records, 150,216 ratings, \\139,530 reviews of 12,350 \textit{u}}\\
    \end{tabu}
    \caption{Description of private datasets utilised in most recent related work with (unofficial) names. Datasets are grouped by their underlying data source if possible. We used the following abbreviations: user(s) \textit{u}, paper(s) \textit{p}, interaction(s) \textit{i}, author(s) \textit{a}, venue(s) \textit{v}, reference(s) \textit{r}, citation(s) \textit{c}, term(s) \textit{t}.}
    \label{tab:datasets_private}
\end{table*}

As the discussed paper recommendation systems utilise different inputs or components of scientific publications and pursue slightly different objectives, datasets to experiment on are also of diverse nature. We do not consider datasets of approaches which do not contain an  evaluation~\cite{DBLP:conf/seke/LiuKCQ19,zarvel} or do not evaluate the actual paper recommendation~\cite{DBLP:journals/elektrik/AhmadA20,DBLP:conf/tencon/ChaudhuriSS19,DBLP:journals/scientometrics/HabibA19,DBLP:conf/iui/RahdariB19,renuka} \textcolor{red}{such as the cosine similarity between a recommended and an initial paper~\cite{DBLP:journals/elektrik/AhmadA20,renuka}, the clustering quality on the constructed features~\cite{DBLP:conf/tencon/ChaudhuriSS19} or the Jensen Shannon Divergence between probability distributions of words between an initial and recommended papers~\cite{DBLP:journals/scientometrics/HabibA19}}. We also do not discuss datasets where only the data sources are mentioned but no remarks are made regarding the size or composition of the dataset~\cite{Bulut2020,DBLP:journals/concurrency/TangLQ21} or ones where we were not able to identify actual numbers~\cite{DBLP:journals/ijiit/Manju}.
Table~\ref{tab:datasets} gives an overview of datasets used in the evaluation of the considered discussed methods. Many of the datasets are unavailable only few years after publication of the approach. Most approaches utilise their own modified version of a public dataset which makes exact replication of experiments hard.
In the following the main underlying data sources and publicly available datasets are discussed. Non-publicly available datasets are briefly described in Table~\ref{tab:datasets_private}.

\subsection{dblp based datasets}

The dblp computer science bibliography (dblp) is a digital library offering metadata on authors, papers and venues from the area of computer science and adjacent fields~\cite{DBLP:journals/pvldb/Ley09}. They provide publicly available short-time stored daily and longer-time stored monthly data dumps\footnote{\url{https://dblp.uni-trier.de/xml/}}. 

The \textit{dblp + Citations v1} dataset~\cite{DBLP:conf/kdd/TangZYLZS08} builds upon a dblp version from 2010 mapped on AMiner. It contains 1,632,442 publications with 2,327,450 citations.

The \textit{dblp + Citations v11} dataset\footnote{\url{https://www.aminer.org/citation}} builds upon dblp. It contains 4,107,340 papers, 245,204 authors, \\16,209 venues and 36,624,464 citations

\textcolor{red}{These datasets do not contain supervised labels provided by human annotators even though the citation information could be used as interaction data.}
%Descriptions of non-public datasets based on dblp (\textit{dblp + IEEE + ACM + Pubmed}, \textit{DBLP paths}, \textit{DBLP-Citation-network f. AMiner}, \textit{dblp}, \textit{DBLP + Citations v8}, \textit{DBLP-REC}, \textit{dblp + AMiner KG}, \textit{dblp + AMiner + venue}) can be found in Table~\ref{tab:datasets_private}.

\subsection{SPRD based datasets}

The Scholarly Paper Recommendation Dataset (abbreviation: SPRD)\footnote{(shortened) \url{shorturl.at/cIQR1}} 
%https://www.db.soc.i.kyoto-u.ac.jp/~sugiyama/SchPaperRecData.html}} 
was constructed by collecting publications written by 50 researchers of different seniority from the area of computer science which are contained in dblp from 2000 to 2006~\cite{DBLP:journals/ccsecis/LiZ19,DBLP:conf/jcdl/SugiyamaK13,DBLP:journals/jodl/SugiyamaK15}. 
The dataset con\-tains 100,351 candidate papers extracted from the ACM Digital Library as well as citations and references for papers. Relevance assessments of papers relevant to their current interests of the 50 researchers are also included.

A subset of SPRD, \textit{SPRD\_Senior}, which contains only the data of senior researchers can also be constructed~\cite{DBLP:conf/jcdl/SugiyamaK10}.

\textcolor{red}{These datasets specifically contain supervised labels provided by human annotators in the form of sets of papers, which researchers found relevant for themselves.}

\subsection{CiteULike based datasets}

CiteULike~\cite{citeulike} was a social bookmarking site for scientific papers. It contained papers and their metadata. Users were able to include priorities, tags or comments for papers on their reading list. There were daily data dumps available from which datasets could be construc\-ted.

\textit{Citeulike-a}~\cite{DBLP:conf/ijcai/WangCL13}\footnote{\url{https://github.com/js05212/citeulike-a}} contains 5,551 users, 16,980 papers with titles and abstracts from 2004 to 2006 and their 204,986 interactions between users and papers. Papers are represented by their title and abstract.

\textit{Citeulike-t}~\cite{DBLP:conf/ijcai/WangCL13}\footnote{\url{https://github.com/js05212/citeulike-t}} contains 7,947 users, 25,975 papers and 134,860 user-paper interactions. Papers are represented by their pre-processed title and abstract.

\textcolor{red}{These datasets contain labelled data as they build upon CiteULike, which provides bookmarked papers of users.}

%The description of a non-public dataset based on CiteULike (\textit{Citeulike\_huge}, \textit{Citeulike\_medium}, \textit{Citeulike\_tiny}) can be found in Table~\ref{tab:datasets_private}.

\subsection{ACM based datasets}

The ACM Digital Library (ACM) is a semi-open digital library offering information on scientific authors, papers, citations and venues from the area of computer science\footnote{\url{https://dl.acm.org/}}. They offer an API to query for information. \textcolor{red}{Datasets building upon this source do not contain supervised labels provided by annotators even though the citation information could be used as interaction data.}

%Descriptions of non-public datasets based on ACM (\textit{ACM paths}, \textit{ACM citation network V8}) can be found in Table~\ref{tab:datasets_private}.

\subsection{Scopus based datasets}

Scopus is a semi-open digital library containing metadata on authors, papers and affiliations in different scientific areas\footnote{\url{https://www.scopus.com/home.uri}}. They offer an API to query for data. \textcolor{red}{Datasets building upon this source usually do not contain labels provided by annotators.}

%Descriptions of non-public datasets based on Scopus (\textit{Scopus\_tiny}, \textit{ScienceDirect + Scopus}, \textit{Scopus}, \textit{Scopus + venue}) can be found in Table~\ref{tab:datasets_private}.

\subsection{AMiner based datasets}

ArnetMiner (AMiner)~\cite{DBLP:conf/kdd/TangZYLZS08} is an open academic search system modelling the academic network consisting of authors, papers and venues from all areas\footnote{\url{https://www.aminer.org/}}. They provide an API to query for information. \textcolor{red}{Datasets building upon this source usually do not contain labelled user interaction data.}

%Descriptions of non-public datasets based on AMiner (\textit{AMiner}, \textit{AMiner + Wanfang}, \textit{AMiner\_tiny}, \textit{AMiner\-\_huge}, \textit{ACM C-D}) can be found in Table~\ref{tab:datasets_private}.

\subsection{AAN based datasets}

The ACL Anthology Network (AAN)~\cite{Radev&al.09,Radev&al.09a,radev} is a networked database containing papers, authors and citations from the area of computational linguistics\footnote{\url{https://aan.how/download/}}. It consists of three networks representing paper-citation relations,~ author-collaboration~ relations~ and~ the~ au\-thor-citation~ relations.~ The~ original~ dataset~ con\-tains 24,766 papers and 124,857 citations~\cite{nair}. \textcolor{red}{Datasets building~ upon~ this~ source~ usually~ do~ not~ contain labelled user interaction data even though the paper-citation,~ author-collaboration~ or~ author-citation relationships could be utilised to replace this data.}

%Descriptions of non-public datasets based on AAN (\textit{AAN\_modified}, \textit{AAN\_tiny}) can be found in Table~\ref{tab:datasets_private}.

\subsection{Sowiport based datasets}

Sowiport was an open digital library containing information on publications from the social sciences and adjacent fields~\cite{sowiport,sowiport2}. \textcolor{red}{The dataset linked papers by their attributes such as authors, publishers, keywords, journals, subjects and citation information. Via author names, keywords and venue titles the network could be traversed by triggering them to start a new search~\cite{sowiport2}.} 
Sowiport co-operated with the recommendation-as-a-service system Mr. DLib~\cite{DBLP:conf/jcdl/CollinsB19}. \textcolor{red}{Datasets building upon this~ source~ usually~ contain~ labelled~ user~ interaction data, the clicked papers of users.}

%Descriptions of non-public datasets based on Sowi\-port (\textit{Sowiport}, \textit{RARD\_tiny}) can be found in Table~\ref{tab:datasets_private}.

\subsection{CiteSeerX based datasets}

CiteSeerX~\cite{citeseer,citeseer2} is a digital library focused on metadata and full-texts of open access literature\footnote{\url{https://citeseerx.ist.psu.edu/index}}. It is the overhauled form of the former digital library CiteSeer. \textcolor{red}{Datasets building upon this source usually do not inherently contain labelled user interaction data.}

%Descriptions of non-public datasets based on CiteSeerX (\textit{CiteSeer}, \textit{CiteSeer\_tiny}, \textit{CiteSeer\_medium}) can be found in Table~\ref{tab:datasets_private}.

\subsection{Patents based datasets}

The Patents dataset provides information on patents and trademarks granted by the United States Patent and Trademark Office\footnote{\url{https://bulkdata.uspto.gov/}}. \textcolor{red}{Datasets building upon this source usually do not contain labelled user interaction data.}

%Descriptions of non-public datasets based on Patents (\textit{Patents\_tiny}, \textit{Patents}, \textit{ACM H-I}) can be found in Table~\ref{tab:datasets_private}.

\subsection{Hep-TH based datasets}

The original unaltered \textit{Hep-TH}~\cite{DBLP:conf/kdd/LeskovecKF05} dataset\footnote{\url{https://snap.stanford.edu/data/cit-HepTh.html}} stems from the area of high energy physics theory. It contains papers in a graph which were published between 1993 and 2003. It was released as part of KDD Cup 2003. \textcolor{red}{Datasets building upon this source usually do not contain labelled user interaction data.}

%Descriptions of non-public datasets based on Hep-TH (\textit{Hep-TH graph}, \textit{arXiv Hep-TH}) can be found in Table~\ref{tab:datasets_private}.

\subsection{MAG based datasets}

The Microsoft Academic Graph (MAG)~\cite{msa} was an open scientific network containing metadata on academic communication activities\footnote{(shortened) \url{shorturl.at/orwXY}}.
%https://docs.microsoft.com/en-us/academic-services/graph/}}. 
Their heterogeneous graph consists of nodes representing fields of study, authors, affiliations, papers and venues. \textcolor{red}{Datasets building upon this source usually do not contain labelled user interaction data besides citation information.}

%Descriptions of non-public datasets based on MAG (\textit{MSA}, \textit{MAG 2017}, \textit{MAG 2018}) can be found in Table~\ref{tab:datasets_private}.

\subsection{Others}

The~ following~ datasets~ have~ no~ common~ underlying data source:
The \textit{BBC}\footnote{\url{http://mlg.ucd.ie/datasets/bbc.html}} dataset contains 2,225 BBC news articles which stem from 5 topics. \textcolor{red}{This dataset does not contain labelled user interaction data.}

\textit{PRSDataset}\footnote{\url{https://sites.google.com/site/tinhuynhuit/dataset}}~ contains~ 2,453~ users,~ 21,940~ items and 35,969 pairs of users and items. \textcolor{red}{This dataset contains user-item interactions.}

%Descriptions of all other non-public datasets can be found in Table~\ref{tab:datasets_private}.

\section{Evaluation}
\label{sec:evaluation}

\textcolor{red}{The performance of a paper recommendation system can be quantified by measuring how well a target value has been approximated by the recommended publications. Relevancy estimations of papers can come from different sources, such as human ratings or datasets. Different interactions 
derived from clicked or liked papers determine the target values which a recommendation system should approximate. The quality of the recommendation can be described by evaluation measures such as precision or MRR.
For example, a dataset could provide information on clicked papers, that are then deemed relevant. The target value which should be approximated with the recommender system are those clicked papers, and the percentage of the recommendations which are contained in the clicked papers could then be reported as the system's precision.}

Due~ to~ the~ vast~ differences~ in~ approaches~ and datasets used to apply the methods, there is also a spectrum of used evaluation measures and objectives. In this section, first we observe different notions of relevance of recommended papers and individual assessment strategies for relevance. Afterwards we analyse commonly used evaluation measures and list ones which are only rarely encountered in evaluation of paper recommendation systems. Lastly we shed light on the different types of evaluation which authors conducted.

In this discussion we again only consider paper recommendation systems which also evaluate their actual approach. We disregard approaches which do evaluate other properties~\cite{DBLP:journals/elektrik/AhmadA20,DBLP:conf/tencon/ChaudhuriSS19,DBLP:journals/scientometrics/HabibA19,DBLP:conf/iui/RahdariB19,renuka,zhang20} or contain no evaluation~\cite{DBLP:conf/seke/LiuKCQ19,zarvel}.~ Thus~ we~ observe~ 54~ different~ approaches in this analysis.

\begin{table}[t]
    \scriptsize
    \resizebox{.47\textwidth}{!}{
        \centering

    \begin{tabu}{l|[1.5pt]l|l|l|[1.5pt]l|l|l|l|l|l|l|[1.5pt]l|l|l|l|l|l|l}
    & \multicolumn{3}{c|[1.5pt]}{\textbf{Relevancy}}  & \multicolumn{7}{c|[1.5pt]}{\textbf{Target value}} & \multicolumn{7}{c}{\textbf{Measures}} \\\hline
         \rotatebox[origin=c]{90}{Work} 	&	 \rotatebox[origin=c]{90}{Human rating}	&	 \rotatebox[origin=c]{90}{Dataset}	&	 \rotatebox[origin=c]{90}{Papers} 	&	 \rotatebox[origin=c]{90}{Clicked} 	&	 \rotatebox[origin=c]{90}{Read}	&	 \rotatebox[origin=c]{90}{Cited}	&	 \rotatebox[origin=c]{90}{Liked}	&	 \rotatebox[origin=c]{90}{Relevancy} 	&	 \rotatebox[origin=c]{90}{Other user}	&	 \rotatebox[origin=c]{90}{Other automatic}	&	  \rotatebox[origin=c]{90}{Precision}	&	 \rotatebox[origin=c]{90}{Recall} 	&	  \rotatebox[origin=c]{90}{F1} 	&	 \rotatebox[origin=c]{90}{nDCG} 	&	 \rotatebox[origin=c]{90}{MRR} 	&	 \rotatebox[origin=c]{90}{MAP}  	&	 \rotatebox[origin=c]{90}{Other} \\\hline
\cite{DBLP:conf/flairs/AfsarCF21} 	&	 	&	 	&	 $\bullet$ 	&	 $\bullet$	&	 	&	 	&	 	&	 	&		&		&	 	&	 	&	 	&	 	&	 	&	 	&	 $\bullet$\\ 
\cite{ahmedi} 	&	 	&	 $\bullet$	&	 	&	 	&	 	&	 	&	 $\bullet$	&	 	&	 	&		&	 	&	 $\bullet$	&	 	&	 	&	 $\bullet$	&	 	&	 \\ 
\cite{DBLP:conf/icmla/AlfarhoodC19} 	&	 	&	 $\bullet$	&	 	&	 	&	 	&	 	&	 $\bullet$	&	 	&	 	&		&	 	&	 $\bullet$	&	 	&	 	&	 	&	 	&	 $\bullet$\\ 
\cite{DBLP:journals/kbs/AliQMAA20} 	&	 	&	 $\bullet$	&	 	&	 	&	 	&	 	&	 	&	 	&		&	 $\bullet$	&	 	&	 $\bullet$	&	 	&	 	&	 $\bullet$	&	 $\bullet$	&	 \\ 
\cite{Bereczki1587420} 	&	 	&	 $\bullet$	&	 	&	 	&	 $\bullet$	&	 	&	 	&	 	&	 	&		&	 	&	 	&	 	&	 	&	 $\bullet$	&	 $\bullet$	&	 \\ 
\cite{Bulut2018APR} 	&	 $\bullet$	&	 	&	 	&	 	&	 	&	 	&	 	&	 $\bullet$	&	 	&		&	 $\bullet$	&	 $\bullet$	&	 $\bullet$	&	 	&	 	&	 	&	 \\ 
\cite{Bulut2020} 	&	 $\bullet$	&	 	&	 	&	 	&	 	&	 $\bullet$	&	 	&	 $\bullet$	&	 	&		&	 $\bullet$	&	 $\bullet$	&	 $\bullet$	&	 	&	 	&	 	&	 $\bullet$\\ 
\cite{DBLP:journals/corr/abs-2107-07831} 	&	 $\bullet$	&	 	&	 	&	 	&	 	&	 	&	 	&	 	&	 $\bullet$	&		&	 $\bullet$	&	 	&	 	&	 	&	 	&	 	&	 $\bullet$\\ 
\cite{DBLP:journals/jodl/ChaudhuriSSS21} 	&	 $\bullet$	&	 	&	 	&	 $\bullet$	&	 	&	 	&	 	&	 	&	 	&		&	 $\bullet$	&	 $\bullet$	&	 	&	 	&	 	&	 $\bullet$	&	 $\bullet$\\ 
\cite{chenban} 	&	 	&	 $\bullet$	&	 	&	 	&	 	&	 	&	 	&	 $\bullet$	&	 	&		&	 	&	 	&	 	&	 $\bullet$	&	 $\bullet$	&	 	&	 \\ 
\cite{DBLP:conf/jcdl/CollinsB19} 	&	 $\bullet$	&	 	&	 	&	 $\bullet$	&	 	&	 	&	 	&	 	&	 	&		&	 	&	 	&	 	&	 	&	 	&	 	&	 $\bullet$\\ 
\cite{DBLP:conf/aiccsa/DuGWHZG20} 	&	 	&	 	&	 $\bullet$ 	&	 	&	 	&	 	&	 	&	 	&		&	 $\bullet$	&	 	&	 $\bullet$	&	 	&	 	&	 $\bullet$	&	 $\bullet$	&	 \\ 
\cite{DBLP:journals/tkde/DuTD21} 	&	 $\bullet$	&	 $\bullet$	&	 	&	 $\bullet$	&	 	&	 	&	 	&	 $\bullet$	&		&	 	&	 	&	 	&	 	&	 $\bullet$	&	 	&	 	&	 \\ 
\cite{DBLP:conf/aaai/GuoCZLDH20} 	&	 	&	 $\bullet$	&	 	&	 	&	 	&	 	&	 $\bullet$	&	 	&		&	 	&	 $\bullet$	&	 	&	 	&	 $\bullet$	&	 $\bullet$	&	 	&	 $\bullet$\\ 
\cite{DBLP:journals/scientometrics/HarunaIQKHMC20} 	&	 	&	 $\bullet$	&	 	&	 	&	 	&	 	&	 	&	 $\bullet$	&		&	 	&	 $\bullet$	&	 $\bullet$	&	 $\bullet$	&	 	&	 $\bullet$	&	 $\bullet$	&	 \\ 
\cite{DBLP:conf/ifip12/HuMLH20} 	&	 	&	 	&	 $\bullet$ 	&	 	&	 	&	 $\bullet$	&	 	&	 	&		&	 	&	 	&	 	&	 	&	 	&	 	&	 	&	 $\bullet$\\ 
\cite{DBLP:conf/wise/HuaCLZZ20} 	&	 $\bullet$	&	 	&	 	&	 	&	 	&	 	&	 	&	 $\bullet$	&		&	 	&	 $\bullet$	&	 $\bullet$	&	 	&	 	&	 	&	 	&	 \\ 
\cite{DBLP:conf/ml4cs/JingY20} 	&	 	&	 	&	 $\bullet$ 	&	 	&	 	&	 $\bullet$	&	 	&	 	&		&	 	&	 	&	 $\bullet$	&	 	&	 $\bullet$	&	 	&	 	&	 \\ 
\cite{kanakia19} 	&	 $\bullet$	&	 	&	 	&	 	&	 	&	 	&	 	&	 $\bullet$	&		&	 	&	 $\bullet$	&	 	&	 	&	 $\bullet$	&	 	&	 	&	 \\ 
\cite{KANG20212020BDP0008} 	&	 $\bullet$	&	 	&	 $\bullet$ 	&	 	&	 	&	 	&	 	&	 $\bullet$	&		&	 	&	 	&	 $\bullet$	&	 	&	 	&	 	&	 	&	 $\bullet$\\ 
\cite{DBLP:journals/tetc/KongMWLX21} 	&	 	&	 	&	 $\bullet$ 	&	 	&	 	&	 $\bullet$	&	 	&	 	&		&	 	&	 $\bullet$	&	 $\bullet$	&	 $\bullet$	&	 $\bullet$	&	 	&	 	&	 \\ 
\cite{DBLP:conf/cscwd/LLP21} 	&	 	&	 $\bullet$	&	 	&	 	&	 	&	 	&	 	&	 	&		&	 $\bullet$	&	 	&	 	&	 	&	 	&	 	&	 	&	 $\bullet$\\ 
\cite{li20} 	&	 	&	 $\bullet$	&	 	&	 $\bullet$	&	 	&	 	&	 	&	 	&		&	 	&	 $\bullet$	&	 	&	 	&	 	&	 	&	 $\bullet$	&	 \\ 
\cite{DBLP:journals/tois/LiCPR19} 	&	 	&	 	&	 $\bullet$ 	&	 	&	 	&	 	&	 	&	 	&		&	 $\bullet$	&	 $\bullet$	&	 $\bullet$	&	 $\bullet$	&	 	&	 	&	 	&	 \\ 
\cite{DBLP:journals/dss/LiWNLL21} 	&	 	&	 	&	 $\bullet$ 	&	 	&	 	&	 $\bullet$	&	 	&	 	&		&	 	&	 $\bullet$	&	 $\bullet$	&	 	&	 	&	 	&	 	&	 \\ 
\cite{lin21} 	&	 	&	 	&	 $\bullet$ 	&	 	&	 	&	 $\bullet$	&	 	&	 	&		&	 	&	 	&	 	&	 	&	 	&	 	&	 $\bullet$	&	 \\ 
\cite{DBLP:journals/complexity/LiuKYQ20} 	&	 	&	 	&	 $\bullet$ 	&	 	&	 	&	 	&	 	&	 	&		&	 $\bullet$	&	 $\bullet$	&	 $\bullet$	&	 $\bullet$	&	 	&	 	&	 	&	 $\bullet$\\ 
\cite{DBLP:conf/cscwd/LuHCP021} 	&	 	&	 $\bullet$	&	 	&	 	&	 	&	 	&	 $\bullet$	&	 	&	 	&		&	 	&	 	&	 	&	 $\bullet$	&	 $\bullet$	&	 	&	 $\bullet$\\ 
\cite{DBLP:journals/access/MaW19a} 	&	 	&	 	&	 $\bullet$ 	&	 	&	 	&	 $\bullet$	&	 	&	 	&	 	&		&	 	&	 	&	 	&	 	&	 	&	 	&	 $\bullet$\\ 
\cite{DBLP:journals/monet/MaZZ19} 	&	 	&	 	&	 $\bullet$ 	&	 	&	 	&	 $\bullet$	&	 	&	 	&	 	&		&	 $\bullet$	&	 $\bullet$	&	 $\bullet$	&	 	&	 	&	 	&	 \\ 
\cite{DBLP:journals/ijiit/Manju} 	&	 $\bullet$	&	 	&	 	&	 	&	 	&	 	&	 	&	 $\bullet$	&	 	&		&	 $\bullet$	&	 	&	 	&	 $\bullet$	&	 $\bullet$	&	 	&	 $\bullet$\\ 
\cite{MohamedHassan} 	&	 	&	 $\bullet$	&	 	&	 	&	 	&	 	&	 $\bullet$	&	 	&	 	&		&	 	&	 $\bullet$	&	 	&	 	&	 	&	 	&	 \\ 
\cite{nair} 	&	 	&	 	&	 $\bullet$ 	&	 	&	 	&	 $\bullet$	&	 	&	 	&	 	&		&	 	&	 	&	 	&	 	&	 	&	 	&	 $\bullet$\\ 
\cite{DBLP:conf/icadl/NishiokaHS19,DBLP:conf/ercimdl/NishiokaHS19,DBLP:journals/peerj-cs/NishiokaHS20} 	&	 $\bullet$	&	 	&	 	&	 	&	 	&	 	&	 	&	 	&	 $\bullet$	&		&	 	&	 	&	 	&	 	&	 	&	 	&	 $\bullet$\\ 
\cite{DBLP:journals/access/SakibAH20} 	&	 	&	 $\bullet$	&	 	&	 	&	 	&	 	&	 	&	 $\bullet$	&	 	&		&	 $\bullet$	&	 $\bullet$	&	 $\bullet$	&	 	&	 $\bullet$	&	 $\bullet$	&	 \\ 
\cite{DBLP:journals/access/SakibAABHHG21} 	&	 	&	 $\bullet$	&	 	&	 	&	 	&	 	&	 	&	 $\bullet$	&	 	&		&	 $\bullet$	&	 $\bullet$	&	 $\bullet$	&	 	&	 $\bullet$	&	 $\bullet$	&	 \\ 
\cite{DBLP:journals/peerj-cs/ShahidAAAA21,shahid21a} 	& $\bullet$	&	 	&	 $\bullet$ 	&	 	&	 	&	 	&	 	&	 $\bullet$	&	& $\bullet$	&	 	&	 	&	 	&	 $\bullet$	&	 	&	 	&	 $\bullet$\\ 
\cite{Sharma} 	&	 	&	 $\bullet$	&	 	&	 	&	 	&	 	&	 	&	 $\bullet$	&	 	&		&	 $\bullet$	&	 $\bullet$	&	 $\bullet$	&	 	&	 	&	 	&	 \\ 
\cite{DBLP:conf/ijcnn/ShiMZJLC20} 	&	 $\bullet$	&	 	&	 	&	 	&	 	&	 	&	 	&	 $\bullet$	&	 	&		&	 $\bullet$	&	 	&	 	&	 	&	 $\bullet$	&	 $\bullet$	&	 \\ 
\cite{subathra} 	&	 	&	 	&	 $\bullet$ 	&	 	&	 	&	 	&	 	&	 	&	 $\bullet$	&		&	 $\bullet$	&	 $\bullet$	&	 	&	 	&	 	&	 	&	 \\ 
\cite{DBLP:journals/concurrency/TangLQ21} 	&	 	&	 $\bullet$	&	 	&	 $\bullet$	&	 	&	 	&	 $\bullet$	&	 	&	 	&		&	 $\bullet$	&	 $\bullet$	&	 $\bullet$	&	 	&	 	&	 	&	 $\bullet$\\ 
\cite{DBLP:conf/bigdataconf/TannerAH19} 	&	 	&	 	&	 $\bullet$ 	&	 	&	 	&	 	&	 	&	 	&		&	 $\bullet$	&	 $\bullet$	&	 $\bullet$	&	 $\bullet$	&	 	&	 	&	 	&	 \\ 
\cite{tao} 	&	 	&	 	&	 $\bullet$ 	&	 	&	 	&	 $\bullet$	&	 	&	 	&		&	 	&	 	&	 	&	 	&	 	&	 	&	 	&	 $\bullet$\\ 
\cite{DBLP:journals/access/WaheedIRMK19} 	&	 $\bullet$	&	 	&	 	&	 	&	 	&	 	&	 	&	 $\bullet$	&		&	 	&	 	&	 	&	 	&	 $\bullet$	&	 $\bullet$	&	 	&	 $\bullet$\\ 
\cite{DBLP:conf/service/WangXTWX20} 	&	 	&	 	&	 $\bullet$ 	&	 	&	 	&	 $\bullet$	&	 	&	 	&		&	 	&	 	&	 	&	 	&	 $\bullet$	&	 	&	 	&	 $\bullet$\\ 
\cite{DBLP:journals/corr/abs-2103-08819} 	&	 $\bullet$ 	&	 $\bullet$	&	 	&	 	&	 	&	 	&	 	&	 	&		&	 $\bullet$	&	 	&	 	&	 	&	 	&	 $\bullet$	&	 	&	 $\bullet$\\ 
\cite{DBLP:journals/asc/WangWYXYY21} 	&	 	&	 $\bullet$	&	 	&	 	&	 	&	 	&	 $\bullet$	&	 	&		&	 	&	 $\bullet$	&	 $\bullet$	&	 	&	 	&	 	&	 $\bullet$	&	 \\ 
\cite{wang21} 	&	 	&	 $\bullet$	&	 	&	 	&	 	&	 	&	 $\bullet$	&	 	&		&	 	&	 $\bullet$	&	 $\bullet$	&	 	&	 	&	 $\bullet$	&	 $\bullet$	&	 \\ 
\cite{DBLP:conf/sigir/XieSB21} 	&	 	&	 	&	 $\bullet$ 	&	 	&	 	&	 $\bullet$	&	 	&	 	&		&	 	&	 	&	 	&	 	&	 $\bullet$	&	 	&	 	&	 \\ 
\cite{xie21} 	&	 	&	 $\bullet$	&	 	&	 	&	 	&	 	&	 	&	 	&		&	 $\bullet$	&	 	&	 	&	 	&	 $\bullet$	&	 	&	 	&	 \\ 
\cite{DBLP:journals/www/YangLLLZZZZ19} 	&	 	&	 	&	 $\bullet$ 	&	 	&	 	&	 $\bullet$	&	 	&	 	&		&	 	&	 $\bullet$	&	 $\bullet$	&	 	&	 	&	 	&	 	&	 \\ 
\cite{DBLP:conf/ksem/YuHLZXLXY19} 	&	 	&	 $\bullet$	&	 	&	 	&	 	&	 	&	 $\bullet$	&	 	&		&	 	&	 	&	 	&	 	&	 $\bullet$	&	 	&	 	&	 $\bullet$\\ 
\cite{ZHANG2019616} 	&	 	&	 	&	 $\bullet$ 	&	 	&	 	&	 	&	 	&	 	&		&	 $\bullet$	&	 	&	 	&	 	&	 	&	 	&	 	&	 $\bullet$\\ 
\cite{DBLP:journals/access/ZhaoKFMN20} 	&	 	&	 $\bullet$	&	 	&	 	&	 	&	 	&	 $\bullet$	&	 	&		&	 	&	 $\bullet$	&	 $\bullet$	&	 $\bullet$	&	 	&	 	&	 	&	 $\bullet$\\ 	
    \end{tabu}
    }
    \caption{Indications whether approaches utilise the specified relevancy definitions, target values of evaluations and evaluation measures.}
    \label{tab:evalmeasures_all}
\end{table}

\subsection{Relevance and Assessment}

%Relevance can be assessed in different ways. Papers which users deem relevant can be expressed by e.g., clicking~\cite{DBLP:journals/concurrency/TangLQ21} or referencing them~\cite{DBLP:journals/dss/LiWNLL21,DBLP:conf/sigir/XieSB21}. 

Relevance of recommended publications can be evaluated against multiple target values: clicked papers~\cite{DBLP:journals/corr/abs-2107-07831,DBLP:journals/tois/LiCPR19,DBLP:journals/concurrency/TangLQ21}, references~\cite{DBLP:conf/ml4cs/JingY20,DBLP:conf/sigir/XieSB21}, references of recently authored papers~\cite{DBLP:journals/dss/LiWNLL21}, papers an author interacted with in the past~\cite{DBLP:conf/cscwd/LLP21}, degree-of-relevancy which is determined by citation strength~\cite{shahid21a}, a ranking based on future citation numbers~\cite{ZHANG2019616} as well as papers accepted~\cite{DBLP:journals/jodl/ChaudhuriSSS21} or deemed relevant by authors~\cite{DBLP:journals/scientometrics/HarunaIQKHMC20,DBLP:journals/access/SakibAABHHG21}. % DBLP:conf/jcdl/SugiyamaK10,DBLP:conf/jcdl/SugiyamaK13,DBLP:journals/jodl/SugiyamaK15,

Assessing the relevance of recommendations can also be conducted in different ways: the top n papers recommended by a system can be judged by either a referee team~\cite{DBLP:journals/corr/abs-2103-08819} or single persons~\cite{DBLP:journals/jodl/ChaudhuriSSS21,DBLP:conf/icadl/NishiokaHS19,DBLP:conf/ercimdl/NishiokaHS19}. Other options for relevance assessment are the usage of a dataset with user ratings~\cite{DBLP:journals/scientometrics/HarunaIQKHMC20,DBLP:journals/access/SakibAABHHG21} or emulation of users and their interests~\cite{DBLP:conf/flairs/AfsarCF21,DBLP:journals/dss/LiWNLL21}.

Table~\ref{tab:evalmeasures_all} holds information on utilised relevance indicators and target values which indicate relevance for the 54 discussed approaches. \textit{Relevancy} describes the method that defines which of the recommended papers are relevant:
\begin{itemize}
    \item Human rating: The approach is evaluated using assessments of real users of results specific to the approach.
    \item Dataset: The approach is evaluated using some type of assessment of a target value which is not specific to the approach but from a dataset. The assessment was either conducted for another approach and re-used or it was collected independent of an approach.
    \item Papers: The approach is evaluated by some type of assessment of a target value which is directly generated from the papers contained in the dataset such as citations or their keywords. 
\end{itemize}

The \textit{target values} in Table~\ref{tab:evalmeasures_all} describe the entities which the approach tried to approximate:
\begin{itemize}
    \item Clicked: The approximated target value is derived from users' clicks on papers.
    \item Read:~ The~ approximated~ target~ value~ is~ derived from users' read papers.
    \item Cited:~ The~ approximated~ target~ value~ is~ derived from cited papers.
    \item Liked:~ The~ approximated~ target~ value~ is~ derived from users' liked papers.
    \item Relevancy: The approximated target value is derived from users' relevance assessment of papers.
    \item \textcolor{red}{Other user: The approximated target value is derived from other entities associated with a user input, e.g. acceptance of users, users' interest and relevancy of the recommended papers' topics.}
    \item \textcolor{red}{Other automatic: The approximated target value is automatically derived from other entities, e.g. user profiles, papers with identical references, degree-of-relevancy, keywords extracted from papers, papers containing~ the~ query~ keywords~ in~ the~ optimal Steiner tree, neighbouring (cited and referencing) papers, included keywords, the classification tag, future citation numbers and an unknown measure derived from a dataset. We refrain from trying to introduce sub-categories for this broad field.}
\end{itemize}

Only~ three~ approaches~ evaluate~ against~ multiple target values~\cite{Bulut2020,DBLP:journals/tkde/DuTD21,DBLP:journals/concurrency/TangLQ21}.
Six approaches (11.11\%) utilise clicks of users, only one approach (1.85\%) uses read papers as target value. Even though cited papers are not the main objective of paper recommendation systems but rather citation recommendation systems, this target was approximated by 13 (24.07\%) of the observed systems. Ten approaches (18.52\%) evaluated against liked papers, 15 (27.78\%) against relevant papers and 13 (24.07\%) against some other target value\textcolor{red}{, either user input (three, 5.55\%) or automatically derived (ten, 18.52\%)}.

\subsection{Evaluation Measures}

We differentiate between commonly used and rarely used evaluation measures for the task of scientific paper recommendation. They are described in the following sections.
Table~\ref{tab:evalmeasures_all} holds indications of utilised evaluation measures for the 54 discussed approaches. 
\textit{Measures} are the methods used to evaluate the approach's ability to approximate the target value which can be of type precision, recall, f1 measure, nDCG, MRR, MAP or another one. %\textcolor{red}{We followed Bai et al.~\cite{DBLP:journals/access/BaiWLYKX19} in observing these regularly used evaluation measures.} 

Out of the observed systems, twelve\footnote{\textcolor{red}{One approach is described in three papers.}} approaches~\cite{DBLP:conf/flairs/AfsarCF21,DBLP:conf/jcdl/CollinsB19,DBLP:journals/tkde/DuTD21,DBLP:conf/cscwd/LLP21,lin21,DBLP:journals/monet/MaZZ19,MohamedHassan,nair,DBLP:conf/icadl/NishiokaHS19,DBLP:conf/ercimdl/NishiokaHS19,DBLP:journals/peerj-cs/NishiokaHS20,tao,xie21,DBLP:conf/sigir/XieSB21} (22.22\%) \\only report one single measure, all others report at least two different ones.

\subsubsection{Commonly Used Evaluation Measures}

\begin{table}[t]
    \centering
   \begin{tabular}{l|l|l|l|l|l|l}
     &  P    & R  & F1  & nDCG  & MRR   & MAP   \\ \hline
    \% &      48.15     &   24.07 &  50      &    25.92   &   27.78    &    22.22   \\
\end{tabular}
    \caption{Common evaluation measures and percentage of observed evaluations of paper recommendation systems in which they were applied. Percentages are rounded to two decimal places.}
    \label{tab:common_eval_measures}
\end{table}

Bai~ et~ al~\cite{DBLP:journals/access/BaiWLYKX19}~ identify~ \textit{precision}~ (P),~ \textit{recall}~ (R),~ \textit{F1}, \textit{nDCG}, \textit{MRR} and \textit{MAP} as evaluation features which have been used regularly in the area of paper recommendation systems. Table~\ref{tab:common_eval_measures} gives usage percentages of each of these measures in observed related work.

Alfarhood and Cheng~\cite{DBLP:conf/icmla/AlfarhoodC19} argue against the use of precision when utilising implicit feedback. If a user gives no feedback for a paper it could either mean disinterest or that a user does not know of the existence of the specific publication.

\subsubsection{Rarely used Evaluation Measures}

\begin{table*}[t]
    \centering
    \scriptsize
    \begin{tabular}{l|l|l}
        Measure & Used by & Description\\\hline
        Average precision & \cite{DBLP:journals/access/WaheedIRMK19} & area under precision-recall curve\\
        Receiver operating characteristic & \cite{ZHANG2019616} & plot of true positives against false positives\\
        AUC & \cite{DBLP:conf/aaai/GuoCZLDH20,DBLP:journals/concurrency/TangLQ21} & area under receiver operating characteristic curve\\
        
        Computation time &  \cite{DBLP:journals/jodl/ChaudhuriSSS21,DBLP:journals/complexity/LiuKYQ20} & time to compute recommendation list\\
        DCG & \cite{DBLP:conf/icmla/AlfarhoodC19} & summed up relevancy divided by logarithm of rank + 1\\
        Click-through-rates & \cite{DBLP:journals/corr/abs-2107-07831,DBLP:conf/jcdl/CollinsB19} & percentage of Clicks on recommendations\\
        Reward & \cite{DBLP:conf/flairs/AfsarCF21,DBLP:conf/cikm/GingstadJB20} & \makecell[tl]{weighted sum of interactions of users with recommendations, e.g. clicked and saved \\papers}\\

        Spearman correlation coefficient & \cite{kanakia19,ZHANG2019616} & correlation between ranks of paper lists\\
        Hit ratio & \cite{DBLP:conf/cscwd/LuHCP021,DBLP:conf/service/WangXTWX20,DBLP:conf/ksem/YuHLZXLXY19} & percentage of relevant items in top $k$ recommendations\\
        Accuracy & \cite{Bulut2020,DBLP:journals/monet/MaZZ19,DBLP:journals/tjs/ShahidAABZYC20} & percentage of relevant papers which the approach identified\\
        Specificity & \cite{Bulut2020} & true negative rate\\
        Mean absolute error & \cite{DBLP:conf/ifip12/HuMLH20} & average difference between real and predicted values\\
        Root mean square error & \cite{DBLP:conf/ifip12/HuMLH20} & expected squared difference between real and predicted values\\
        Fallout & \cite{DBLP:journals/ijiit/Manju} & percentage of irrelevant recommendations out of all irrelevant papers\\
        Support & \cite{nair} & frequency of occurrences of set\\
        TopN & \cite{DBLP:journals/corr/abs-2103-08819} & probability that target keywords are encountered in first n recommended papers\\
        FindN & \cite{DBLP:journals/corr/abs-2103-08819} & number of target keywords which are encountered in first n recommended papers\\
        Coverage & \cite{DBLP:journals/access/ZhaoKFMN20} & method's ability to discover the long tail of papers\\
        Popularity & \cite{DBLP:journals/access/ZhaoKFMN20} & \makecell[tl]{average logarithm of the number of ratings of papers in recommendation, \\indicates novelty of results}\\
        Average paper popularity & \cite{DBLP:journals/complexity/LiuKYQ20} & paper popularity divided by number of recommendations\\

        Intra-list similarity & \cite{DBLP:journals/access/ZhaoKFMN20} & \makecell[tl]{dissimilarity between recommended papers, smaller value indicates more diverse \\recommendation}\\
        Serendipity score & \cite{DBLP:conf/icadl/NishiokaHS19,DBLP:conf/ercimdl/NishiokaHS19,DBLP:journals/peerj-cs/NishiokaHS20}& summed up usefulness divided by unexpectedness of recommended papers\\
        Success rate & \cite{DBLP:journals/complexity/LiuKYQ20} & number of recommendations < 2 $\times$ number of keywords\\
        Number of recommended papers & \cite{DBLP:journals/complexity/LiuKYQ20} & size of set of recommended papers\\
    \end{tabular}
    \caption{Overview of rare existing measures used in evaluations of observed approaches.}
    \label{tab:eval_rare_existing_measures}
\end{table*}

We found a plethora of rarer used evaluation measures which have either been utilised only by the work they were introduced in or to evaluate few approaches. Our analysis in this aspect might be highly influenced by the narrow time frame we observe. Novel measures might require more time to be adopted by a broader audience. Thus we differentiate between novel rarely used evaluation measures and ones where authors do not explicitly claim they are novel. A list of rare but already defined evaluation measures can be found in Table~\ref{tab:eval_rare_existing_measures}. In total 25 approaches (46.3\%) did use an evaluation measure not considered common.

\textbf{Novel rarely used Evaluation Measures.}
In our considered approaches we only encountered three novel evaluation measures:
\textit{Recommendation quality} as defined by Chaudhuri et al.~\cite{DBLP:journals/jodl/ChaudhuriSSS21} is the acceptance of recommendations by users rated on a Likert scale from 1 to 10.

\textit{TotNP\_EU} is a measure defined by Manju et al.~\cite{DBLP:journals/ijiit/Manju} specifically introduced for measuring performance of approaches regarding the cold start problem. It indicates the number of new publications suggested to users with a prediction value above a certain threshold.

\textit{TotNP\_AVG} is another measure defined by Manju et al.~\cite{DBLP:journals/ijiit/Manju} for measuring performance of approaches regarding the cold start problem. It indicates the average number of new publications suggested to users with a prediction value above a certain threshold.

\subsection{Evaluation Types}
\label{eval:eval_types}

Evaluations can be classified into different categories. We follow the notion of Beel and Langer~\cite{DBLP:conf/ercimdl/BeelL15} who differentiate between user studies, online evaluations and offline evaluations. They define \textit{user studies} as ones where users' satisfaction with recommendation results is measured by collecting explicit ratings. \textit{Online evaluations} are ones where users do not explicitly rate the recommendation results; relevancy is derived from e.g. clicks.
In \textit{offline evaluations} a ground truth is used to evaluate the approach. 

%As papers with user studies we only consider those which in some way actively acquired users to evaluate paper recommendations. Here we do not consider utilisation of datasets from possible previous evaluations, emulated users or the collection of click data only. We also do not discuss studies which do not mention the number of participants or the composition of the user group, e.g., as seen with Shi et al.~\cite{DBLP:conf/ijcnn/ShiMZJLC20}. From the 54 of approaches we observed only \textcolor{red}{this} (. \%) did conduct a user study which corresponds to this definition.

From the 54 observed approaches we found four using multiple evaluation types~\cite{DBLP:conf/aiccsa/DuGWHZG20,KANG20212020BDP0008,DBLP:journals/tjs/ShahidAABZYC20,shahid21a,DBLP:journals/corr/abs-2103-08819}.
Twelve (22.22\%) were conducting user studies which describe the size and composition of the participant group.\footnote{Shi et al.~\cite{DBLP:conf/ijcnn/ShiMZJLC20} also conduct a user study but do not describe their participants.}
Only two approaches~\cite{DBLP:conf/jcdl/CollinsB19,DBLP:journals/ijiit/Manju} (3.7\%) in the observed papers were evaluated with an online evaluation.
We found 44 approaches (81.48\%) providing an offline evaluation.~
Offline~ evaluations~ being~ the~ most~ common form of evaluation is unsurprising as this tendency has also been observed in an evaluation of general scientific recommender systems~\cite{DBLP:journals/kais/ChampiriAS19}. 
Offline evaluations are fast and do not require users~\cite{DBLP:journals/kais/ChampiriAS19}. Nevertheless the margin by which this form of evaluation is conducted could be rather surprising.

A distinction in \textit{lab-based} vs. \textit{real world} user studies can be conducted~\cite{DBLP:journals/jodl/BeelGLB16,DBLP:conf/ercimdl/BeelL15}. User studies where participants rate recommendations according to some criteria and are aware of the study are lab-based, all others are considered real world studies. Living labs~\cite{DBLP:conf/cikm/GingstadJB20,DBLP:conf/ecir/BeelCKDK19,DBLP:conf/clef/SchaerBCWST21} for example enable real world user studies.
On average the lab-based user studies were conducted with 17.83 users. Table~\ref{tab:user_study} holds information on the number of participants for all studies as well as the composition of groups in terms of seniority. 

For offline evaluation, they can either be ones with an \textit{explicit} ground truth given by a dataset containing user rankings, \textit{implicit} ones by deriving user interactions such as liked or cited papers or \textit{expert} ones with manually collected expert ratings~\cite{DBLP:conf/ercimdl/BeelL15}. We found 22 explicit offline evaluations (40.74\%) corresponding to ones using datasets to estimate relevance (see Table~\ref{tab:evalmeasures_all}) and 21 implicit offline evaluations (38.89\%) corresponding to ones using paper information to identify relevant recommendations (see Table~\ref{tab:evalmeasures_all}). We did not find any expert offline evaluations.

\begin{table*}[t]
\scriptsize
    \centering
    \begin{tabular}{l|l|l}
        Work & \#P & Composition \\ \hline
        Bulut et al.~\cite{Bulut2018APR}& 50 & \makecell[tl]{PhD students studying in Turkey in 2019}\\
        Bulut et al.~\cite{Bulut2020} & 10 + 30 & researchers\\
        Chaudhuri et al.~\cite{DBLP:journals/corr/abs-2107-07831} & 50 & NA\\
        Chaudhuri et al.~\cite{DBLP:journals/jodl/ChaudhuriSSS21} & 45 & \makecell[tl]{from 9 different areas, different seniority levels: 12 faculty members, 20 postgraduate students, \\13 undergraduate students}\\
        %Collins and Beel~\cite{DBLP:conf/jcdl/CollinsB19} & NA & users of Sowiport or Jabref\\
        Du et al.~\cite{DBLP:journals/tkde/DuTD21} & NA & college students or patent analysis experts\\
        Hua et al.~\cite{DBLP:conf/wise/HuaCLZZ20} & 10 & experts\\
        Kanakia et al.~\cite{kanakia19} & 40 &  \makecell[tl]{full-time computer science researchers at Microsoft Research}\\
        %Manju et al.~\cite{DBLP:journals/ijiit/Manju} & 4200 & Mendeley users between March and May 2015\\
        Kang et al.~\cite{KANG20212020BDP0008} & 12 & postgraduates\\
        Nishioka et al.~\cite{DBLP:conf/icadl/NishiokaHS19,DBLP:conf/ercimdl/NishiokaHS19,DBLP:journals/peerj-cs/NishiokaHS20} & 22 & \makecell[tl]{seniority based on highest degree: 2 Master's, 13 PhD, 7 lecturers/professors; 2 female, 20 \\male; 17 working in academia, 3 working in industry}\\
        %Rahdari and Brusilovsky~\cite{DBLP:conf/iui/RahdariB19} & 15 & attendees of the EC-TEL 2018 conference\\
        Shahid et al.~\cite{DBLP:journals/peerj-cs/ShahidAAAA21} & 20 & post-graduate students\\
        Waheed et al.~\cite{DBLP:journals/access/WaheedIRMK19} & 20 & researchers \\
        Wang et al.~\cite{DBLP:journals/corr/abs-2103-08819} & 5 & 1 doctoral supervisor, 2 master supervisors, 2 graduate students\\
    \end{tabular}
    \caption{For all observed works with user studies we list their number of participants (\#P) and their composition. NA indicates that \#P or compositions were not described in a specific user study.}
    \label{tab:user_study}
\end{table*}

\section{Changes compared to 2016}

This chapter briefly summarises some of the changes in the set of papers we observed when compared to the study by Beel et al.~\cite{DBLP:journals/jodl/BeelGLB16}. Before we start the comparison, we want to point to the fact that we observed papers from two years in which the publication process could have been massively affected by the COVID-19 pandemic.

\subsection{Number of Papers per Year and Publication Medium}

Beel et al.~\cite{DBLP:journals/jodl/BeelGLB16} studied works between and including 1998 and 2013 while we observed works which appeared between January 2019 and October 2021. While the previous study did include all 185 papers (of which 96 were paper recommendation approaches) in their discussion of papers per year which were published in the area of the topic paper or citation recommendation but later on only studied 62 papers for an in-depth review, we generally only studied 65 publications which present novel paper recommendation approaches (see Section~\ref{sec:litReview_prs}) in this aspect.
Compared to the time frame observed in this previous literature review, we encountered fewer papers being published on the actual topic of scientific paper recommendation per year. In the former work, the published number of papers was rising and hitting 40 in 2013. We found this number being stuck on a constant level between 21 and 23 in the three years we observed. This could hint at differing interest in this topic over time, with a current demise or the trend to work in this area having surpassed its zenith. 

While Beel et al.~\cite{DBLP:journals/jodl/BeelGLB16} found 59\% of conference papers and 16\% of journal articles, we found 54.85\% of conference papers and 41.54\% of journal articles. The shift to journal articles could stem from a general shift towards journal articles in computer science\footnote{Compare the 99.363 journal articles and 151.617 conference papers published in 2013 to the 187.263 journal articles and 157.460 conference articles in 2021 in dblp.}.

\subsection{Classification}

While Beel et al.~\cite{DBLP:journals/jodl/BeelGLB16} found 55\% of their studied 62 papers applying methods from content-based filtering, we found only found 7.69\% (5) of our 65 papers identifying as content-based approaches. Beel et al.~\cite{DBLP:journals/jodl/BeelGLB16} report 18\% of approaches applied collaborative filtering. We encountered 4.62\% (three) having this component as part of their self-defined classification. As for graph-based recommendation approach, Beel et al.~\cite{DBLP:journals/jodl/BeelGLB16} found 16\% while we only encountered 7.69\% (five) of papers with this description. In terms of hybrid approaches, Beel et al.~\cite{DBLP:journals/jodl/BeelGLB16} encountered five (8.06\%) truly hybrid ones. In our study, we found 18 approaches (27.69\%) labelling themselves as hybrid recommendation systems.\footnote{Note that not all approaches classified their type of paper recommendation and several papers did not classify themselves in the wide-spread categorisation (see Section~\ref{sec:litReview_former}).}

\subsection{Evaluation}

\begin{table}[t]
    \centering
\begin{tabular}{l|l|l|l|l}
     & Offline & Online & User quant. & User qual.\\ \hline
    \cite{DBLP:journals/jodl/BeelGLB16} & 71 & 7 & 25 & 3\\
    current& 81.48 & 3.7 & 24.07 & 0\\
\end{tabular}
    \caption{Percentage of studies using the different methods. Some studies utilised multiple methods, thus the percentages do not add up to 100\%.}
    \label{tab:evalcomp}
\end{table}

Table~\ref{tab:evalcomp} shows the comparison of the distributions of the different types of evaluations between our study observing 54 papers with evaluations and the one conducted by Beel et al.~\cite{DBLP:journals/jodl/BeelGLB16}, which regards 75 papers for this aspect. The percentage of quantitative user studies (User quant) is comparable for both studies. A peculiar difference is the percentage of offline evaluations, which is much higher in our current time frame. 

When observing the evaluation measures, we found some differences compared to the previous study. While 48.15\% of papers with an evaluation report precision in our case, in Beel et al.'s~\cite{DBLP:journals/jodl/BeelGLB16} 72\% of approaches with an evaluation report this value. As a contrast, we found 50\% of papers reporting F1 while only 11\% of papers reported this measure according to Beel et al.~\cite{DBLP:journals/jodl/BeelGLB16}. This might hint at a shift away from precision (which Beel et al.~\cite{DBLP:journals/jodl/BeelGLB16} did describe as a problematic measure) to focus more on also incorporating recall into the quality assessment of recommendation systems.  

\subsection{Discussion}

In general, the two reviews regard different time frames. We encounter non-marginal differences in the three dimensions discussed in this Section. A more concise comparison could be made if a time slice would be regarded for both studies, such that the research output and shape could be observed from three years each. We cannot clearly identify emerging trends (as with the offline evaluation) as we do not know if it has been conducted in this percentage of papers since the 2010s or if it only just picked up to be a more wide-spread evaluation form.

\section{Open Challenges and Objectives}
\label{sec:openChallenges}

All paper recommendation approaches which were considered in this survey could have been improved in some way or another. Some papers did not conduct evaluations which would satisfy a critical reader, others could be more convincing if they compared their methods to appropriate competitors. 
The possible problems we encountered within the papers can be summarised in different open challenges, which papers should strive to overcome.
We separate our analysis and discussion of open challenges in those which have already been described by previous literature reviews (see Section~\ref{sec:chall_prev}) and ones we identify as new or emerging problems (see Section~\ref{sec:chall_new}). Lastly we briefly discuss the presented challenges (see Section~\ref{sec:chall_dis}).

\subsection{Challenges Highlighted in Previous Works}
\label{sec:chall_prev}

In the following we will explain possible shortcomings which were already explicitly discussed in previous literature reviews~\cite{DBLP:journals/access/BaiWLYKX19,DBLP:journals/jodl/BeelGLB16,DBLP:journals/tjs/ShahidAABZYC20}. We regard these challenges in light of current paper recommendation systems to identify problems which are nowadays still encountered.

\subsubsection{Neglect of User Modelling}

Neglect of user modelling has been described by Beel et al.~\cite{DBLP:journals/jodl/BeelGLB16} as identification of target audiences' information needs. They describe the trade-off between specifying keywords which brings recommendation systems closer to search engines and utilising user profiles as input.

Currently only some approaches consider users of systems to influence the recommendation outcome, as seen with Table~\ref{tab:dimensions_cat} users are not always part of the input to systems. Instead many paper recommendation systems assume that users do not state their information needs explicitly but only enter keywords or a paper.
With paper recommendation systems where users are not considered, the problem of neglecting user modelling still holds.

\subsubsection{Focus on Accuracy}

Focus on accuracy as a problem is described by Beel et al.~\cite{DBLP:journals/jodl/BeelGLB16}. They state putting users' satisfaction with recommendations on a level with accuracy of approaches does not depict reality. More factors should be considered.

Only over one fourth of current approaches do not only report precision or accuracy but also observe more diversity focused measures such as MMR. We also found usage of less widespread measures to capture different aspects such as popularity, serendipity or click-through-rate.

\subsubsection{Translating Research into Practice}

The missing translation of research into practice is described by Beel et al.~\cite{DBLP:journals/jodl/BeelGLB16}. They mention the small percentage of approaches which are available as prototype as well as the discrepancy between real world systems and methods described in scientific papers.

Only \textcolor{red}{five} of our observed approaches definitively must have been available online at any point in time~\cite{DBLP:conf/jcdl/CollinsB19,kanakia19,DBLP:journals/ijiit/Manju,DBLP:conf/iui/RahdariB19,zarvel}. We did not encounter any of the more complex approaches being used in widespread paper recommendation systems.

\subsubsection{Persistence and Authority}

\begin{table}[t]
    \centering
    \scriptsize
    \begin{tabular}{l|l}
        Group & Papers \\\hline
        Capital University of Science and Technology & \cite{DBLP:journals/elektrik/AhmadA20,DBLP:journals/scientometrics/HabibA19}\\
        Fırat University & \cite{Bulut2018APR,Bulut2020}\\
        IIT Kharagpur & \cite{DBLP:conf/tencon/ChaudhuriSS19,DBLP:journals/corr/abs-2107-07831,DBLP:journals/jodl/ChaudhuriSSS21}\\
        Qufu Normal University & \cite{DBLP:conf/seke/LiuKCQ19,DBLP:journals/complexity/LiuKYQ20}\\
        Kyoto-Kiel-Essex & \cite{DBLP:conf/icadl/NishiokaHS19,DBLP:conf/ercimdl/NishiokaHS19,DBLP:journals/peerj-cs/NishiokaHS20}\\
        University of Malaya-Bayero University & \cite{DBLP:journals/access/SakibAH20,DBLP:journals/access/SakibAABHHG21}\\
        Pakistan & \cite{DBLP:journals/peerj-cs/ShahidAAAA21,shahid21a}\\
        Hefei University of Technology & \cite{DBLP:journals/asc/WangWYXYY21,wang21}\\
        Shandong University & \cite{DBLP:conf/sigir/XieSB21,xie21}\\
        Australia & \cite{ZHANG2019616,zhang20}\\
    \end{tabular}
    \caption{Overview of research groups with multiple papers.}
    \label{tab:persistence}
\end{table}

Beel et al.~\cite{DBLP:journals/jodl/BeelGLB16} describe the lack of persistence and authority in the field of paper recommendation systems as one of the main reasons why research is not adapted in practice. 

The analysis of this possible shortcoming of current work could be highly affected by the short time period from which we observed works. We found several groups publishing multiple papers as seen in Table~\ref{tab:persistence} which corresponds to 29.69\% of approaches. The most papers a group published was three so this amount still cannot fully mark a research group as authority in the area.

\subsubsection{Cooperation}

Problems with cooperation are described by Beel et al.~\cite{DBLP:journals/jodl/BeelGLB16}. They state even though approaches have been proposed by multiple authors building upon prior work is rare. Corporations between different research groups are also only encountered sporadically.

\begin{table}[t]
    \centering
    \scriptsize
    \begin{tabular}{l|l|l|l|l|l|l|l}
        \# & 2 & 3 & 4 & 5 & 6 & 7 & 8\\\hline
        \% & 14.06 & 31.25 & 14.06 & 23.44 & 7.81 & 3.13 & 3.13
    \end{tabular}
    \caption{Percentage of the 64 considered papers with different numbers of authors (\#). Publications with 1 and 10 authors were encountered only once (1.56\% each).}
    \label{tab:aut_perc}
\end{table}

Here again we want to point to the fact that our observed time frame of less than three years might be too short to make substantive claims regarding this aspect. Table~\ref{tab:aut_perc} holds information on the different numbers of authors for papers and the percentage of papers out of the 64 observed ones which are authored by groups of this size. We only encountered little cooperation between different co-author groups (see Haruna et al.~\cite{DBLP:journals/scientometrics/HarunaIQKHMC20} and Sakib et al.~\cite{DBLP:journals/access/SakibAABHHG21} for an exception). There were several groups not extending their previous work~\cite{ZHANG2019616,zhang20}. We refrain from analysing citations of related previous approaches as our considered period of less than three years is too short for all publications to have been able to be recognised by the wider scientific community.

\subsubsection{Information Scarcity}

Information scarcity is described by Beel et al.~\cite{DBLP:journals/jodl/BeelGLB16} as researchers' tendency to only provide insufficient detail to re-implement their approaches. This leads to problems with reproducibility.

Many of the approaches we encountered did not provide sufficient information to make a re-implementation possible: with Afsar et al.~\cite{DBLP:conf/flairs/AfsarCF21} it is unclear how the knowledge graph and categories were formed, Collins and Beel~\cite{DBLP:conf/jcdl/CollinsB19} do not describe their Doc2Vec enough, Liu et al.~\cite{DBLP:journals/complexity/LiuKYQ20} do not specify the extraction of keywords for papers in the graph and Tang et al.~\cite{DBLP:journals/concurrency/TangLQ21} do not clearly describe their utilisation of Word2Vec. In general oftentimes details are missing~\cite{ahmedi,DBLP:conf/icmla/AlfarhoodC19,DBLP:conf/seke/LiuKCQ19,DBLP:journals/www/YangLLLZZZZ19}. 
Exceptions to these observations are e.g. found with Bereczki~\cite{Bereczki1587420}, Nishioka et al.~\cite{DBLP:conf/icadl/NishiokaHS19,DBLP:conf/ercimdl/NishiokaHS19,DBLP:journals/peerj-cs/NishiokaHS20} and Sakib et al.~\cite{DBLP:journals/access/SakibAABHHG21}.

We did not find a single paper's code e.g. provided as a link to GitHub.

\subsubsection{Cold Start}

Pure collaborative filtering systems encounter the cold start problem as described by Bai et al.~\cite{DBLP:journals/access/BaiWLYKX19} and Shahid et al.~\cite{DBLP:journals/tjs/ShahidAABZYC20}. If new users are considered, no historical data is available, they cannot be compared to other users to find relevant recommendations. 

While this problem still persists, most current approaches are no pure collaborative filtering based recommendation systems (see Section~\ref{sec:litReview_former}). Systems using deep learning could overcome this issue~\cite{DBLP:journals/ccsecis/LiZ19}. There are approaches specifically targeting this problem~\cite{lin21,DBLP:conf/ijcnn/ShiMZJLC20}, some~\cite{lin21} also introduced specific evaluation measures (totNP\_EU and avgNP\_EU) to quantify systems' ability to overcome the cold start problem.

\subsubsection{Sparsity or Reduce Coverage}

Bai et al.~\cite{DBLP:journals/access/BaiWLYKX19} state the user-paper-matrix being sparse for collaborative filtering based approaches. Shahid et al.~\cite{DBLP:journals/tjs/ShahidAABZYC20} also mention this problem as the \textit{reduce coverage problem}. This trait makes it hard for approaches to learn relevancy of infrequently rated papers.

Again, while this problem is still encountered, current approaches mostly are no longer pure collaborative filtering based systems but instead utilise more information (see Section~\ref{sec:litReview_former}). Using deep learning in the recommendation process might reduce the impact of this problem~\cite{DBLP:journals/ccsecis/LiZ19}.

\subsubsection{Scalability}
The~ problem~ of~ scalability~ was~ described~ by~ Bai~ et al.~\cite{DBLP:journals/access/BaiWLYKX19}.~ They~ state~ paper~ recommendation~ systems should be able to work in huge, ever expanding environments where new users and papers are added regularly.

A~ few~ approaches~\cite{DBLP:journals/scientometrics/HabibA19,KANG20212020BDP0008,DBLP:journals/access/SakibAABHHG21,DBLP:journals/corr/abs-2103-08819}~ contain~ a~ web crawling step which directly tackles challenges related to outdated or missing data.
Some approaches~\cite{DBLP:journals/jodl/ChaudhuriSSS21,DBLP:journals/complexity/LiuKYQ20} evaluate the time it takes to compute paper recommendations which also indicates their focus on this general problem.
But most times scalability is not explicitly mentioned by current paper recommendation systems. There are several works~\cite{DBLP:conf/wise/HuaCLZZ20,kanakia19,DBLP:conf/ijcnn/ShiMZJLC20,DBLP:journals/access/WaheedIRMK19,xie21} evaluating on bigger datasets with over 1 million papers and which thus are able to handle big amounts of data. Sizes of current relevant real-world data collections exceed this threshold many times over (see e.g. PubMed with over 33 million papers\footnote{\url{https://pubmed.ncbi.nlm.nih.gov/}} or SemanticScholar with over 203 million papers\footnote{\url{https://www.semanticscholar.org/product/api}}).
Kanakia et al.~\cite{kanakia19} explicitly state scalability as a problem their approach is able to overcome. Instead of comparing each paper to all other papers they utilise clustering to reduce the number of required computations. They present the only approach running on several hundred million publications. Nair et al.~\cite{nair} mention scalability issues they encountered even when only considering around 25,000 publications and their citation relations.

\subsubsection{Privacy}

The problem of privacy in personalised paper recommendation~ is~ described~ by~ Bai~ et~ al.~\cite{DBLP:journals/access/BaiWLYKX19}.~ Shahid~ et al.~\cite{DBLP:journals/tjs/ShahidAABZYC20} also mention this as a problem occurring in collaborative filtering approaches. An issue is encountered when sensitive information such as habits or weaknesses that users might not want to disclose is used by a system. This leads to users' having negative impressions of systems. Keeping sensitive information private should therefore be a main goal.

In the current approaches, we did not find a discussion of privacy concerns.
Some approach even explicitly utilise likes~\cite{DBLP:conf/iui/RahdariB19} or association rules~\cite{ahmedi} of other users while failing to mention privacy altogether. In approaches not incorporating any user data, this issue does not arise at all.

\subsubsection{Serendipity}

Serendipity is described by Bai et al.~\cite{DBLP:journals/access/BaiWLYKX19} as an attribute often encountered in collaborative filtering~\cite{DBLP:journals/jodl/BeelGLB16}. Usually paper recommender systems focus on identification of relevant papers even though also including not obviously relevant ones might enhance the overall recommendation. Junior researchers could profit from stray recommendations to broaden their horizon, senior researchers might be able to gain knowledge to enhance their research. The ratio between clearly relevant and serendipitous papers is crucial to prevent users from losing trust in the recommender system.

A main objective of the works of Nishioka et al.~\cite{DBLP:conf/icadl/NishiokaHS19,DBLP:conf/ercimdl/NishiokaHS19,DBLP:journals/peerj-cs/NishiokaHS20} is serendipity. Other approaches do not mention this aspect.

\subsubsection{Unified Scholarly Data Standards}

Different data formats of data collections is mentioned as a problem by Bai et al.~\cite{DBLP:journals/access/BaiWLYKX19}. They mention digital libraries containing relevant information which needs to be unified in order to use the data in a paper recommendation system. Additionally the combination of datasets could also lead to problems.

Many of the approaches we observe do not consider data collection or preparation as part of the approach, they often only mention the combination of different datasets as part of the evaluation (see e.g. Du et al.~\cite{DBLP:conf/aiccsa/DuGWHZG20}, Li et al.~\cite{DBLP:journals/tois/LiCPR19} or Xie et al.~\cite{DBLP:conf/sigir/XieSB21}).
An exception to this general rule are systems which contain a web crawling step for data (see e.g. Ahmad and Afzal~\cite{DBLP:journals/elektrik/AhmadA20} or Sakib et al.~\cite{DBLP:journals/access/SakibAABHHG21}). Even with this type of approaches the combination of datasets and their diverse data formats is not identified as a problem.

\subsubsection{Synonymy}

Shahid et al.~\cite{DBLP:journals/tjs/ShahidAABZYC20} describe the problem of synonymy encountered in collaborative filtering approaches. They define this problem as different words having the same meaning. 

Even though there are still approaches (not necessarily CF ones) utilising basic TF-IDF representations of papers~\cite{DBLP:journals/elektrik/AhmadA20,DBLP:conf/wise/HuaCLZZ20,renuka,Sharma}, nowadays this problem can be bypassed by using a text embedding method such as Doc2Vec or BERT.

\subsubsection{Gray Sheep}

Gray sheep is a problem described by Shahid et al.~\cite{DBLP:journals/tjs/ShahidAABZYC20} as an issue encountered in collaborative filtering approaches. They describe it as some users not consistently (dis)agreeing with any reference group.

We did not find any current approach mentioning this problem.

\subsubsection{Black Sheep}

Black sheep is a problem described by Shahid et al.~\cite{DBLP:journals/tjs/ShahidAABZYC20} as an issue encountered in collaborative filtering approaches.~ They describe it as some users not (dis)agree-ing with any reference group.

We did not find any current approach mentioning this problem.

\subsubsection{Shilling attack}

Shilling attacks are described by Shahid et al.~\cite{DBLP:journals/tjs/ShahidAABZYC20} as a problem~ encountered~ in~ collaborative~ filtering~ ap\-proaches. They define this problem as users being able to manually enhance visibility of their own research by rating authored papers as relevant while negatively rating any other recommendations.

Although we did not find any current approach mentioning this problem we assume maybe it is no longer highly relevant as most approaches are no longer pure collaborative filtering ones. Additionally from the considered collaborative filtering approaches no one explicitly stated to feed relevance ratings back into the system.

\subsection{Emerging Challenges}
\label{sec:chall_new}

In addition to the open challenges discussed in former literature reviews by Bai et al.~\cite{DBLP:journals/access/BaiWLYKX19}, Beel et al.~\cite{DBLP:journals/jodl/BeelGLB16} and Shahid et al.~\cite{DBLP:journals/tjs/ShahidAABZYC20} we identified the following problems and derive desirable goals for future approaches from them.

\subsubsection{User Evaluation} 

Paper recommendation is always targeted at human users. But oftentimes an evaluation with real users to quantify users' satisfaction with recommended publications is simply not conducted~\cite{DBLP:conf/iui/RahdariB19}. Conducting huge user~ studies~ is~ not~ feasible~\cite{DBLP:journals/scientometrics/HabibA19}.~ So~ sometimes~ user data~ to~ evaluate~ with~ is~ fetched~ from~ the~ presented datasets~\cite{DBLP:journals/scientometrics/HarunaIQKHMC20,DBLP:journals/access/SakibAABHHG21} or user behaviour is artificially emulated~\cite{DBLP:conf/flairs/AfsarCF21,Bereczki1587420,DBLP:journals/dss/LiWNLL21}.
Noteworthy counter-examples\footnote{For a full list of approaches conducting user studies see Table~\ref{tab:user_study}.} are the studies by Bulut et al.~\cite{Bulut2018APR} who emailed 50 researchers to rate relevancy of recommended articles or Chaudhuri et al.~\cite{DBLP:journals/jodl/ChaudhuriSSS21} who asked 45 participants to rate their acceptance of recommended publications.
Another option to overcome this issue is utilisation of living labs as seen with ArXivDigest~\cite{DBLP:conf/cikm/GingstadJB20}, Mr. DLib's living lab~\cite{DBLP:conf/ecir/BeelCKDK19} or LiLAS for the related tasks of dataset recommendation for scientific publications and multi-lingual document retrieval~\cite{DBLP:conf/clef/SchaerBCWST21}.

\textbf{Desirable goal.} Paper recommendation systems targeted at users should always contain a user evaluation with a description of the composition of participants.

\subsubsection{Target audience}

Current works mostly fail to clearly characterise the intended users of a system  altogether and the varying interests of different types of users are not examined in their evaluations. There are some noteworthy counter-examples: Afsar et al.~\cite{DBLP:conf/flairs/AfsarCF21} mention cancer patients and their close relatives as intended target audience.~ Bereczki~\cite{Bereczki1587420}~ identifies~ new~ users~ as~ a special group they want to recommend papers to. 
Hua et al.~\cite{DBLP:conf/wise/HuaCLZZ20} consider users who start diving into a topic which they have not yet researched before. Sharma et al.~\cite{Sharma} name subject~ matter~ experts~ incorporating~ articles~ into~ a medical knowledge base as their target audience. Shi et al.~\cite{DBLP:conf/ijcnn/ShiMZJLC20} clearly state use cases for their approach which always target users which are unaware of a topic but already have one interesting paper from the area. They strive to recommend more papers similar to the first one.

User characteristics such as registration status of users are already mentioned by Beel et al.~\cite{DBLP:journals/jodl/BeelGLB16} as a factor which is disregarded in evaluations. We want to extend on this point and highlight the oftentimes missing or inadequate descriptions of intended users of paper recommendation systems. Traits of users and their information needs are not only important for experiments but should also be regarded in the construction of an approach. The targeted audience of a paper recommendation system should influence its suggestions. Bai et al.~\cite{DBLP:journals/access/BaiWLYKX19} highlight different needs of junior researchers which should be recommended a broad variety of papers as they still have to figure out their direction. They state recommendations for senior researchers should be more in line with their already established interests. Sugiyama and Kan~\cite{DBLP:conf/jcdl/SugiyamaK11} describe the need to help discover interdisciplinary research for this experienced user group.
Most works do not recognise possible different functions of paper recommendation systems for users depending on their level of seniority. If papers include an evaluation with real persons, they e.g. mix Master's students with professors but do not address their different goals or expectations from paper recommendation~\cite{DBLP:conf/icadl/NishiokaHS19}. Chaudhuri et al.~\cite{DBLP:journals/jodl/ChaudhuriSSS21} have junior, experienced and expert users as participants of their study and give individual ratings but do not calculate evaluation scores per user group.
In some studies the exact composition of test users is not even mentioned (see Table~\ref{tab:user_study}).

\textbf{Desirable goal.} Definition and consideration of a specific target audience for an approach and evaluation with members of this audience. If there is no specific person group a system should suit best, this should be discussed, executed and evaluated accordingly.

\subsubsection{Recommendation Scenario}

Suggested papers from an approach should either be ones to read~\cite{DBLP:journals/corr/abs-2103-08819,DBLP:conf/ml4cs/JingY20}, to cite or fulfil another specified information need such as help patients in cancer treatment decision making~\cite{DBLP:conf/flairs/AfsarCF21}. Most work does not clearly state which is the case. Instead recommended papers are only said to be related~\cite{DBLP:conf/icmla/AlfarhoodC19,DBLP:conf/jcdl/CollinsB19}, relevant~\cite{DBLP:conf/icmla/AlfarhoodC19,DBLP:journals/kbs/AliQMAA20,DBLP:journals/jodl/ChaudhuriSSS21,chenban,DBLP:journals/scientometrics/HabibA19,DBLP:conf/wise/HuaCLZZ20,kanakia19,DBLP:journals/tetc/KongMWLX21,DBLP:journals/tois/LiCPR19,DBLP:journals/dss/LiWNLL21,DBLP:conf/kdd/TangZYLZS08,DBLP:conf/sigir/XieSB21,DBLP:journals/www/YangLLLZZZZ19}, satisfactory~\cite{DBLP:conf/wise/HuaCLZZ20,DBLP:journals/complexity/LiuKYQ20}, suitable~\cite{Bulut2020}, appropriate and useful~\cite{Bulut2018APR,DBLP:journals/access/SakibAABHHG21} or a description which scenario is tackled is skipped altogether~\cite{ahmedi,DBLP:conf/aaai/GuoCZLDH20,DBLP:journals/scientometrics/HarunaIQKHMC20,DBLP:conf/iui/RahdariB19}.

In rare cases if the recommendation scenario is mentioned there is the possibility of it not perfectly fitting the evaluated scenario. This can e.g. be seen in the work of Jing and Yu~\cite{DBLP:conf/ml4cs/JingY20} where they propose paper recommendation for papers to read but evaluate papers which were cited. Cited papers should always be ones which have been read beforehand but the decision to cite papers can be influenced by multiple aspects~\cite{garfield}.

\textbf{Desirable goal.} The clear description of the recommendation scenario is important for comparability of approaches as well as the validity of the evaluation.

\subsubsection{Fairness/Diversity}

Anand et al~\cite{DBLP:conf/ecir/Anand0D17} define fairness as the balance between relevance~ and~ diversity~ of~ recommendation~ results. Only focusing on fit between the user or input paper and suggestions would lead to highly similar results which might not be vastly different from each other. Having diverse recommendation results can help cover multiple aspects of a user query instead of only satisfying the most prominent feature of the query~\cite{DBLP:conf/ecir/Anand0D17}. 
In general more diverse recommendations provide greater utility for users~\cite{DBLP:journals/peerj-cs/NishiokaHS20}.
\textcolor{red}{Ekstrand et al.~\cite{DBLP:journals/corr/abs-2105-05779} give a detailed overview of current constructs for measuring algorithmic fairness in information access and describe possibly arising problems in this context.}

Most of the current paper recommendation systems do not consider fairness but some approaches specifically mention diversity~\cite{DBLP:journals/jodl/ChaudhuriSSS21,DBLP:conf/icadl/NishiokaHS19,DBLP:conf/ercimdl/NishiokaHS19,DBLP:journals/peerj-cs/NishiokaHS20} while striving to recommend relevant publications. Thus these systems consider fairness.

Over one fourth of considered approaches with an evaluation report MMR as a measure of their system's quality. This at least seems to show researchers' awareness of the general problem of diverse recommendation results.

\textbf{Desirable Goal.} Diversification of suggested papers to ensure fairness of the approach.

\subsubsection{Complexity}

Paper recommendation systems tend to become more complex,~ convoluted~ or~ composed~ of~ multiple~ parts. We observed this trend by regarding the classification of current systems compared to previous literature reviews (see Section~\ref{sec:litReview_former}). While systems' complexity increases, users' interaction with the systems should not become more complex. If an approach requires user interaction at all, it should be as simple as possible. Users should not be required to construct sophisticated knowledge graphs~\cite{DBLP:journals/corr/abs-2103-08819} or enter multiple rounds of keywords for an approach to learn their user profile~\cite{DBLP:journals/corr/abs-2107-07831}.
  
\textbf{Desirable Goal.} Maintain simplicity of usage even if approaches become more complex.

\subsubsection{Explainability}

Confidence in the recommendation system has already been mentioned by Beel et al.~\cite{DBLP:journals/jodl/BeelGLB16} as an example of what could enhance users' satisfaction but what is overlooked in approaches in favour of accuracy. This aspect should be considered with more vigour as the general research area of explainable recommendation has gained immense traction~\cite{DBLP:journals/ftir/ZhangC20}.
Gingstad et al.~\cite{DBLP:conf/cikm/GingstadJB20} regard explainability as a core component of paper recommendation systems. Xie et al.~\cite{xie21} mention explainability as a key feature of their approach but do not state how they achieve it or if their explanations satisfy users.
Suggestions of recommendation systems should be explainable to enhance their trustworthiness and make them more engaging~\cite{DBLP:conf/recsys/McInerneyLHHBGM18}.
Here, different explanation goals such as effectiveness, efficiency, transparency or trust and their influence on each other should be considered~\cite{DBLP:conf/sigir/BalogR20}.
If an approach uses neural networks~\cite{DBLP:journals/corr/abs-2107-07831,DBLP:conf/aaai/GuoCZLDH20,DBLP:conf/cscwd/LLP21,DBLP:journals/tois/LiCPR19} it is oftentimes impossible to explain why the system learned, that a specific suggested paper might be relevant.

Lee et al.~\cite{lee} introduce a general approach which could be applied to any paper recommendation system to generate explanations for recommendations. Even though this option seems to help solve the described problem it is not clear how valuable post-hoc explanations are compared to systems which construct them directly.

\textbf{Desirable Goal.} The conceptualisation of recommendation systems which comprehensibly explain their users why a specific paper is suggested.

\subsubsection{Public Dataset}

Current approaches utilise many different datasets (see Table~\ref{tab:datasets}). A large portion of them are built by the authors such that they are not publicly available for others to use as well~\cite{DBLP:conf/flairs/AfsarCF21,DBLP:journals/tkde/DuTD21,wang21}. Part of the approaches already use open datasets in their evaluation but a large portion still does not seem to regard this as a priority (see Table~\ref{tab:datasets_private}).
Utilisation of already public data sources or construction of datasets which are also published and remain available thus should be a priority in order to support reproducibility of approaches.

\textbf{Desirable Goal} Utilisation of publicly available datasets in the evaluation of paper recommendation systems.

\subsubsection{Comparability}

From~ the~ approaches~ we~ observed,~ many~ identified themselves as paper recommendation ones but only evaluated against systems, which are more general recommendation systems or ones utilising some same methodologies but not from the sub-domain of paper recommendation (seen with e.g. Guo et al~\cite{DBLP:conf/aaai/GuoCZLDH20}, Tanner et al.~\cite{DBLP:conf/bigdataconf/TannerAH19} or Yang et al.~\cite{DBLP:journals/www/YangLLLZZZZ19}).
While some of the works might claim to only be applied on paper recommendation and be of more general applicability (see e.g. the works by Ahmedi et al.~\cite{ahmedi} or Alfarhood and Cheng~\cite{DBLP:conf/icmla/AlfarhoodC19}) we state that they should still be compared to ones, which mainly identify as paper recommendation systems as seen in the work of Chaudhuri et al.~\cite{DBLP:journals/corr/abs-2107-07831}. Only if a more general approach is compared to a paper recommendation approach, its usefulness for the area of paper recommendation can be fully assessed.

Several times, the baselines to evaluate against are not even other works but artificially constructed ones~\cite{DBLP:journals/elektrik/AhmadA20,DBLP:journals/scientometrics/HabibA19} or no other approach at all~\cite{Bulut2018APR}. 

\textbf{Desirable Goal.} Evaluation of paper recommendation approaches, even those which are applicable in a wider context, should always be against at least one paper recommendation system to clearly report relevance of the proposed method in the claimed context.

\subsection{Discussion and Outlook}
\label{sec:chall_dis}

From the already existing problems, several of them are still encountered in current paper recommendation approaches. Users are not always part of the approaches so users are not always modelled but this also prevents privacy issues. Accuracy seems to still be the main focus of recommendation systems. Novel techniques proposed in papers are not available online or applied by existing paper recommendation systems. Approaches do not provide enough details to enable re-implementation.
\textcolor{red}{Providing the code online or in a living lab environment could help overcome many of these issues.}

Other problems mainly encountered in pure collaborative filtering systems such as the cold start problem, sparsity, synonymy, gray sheep, black sheep and shilling attacks do not seem to be as relevant anymore. We observed a trend towards hybrid models, this recommendation system type can overcome these issues. These hybrid models should also be able to produce serendipitous recommendations.

Unifying data sources is conducted often but nowadays it does not seem to be regarded as a problem. With scalability we encountered the same. Approaches are oftentimes able to handle millions of papers, here they do not specifically mention scalability as a problem they overcome but they also mostly do not consider huge datasets with several hundreds of millions of publications. 

Due to the limited scope of our survey we are not able to derive substantive claims regarding cooperation and persistence. We found around 30\% of approaches published by groups which authored multiple papers and very few collaborations between different author groups.

As for the newly introduced problems, part of the observed approaches conducted evaluations with users, on publicly available datasets and against other paper recommendation systems. Many works considered a low complexity for users. 
\textcolor{red}{Even though user evaluations are desirable, they come with high costs. Usage of evaluation datasets with real human annotations could help overcome this issue partially, another straightforward solution would be the incorporation in a living lab. The second option would also help with comparability of approaches. Usage of available datasets can become increasingly complicated if approaches use new data which is currently not contained in existing datasets.}\footnote{\textcolor{red}{We did not encounter many papers utilising types of data as part of their approach, which is not typically included in existing datasets; one of the noteworthy exceptions could be the approach by Nishioka et al.~\cite{DBLP:conf/icadl/NishiokaHS19,DBLP:conf/ercimdl/NishiokaHS19,DBLP:journals/peerj-cs/NishiokaHS20}, which utilised Tweets of users.}}

Target audiences in general were rarely defined, the recommendation scenario was mostly not described. Diversity was considered by few. Overall the explainability of recommendations was dismissed. \textcolor{red}{The first two of these issues are ones which could be comparatively easily fixed or addressed in the papers without changing the approach. As for diversity and explainability, the approaches would need to be modelled specifically such that these attributes could be satisfied.}

To conclude, there are many challenges which are not constantly considered by current approaches. They define the requirements for future works in the area of paper recommendation systems.

\section{Conclusion}
\label{sec:conclusion}

This literature review of publications targeting paper recommendation between January 2019 and October 2021 provided comprehensive overviews of their methods, datasets and evaluation measures. We showed the need for a richer multi-dimensional characterisation of paper recommendation as former ones no longer seem sufficient in classifying the increasingly complex approaches.
We also revisited known open challenges in the current time frame and highlighted possibly under-observed problems which future works could focus on.

Efforts should be made to standardise or better differentiate between the varying notions of relevancy and recommendation scenarios when it comes to paper recommendation.
Future work could try revaluate already existing methods with real humans and against other paper recommendation systems. This could for example be realised in an extendable paper recommendation benchmarking system similar to the in a living lab environments ArXivDigest~\cite{DBLP:conf/cikm/GingstadJB20}, Mr. DLib's living lab~\cite{DBLP:conf/ecir/BeelCKDK19} or LiLAS~\cite{DBLP:conf/clef/SchaerBCWST21} but with the additional property that it also provides build-in offline evaluations. As fairness and explainability of current paper recommendation systems have not been tackled widely, those aspects should be further explored.
Another direction could be the comparison of multiple rare evaluation measures on the same system to help identify those which should be focused on in the future.
As we observed a vast variety in datasets utilised for evaluation of the approaches (see Table~\ref{tab:datasets}), construction of publicly available and widely reusable ones would be worthwhile.

\bibliographystyle{spmpsci}
\bibliography{bibliography}

\begin{thebibliography}{100}
\providecommand{\url}[1]{{#1}}
\providecommand{\urlprefix}{URL }
\expandafter\ifx\csname urlstyle\endcsname\relax
  \providecommand{\doi}[1]{DOI~\discretionary{}{}{}#1}\else
  \providecommand{\doi}{DOI~\discretionary{}{}{}\begingroup
  \urlstyle{rm}\Url}\fi

\bibitem{DBLP:conf/flairs/AfsarCF21}
Afsar, M.M., Crump, T., Far, B.H.: {An Exploration On-demand Article
  Recommender System for Cancer Patients Information Provisioning}.
\newblock In: FLAIRS Conference'21 (2021).
\newblock \urlprefix\url{https://doi.org/10.32473/flairs.v34i1.128339}

\bibitem{DBLP:journals/elektrik/AhmadA20}
Ahmad, S., Afzal, M.T.: {Combining metadata and co-citations for recommending
  related papers}.
\newblock Turkish J. Electr. Eng. Comput. Sci. \textbf{28}(3), 1519--1534
  (2020).
\newblock \urlprefix\url{https://doi.org/10.3906/elk-1908-19}

\bibitem{ahmedi}
Ahmedi, L., Rexhepi, E., Bytyçi, E.: Using association rule mining to enrich
  user profiles with research paper recommendation.
\newblock Int. J. Com. Dig. Sys.  (2021).
\newblock \urlprefix\url{https://doi.org/10.12785/ijcds/110192}

\bibitem{DBLP:conf/icmla/AlfarhoodC19}
Alfarhood, M., Cheng, J.: {Collaborative Attentive Autoencoder for Scientific
  Article Recommendation}.
\newblock In: {ICMLA}'19, pp. 168--174. {IEEE} (2019).
\newblock \urlprefix\url{https://doi.org/10.1109/ICMLA.2019.00034}

\bibitem{DBLP:journals/kbs/AliQMAA20}
Ali, Z., Qi, G., Muhammad, K., Ali, B., Abro, W.A.: Paper recommendation based
  on heterogeneous network embedding.
\newblock Knowl. Based Syst. \textbf{210}, 106438 (2020).
\newblock \urlprefix\url{https://doi.org/10.1016/j.knosys.2020.106438}

\bibitem{DBLP:conf/lwa/AlzoghbiA0L15}
Alzoghbi, A., Ayala, V.A.A., Fischer, P.M., Lausen, G.: {PubRec: Recommending
  Publications Based on Publicly Available Meta-Data}.
\newblock In: {LWA}'15, \emph{{CEUR} Workshop Proceedings}, vol. 1458, pp.
  11--18. CEUR-WS.org (2015).
\newblock \urlprefix\url{http://ceur-ws.org/Vol-1458/D01\_CRC69\_Alzoghbi.pdf}

\bibitem{DBLP:conf/webi/AmamiFSP17}
Amami, M., Faiz, R., Stella, F., Pasi, G.: A graph based approach to scientific
  paper recommendation.
\newblock In: WI'17, pp. 777--782. {ACM} (2017).
\newblock \urlprefix\url{https://doi.org/10.1145/3106426.3106479}

\bibitem{DBLP:conf/ecir/Anand0D17}
Anand, A., Chakraborty, T., Das, A.: {FairScholar: Balancing Relevance and
  Diversity for Scientific Paper Recommendation}.
\newblock In: {ECIR}'17, \emph{LNCS}, vol. 10193, pp. 753--757 (2017).
\newblock \urlprefix\url{https://doi.org/10.1007/978-3-319-56608-5\_76}

\bibitem{DBLP:journals/access/BaiWLYKX19}
Bai, X., Wang, M., Lee, I., Yang, Z., Kong, X., Xia, F.: {Scientific Paper
  Recommendation: A Survey}.
\newblock {IEEE} Access \textbf{7}, 9324--9339 (2019).
\newblock \urlprefix\url{https://doi.org/10.1109/ACCESS.2018.2890388}

\bibitem{DBLP:conf/sigir/BalogR20}
Balog, K., Radlinski, F.: {Measuring Recommendation Explanation Quality: The
  Conflicting Goals of Explanations}.
\newblock In: {SIGIR}'20, pp. 329--338. {ACM} (2020).
\newblock \urlprefix\url{https://doi.org/10.1145/3397271.3401032}

\bibitem{barolli}
Barolli, L., Di~Cicco, F., Fonisto, M.: {An Investigation of Covid-19 Papers
  for a Content-Based Recommendation System}.
\newblock In: 3PGCIC'22, pp. 156--164. Springer International Publishing
  (2022).
\newblock \urlprefix\url{https://doi.org/10.1007/978-3-030-89899-1_16}

\bibitem{distiller}
Basaldella, M., Nart, D.D., Tasso, C.: {Introducing Distiller: A Unifying
  Framework for Knowledge Extraction}.
\newblock In: IT@LIA@AI*IA'15, \emph{{CEUR} Workshop Proceedings}, vol. 1509.
  CEUR-WS.org (2015).
\newblock \urlprefix\url{http://ceur-ws.org/Vol-1509/ITALIA2015\_paper\_4.pdf}

\bibitem{rard}
Beel, J., Carevic, Z., Schaible, J., Neusch, G.: {RARD: The Related-Article
  Recommendation Dataset}.
\newblock D Lib Mag. \textbf{23}(7/8) (2017).
\newblock \urlprefix\url{https://doi.org/10.1045/july2017-beel}

\bibitem{DBLP:conf/ecir/BeelCKDK19}
Beel, J., Collins, A., Kopp, O., Dietz, L.W., Knoth, P.: {Online Evaluations
  for Everyone: Mr. DLib's Living Lab for Scholarly Recommendations}.
\newblock In: {ECIR}'19, \emph{LNCS}, vol. 11438, pp. 213--219. Springer
  (2019).
\newblock \urlprefix\url{https://doi.org/10.1007/978-3-030-15719-7\_27}

\bibitem{sowiport}
Beel, J., Dinesh, S., Mayr, P., Carevic, Z., Jain, R.: {Stereotype and
  Most-Popular Recommendations in the Digital Library Sowiport}.
\newblock In: {ISI}'17, \emph{Schriften zur Informationswissenschaft}, vol.~70,
  pp. 96--108. Verlag Werner H{\"{u}}lsbusch (2017).
\newblock \urlprefix\url{https://doi.org/10.18452/1441}

\bibitem{DBLP:journals/jodl/BeelGLB16}
Beel, J., Gipp, B., Langer, S., Breitinger, C.: Research-paper recommender
  systems: a literature survey.
\newblock Int. J. Digit. Libr. \textbf{17}(4), 305--338 (2016).
\newblock \urlprefix\url{https://doi.org/10.1007/s00799-015-0156-0}

\bibitem{DBLP:conf/ercimdl/BeelL15}
Beel, J., Langer, S.: {A Comparison of Offline Evaluations, Online Evaluations,
  and User Studies in the Context of Research-Paper Recommender Systems}.
\newblock In: {TPDL}'15, \emph{LNCS}, vol. 9316, pp. 153--168. Springer (2015).
\newblock \urlprefix\url{https://doi.org/10.1007/978-3-319-24592-8\_12}

\bibitem{DBLP:journals/jodl/BeierleACB20}
Beierle, F., Aizawa, A., Collins, A., Beel, J.: Choice overload and
  recommendation effectiveness in related-article recommendations.
\newblock Int. J. Digit. Libr. \textbf{21}(3), 231--246 (2020).
\newblock \urlprefix\url{https://doi.org/10.1007/s00799-019-00270-7}

\bibitem{Bereczki1587420}
Bereczki, M.: {Graph Neural Networks for Article Recommendation based on
  Implicit User Feedback and Content}.
\newblock Master's thesis, KTH, School of Electrical Engineering and Computer
  Science (EECS) (2021)

\bibitem{citeulike}
Bogers, T., van~den Bosch, A.: Recommending scientific articles using
  citeulike.
\newblock In: RecSys'08, pp. 287--290. {ACM} (2008).
\newblock \urlprefix\url{https://doi.org/10.1145/1454008.1454053}

\bibitem{Bulut2020}
Bulut, B., G{\"u}ndo{\u{g}}an, E., Kaya, B., Alhajj, R., Kaya, M.: {User's
  Research Interests Based Paper Recommendation System: A Deep Learning
  Approach}, pp. 117--130.
\newblock Springer International Publishing (2020).
\newblock \urlprefix\url{https://doi.org/10.1007/978-3-030-33698-1_7}

\bibitem{Bulut2018APR}
Bulut, B., Kaya, B., Kaya, M.: {A Paper Recommendation System Based on User
  Interest and Citations}.
\newblock In: UBMYK'19, pp. 1--5 (2019).
\newblock \urlprefix\url{https://doi.org/10.1109/UBMYK48245.2019.8965533}

\bibitem{DBLP:journals/kais/ChampiriAS19}
Champiri, Z.D., Asemi, A., Salim, S.S.B.: Meta-analysis of evaluation methods
  and metrics used in context-aware scholarly recommender systems.
\newblock Knowl. Inf. Syst. \textbf{61}(2), 1147--1178 (2019).
\newblock \urlprefix\url{https://doi.org/10.1007/s10115-018-1324-5}

\bibitem{DBLP:journals/corr/abs-2107-07831}
Chaudhuri, A., Samanta, D., Sarma, M.: {Modeling User Behaviour in Research
  Paper Recommendation System}.
\newblock CoRR \textbf{abs/2107.07831} (2021).
\newblock \urlprefix\url{https://arxiv.org/abs/2107.07831}

\bibitem{DBLP:conf/tencon/ChaudhuriSS19}
Chaudhuri, A., Sarma, M., Samanta, D.: {Advanced Feature Identification towards
  Research Article Recommendation: {A} Machine Learning Based Approach}.
\newblock In: {TENCON}'19, pp. 7--12. {IEEE} (2019).
\newblock \urlprefix\url{https://doi.org/10.1109/TENCON.2019.8929386}

\bibitem{DBLP:journals/jodl/ChaudhuriSSS21}
Chaudhuri, A., Sinhababu, N., Sarma, M., Samanta, D.: Hidden features
  identification for designing an efficient research article recommendation
  system.
\newblock Int. J. Digit. Libr. \textbf{22}(2), 233--249 (2021).
\newblock \urlprefix\url{https://doi.org/10.1007/s00799-021-00301-2}

\bibitem{chenban}
Chen, J., Ban, Z.: {Academic Paper Recommendation Based on Clustering and
  Pattern Matching}, pp. 171--182.
\newblock Springer (2019).
\newblock \urlprefix\url{https://doi.org/10.1007/978-981-32-9298-7_14}

\bibitem{DBLP:conf/jcdl/CollinsB19}
Collins, A., Beel, J.: {Document Embeddings vs. Keyphrases vs. Terms for
  Recommender Systems: A Large-Scale Online Evaluation}.
\newblock In: {JCDL}'19, pp. 130--133. {IEEE} (2019).
\newblock \urlprefix\url{https://doi.org/10.1109/JCDL.2019.00027}

\bibitem{DBLP:conf/aiccsa/DuGWHZG20}
Du, N., Guo, J., Wu, C.Q., Hou, A., Zhao, Z., Gan, D.: {Recommendation of
  Academic Papers based on Heterogeneous Information Networks}.
\newblock In: {AICCSA}'20, pp. 1--6. {IEEE} (2020).
\newblock \urlprefix\url{https://doi.org/10.1109/AICCSA50499.2020.9316516}

\bibitem{DBLP:journals/tkde/DuTD21}
Du, Z., Tang, J., Ding, Y.: {POLAR++: Active One-Shot Personalized Article
  Recommendation}.
\newblock {IEEE} Trans. Knowl. Data Eng. \textbf{33}(6), 2709--2722 (2021).
\newblock \urlprefix\url{https://doi.org/10.1109/TKDE.2019.2953721}

\bibitem{DBLP:journals/corr/abs-2105-05779}
Ekstrand, M.D., Das, A., Burke, R., Diaz, F.: {Fairness and Discrimination in
  Information Access Systems}.
\newblock CoRR \textbf{abs/2105.05779} (2021).
\newblock \urlprefix\url{https://arxiv.org/abs/2105.05779}

\bibitem{DBLP:journals/jodl/FarberJ20}
F{\"{a}}rber, M., Jatowt, A.: Citation recommendation: approaches and datasets.
\newblock Int. J. Digit. Libr. \textbf{21}(4), 375--405 (2020).
\newblock \urlprefix\url{https://doi.org/10.1007/s00799-020-00288-2}

\bibitem{news1}
Feng, S., Meng, J., Zhang, J.: {News Recommendation Systems in the Era of
  Information Overload}.
\newblock J. Web Eng. \textbf{20}(2), 459--470 (2021).
\newblock \urlprefix\url{https://doi.org/10.13052/jwe1540-9589.20210}

\bibitem{garfield}
Garfield, E.: Can citation indexing be automated? (1964)

\bibitem{citeseer}
Giles, C.L., Bollacker, K.D., Lawrence, S.: {CiteSeer: An Automatic Citation
  Indexing System}.
\newblock In: {ACM} DL'98, pp. 89--98. {ACM} (1998).
\newblock \urlprefix\url{https://doi.org/10.1145/276675.276685}

\bibitem{DBLP:conf/cikm/GingstadJB20}
Gingstad, K., Jekteberg, {\O}., Balog, K.: {ArXivDigest: {A} Living Lab for
  Personalized Scientific Literature Recommendation}.
\newblock In: {CIKM}'20, pp. 3393--3396. {ACM} (2020).
\newblock \urlprefix\url{https://doi.org/10.1145/3340531.3417417}

\bibitem{DBLP:conf/aaai/GuoCZLDH20}
Guo, G., Chen, B., Zhang, X., Liu, Z., Dong, Z., He, X.: {Leveraging
  Title-Abstract Attentive Semantics for Paper Recommendation}.
\newblock In: {IAAI}'20, pp. 67--74. {AAAI} Press (2020).
\newblock \urlprefix\url{https://aaai.org/ojs/index.php/AAAI/article/view/5335}

\bibitem{DBLP:journals/scientometrics/HabibA19}
Habib, R., Afzal, M.T.: Sections-based bibliographic coupling for research
  paper recommendation.
\newblock Scientometrics \textbf{119}(2), 643--656 (2019).
\newblock \urlprefix\url{https://doi.org/10.1007/s11192-019-03053-8}

\bibitem{DBLP:journals/scientometrics/HarunaIQKHMC20}
Haruna, K., Ismail, M.A., Qazi, A., Kakudi, H.A., Hassan, M., Muaz, S.A.,
  Chiroma, H.: Research paper recommender system based on public contextual
  metadata.
\newblock Scientometrics \textbf{125}(1), 101--114 (2020).
\newblock \urlprefix\url{https://doi.org/10.1007/s11192-020-03642-y}

\bibitem{sowiport2}
Hienert, D., Sawitzki, F., Mayr, P.: {Digital Library Research in Action:
  Supporting Information Retrieval in Sowiport}.
\newblock D Lib Mag. \textbf{21}(3/4) (2015).
\newblock \urlprefix\url{https://doi.org/10.1045/march2015-hienert}

\bibitem{DBLP:conf/ifip12/HuMLH20}
Hu, D., Ma, H., Liu, Y., He, X.: {Scientific Paper Recommendation Using
  Author's Dual Role Citation Relationship}.
\newblock In: Intelligent Information Processing'20, \emph{{IFIP} Advances in
  Information and Communication Technology}, vol. 581, pp. 121--132. Springer
  (2020).
\newblock \urlprefix\url{https://doi.org/10.1007/978-3-030-46931-3\_12}

\bibitem{DBLP:conf/wise/HuaCLZZ20}
Hua, S., Chen, W., Li, Z., Zhao, P., Zhao, L.: {Path-Based Academic Paper
  Recommendation}.
\newblock In: {WISE}'20, \emph{LNCS}, vol. 12343, pp. 343--356. Springer
  (2020).
\newblock \urlprefix\url{https://doi.org/10.1007/978-3-030-62008-0\_24}

\bibitem{news2}
Ji, Z., Wu, M., Yang, H., Armend{\'{a}}riz{-}I{\~{n}}igo, J.E.: Temporal
  sensitive heterogeneous graph neural network for news recommendation.
\newblock Future Gener. Comput. Syst. \textbf{125}, 324--333 (2021).
\newblock \urlprefix\url{https://doi.org/10.1016/j.future.2021.06.007}

\bibitem{DBLP:conf/ml4cs/JingY20}
Jing, S., Yu, S.: {Research of Paper Recommendation System Based on Citation
  Network Model}.
\newblock In: {ML4CS}'20, \emph{LNCS}, vol. 12488, pp. 237--247. Springer
  (2020).
\newblock \urlprefix\url{https://doi.org/10.1007/978-3-030-62463-7\_22}

\bibitem{kanakia19}
Kanakia, A., Shen, Z., Eide, D., Wang, K.: {A Scalable Hybrid Research Paper
  Recommender System for Microsoft Academic}.
\newblock CoRR \textbf{abs/1905.08880} (2019).
\newblock \urlprefix\url{http://arxiv.org/abs/1905.08880}

\bibitem{KANG20212020BDP0008}
Kang, Y., Hou, A., Zhao, Z., Gan, D.: {A Hybrid Approach for Paper
  Recommendation}.
\newblock IEICE Transactions on Information and Systems \textbf{E104.D}(8),
  1222--1231 (2021).
\newblock \urlprefix\url{https://doi.org/10.1587/transinf.2020BDP0008}

\bibitem{DBLP:conf/clef/KellerM21}
Keller, J., Munz, L.P.M.: {TEKMA} at {CLEF-2021:} {BM-25} based rankings for
  scientific publication retrieval and data set recommendation.
\newblock In: {CLEF}'21, \emph{{CEUR} Workshop Proceedings}, vol. 2936, pp.
  1700--1711. CEUR-WS.org (2021).
\newblock \urlprefix\url{http://ceur-ws.org/Vol-2936/paper-144.pdf}

\bibitem{DBLP:journals/tetc/KongMWLX21}
Kong, X., Mao, M., Wang, W., Liu, J., Xu, B.: {VOPRec: Vector Representation
  Learning of Papers with Text Information and Structural Identity for
  Recommendation}.
\newblock {IEEE} Trans. Emerg. Top. Comput. \textbf{9}(1), 226--237 (2021).
\newblock \urlprefix\url{https://doi.org/10.1109/TETC.2018.2830698}

\bibitem{DBLP:conf/cscwd/LLP21}
L, H., Liu, S., Pan, L.: {Paper Recommendation Based on Author-paper Interest
  and Graph Structure}.
\newblock In: {CSCWD}'21, pp. 256--261. {IEEE} (2021).
\newblock \urlprefix\url{https://doi.org/10.1109/CSCWD49262.2021.9437743}

\bibitem{DBLP:conf/recsys/LeKD19}
Le, M., Kayal, S., Douglas, A.: {The Impact of Recommenders on Scientific
  Article Discovery: The Case of Mendeley Suggest}.
\newblock In: ImpactRS@RecSys'19, \emph{{CEUR} Workshop Proceedings}, vol.
  2462. CEUR-WS.org (2019).
\newblock \urlprefix\url{http://ceur-ws.org/Vol-2462/paper5.pdf}

\bibitem{lee}
Lee, B.C.G., Lo, K., Downey, D., Weld, D.S.: {Explanation-Based Tuning of
  Opaque Machine Learners with Application to Paper Recommendation}.
\newblock CoRR \textbf{abs/2003.04315} (2020).
\newblock \urlprefix\url{https://arxiv.org/abs/2003.04315}

\bibitem{DBLP:journals/corr/abs-1304-5457}
Lee, J., Lee, K., Kim, J.G.: {Personalized Academic Research Paper
  Recommendation System}.
\newblock CoRR \textbf{abs/1304.5457} (2013).
\newblock \urlprefix\url{http://arxiv.org/abs/1304.5457}

\bibitem{DBLP:conf/kdd/LeskovecKF05}
Leskovec, J., Kleinberg, J.M., Faloutsos, C.: Graphs over time: densification
  laws, shrinking diameters and possible explanations.
\newblock In: {SIGKDD}'05, pp. 177--187. {ACM} (2005).
\newblock \urlprefix\url{https://doi.org/10.1145/1081870.1081893}

\bibitem{DBLP:journals/pvldb/Ley09}
Ley, M.: {DBLP - Some Lessons Learned}.
\newblock Proc. {VLDB} Endow. \textbf{2}(2), 1493--1500 (2009).
\newblock \urlprefix\url{https://doi.org/10.14778/1687553.1687577}

\bibitem{li20}
Li, W., Chang, C., He, C., Wu, Z., Guo, J., Peng, B.: {Academic Paper
  Recommendation Method Combining Heterogeneous Network and Temporal
  Attributes}.
\newblock In: ChineseCSCW'21, pp. 456--468. Springer Singapore (2021).
\newblock \urlprefix\url{https://doi.org/10.1007/978-981-16-2540-4_33}

\bibitem{DBLP:journals/tois/LiCPR19}
Li, X., Chen, Y., Pettit, B., de~Rijke, M.: {Personalised Reranking of Paper
  Recommendations Using Paper Content and User Behavior}.
\newblock {ACM} Trans. Inf. Syst. \textbf{37}(3), 31:1--31:23 (2019).
\newblock \urlprefix\url{https://doi.org/10.1145/3312528}

\bibitem{DBLP:journals/dss/LiWNLL21}
Li, Y., Wang, R., Nan, G., Li, D., Li, M.: A personalized paper recommendation
  method considering diverse user preferences.
\newblock Decis. Support Syst. \textbf{146}, 113546 (2021).
\newblock \urlprefix\url{https://doi.org/10.1016/j.dss.2021.113546}

\bibitem{DBLP:journals/ccsecis/LiZ19}
Li, Z., Zou, X.: {A Review on Personalized Academic Paper Recommendation}.
\newblock Comput. Inf. Sci. \textbf{12}(1), 33--43 (2019).
\newblock \urlprefix\url{https://doi.org/10.5539/cis.v12n1p33}

\bibitem{lin21}
Lin, S.j., Lee, G., Peng, S.L.: {Academic Article Recommendation by Considering
  the Research Field Trajectory}, pp. 447--454.
\newblock Springer (2021).
\newblock \urlprefix\url{https://doi.org/10.1007/978-3-030-65407-8_39}

\bibitem{DBLP:conf/seke/LiuKCQ19}
Liu, H., Kou, H., Chi, X., Qi, L.: Combining time, keywords and authors
  information to construct papers correlation graph {(S)}.
\newblock In: {SEKE}'19, pp. 11--19. {KSI} Research Inc. and Knowledge Systems
  Institute Graduate School (2019).
\newblock \urlprefix\url{https://doi.org/10.18293/SEKE2019-161}

\bibitem{DBLP:journals/complexity/LiuKYQ20}
Liu, H., Kou, H., Yan, C., Qi, L.: {Keywords-Driven and Popularity-Aware Paper
  Recommendation Based on Undirected Paper Citation Graph}.
\newblock Complex. \textbf{2020}, 2085638:1--2085638:15 (2020).
\newblock \urlprefix\url{https://doi.org/10.1155/2020/2085638}

\bibitem{DBLP:conf/cscwd/LuHCP021}
Lu, Y., He, Y., Cai, Y., Peng, Z., Tang, Y.: {Time-aware Neural Collaborative
  Filtering with Multi-dimensional Features on Academic Paper Recommendation}.
\newblock In: {CSCWD}'21, pp. 1052--1057. {IEEE} (2021).
\newblock \urlprefix\url{https://doi.org/10.1109/CSCWD49262.2021.9437673}

\bibitem{DBLP:journals/access/MaW19a}
Ma, X., Wang, R.: {Personalized Scientific Paper Recommendation Based on
  Heterogeneous Graph Representation}.
\newblock {IEEE} Access \textbf{7}, 79887--79894 (2019).
\newblock \urlprefix\url{https://doi.org/10.1109/ACCESS.2019.2923293}

\bibitem{DBLP:journals/monet/MaZZ19}
Ma, X., Zhang, Y., Zeng, J.: {Newly Published Scientific Papers Recommendation
  in Heterogeneous Information Networks}.
\newblock Mob. Networks Appl. \textbf{24}(1), 69--79 (2019).
\newblock \urlprefix\url{https://doi.org/10.1007/s11036-018-1133-9}

\bibitem{DBLP:journals/ijiit/Manju}
{Manju G.}, {Abhinaya P.}, {Hemalatha M. R.}, {Manju Ganesh G.}, {Manju G. G.}:
  {Cold Start Problem Alleviation in a Research Paper Recommendation System
  Using the Random Walk Approach on a Heterogeneous User-Paper Graph}.
\newblock Int. J. Intell. Inf. Technol. \textbf{16}(2), 24--48 (2020).
\newblock \urlprefix\url{https://doi.org/10.4018/IJIIT.2020040102}

\bibitem{DBLP:conf/recsys/McInerneyLHHBGM18}
McInerney, J., Lacker, B., Hansen, S., Higley, K., Bouchard, H., Gruson, A.,
  Mehrotra, R.: Explore, exploit, and explain: personalizing explainable
  recommendations with bandits.
\newblock In: RecSys'18, pp. 31--39. {ACM} (2018).
\newblock \urlprefix\url{https://doi.org/10.1145/3240323.3240354}

\bibitem{DBLP:journals/cit/MedicS20}
Medic, Z., Snajder, J.: {A Survey of Citation Recommendation Tasks and
  Methods}.
\newblock J. Comput. Inf. Technol. \textbf{28}(3), 183--205 (2020).
\newblock \urlprefix\url{http://cit.fer.hr/index.php/CIT/article/view/5160}

\bibitem{DBLP:conf/recsys/HassanSGMB19}
Mohamed~Hassan, H.A., Sansonetti, G., Gasparetti, F., Micarelli, A., Beel, J.:
  {BERT, ELMo, {USE} and InferSent Sentence Encoders: The Panacea for
  Research-Paper Recommendation?}
\newblock In: RecSys'19, \emph{{CEUR} Workshop Proceedings}, vol. 2431, pp.
  6--10. CEUR-WS.org (2019).
\newblock \urlprefix\url{http://ceur-ws.org/Vol-2431/paper2.pdf}

\bibitem{MohamedHassan}
{Mohamed Hassan}, H.A., Sansonetti, G., Micarelli, A.: {Tag-Aware Document
  Representation for Research Paper Recommendation} (2020).
\newblock
  \urlprefix\url{https://www.researchgate.net/publication/343319230_Tag-Aware_Document_Representation_for_Research_Paper_Recommendation}

\bibitem{wikipedia2}
Moskalenko, O., S{\'{a}}ez{-}Trumper, D., Parra, D.: {Scalable Recommendation
  of Wikipedia Articles to Editors Using Representation Learning}.
\newblock In: RecSys'20, \emph{{CEUR} Workshop Proceedings}, vol. 2697.
  CEUR-WS.org (2020).
\newblock \urlprefix\url{http://ceur-ws.org/Vol-2697/paper1\_complexrec.pdf}

\bibitem{nair}
Nair, A.M., Benny, O., George, J.: {Content Based Scientific Article
  Recommendation System Using Deep Learning Technique}.
\newblock In: Inventive Systems and Control, pp. 965--977. Springer Singapore
  (2021).
\newblock \urlprefix\url{https://doi.org/10.1007/978-981-16-1395-1_70}

\bibitem{DBLP:conf/ictai/Ng20}
Ng, Y.: {Research Paper Recommendation Based on Content Similarity, Peer
  Reviews, Authority, and Popularity}.
\newblock In: {ICTAI}'20, pp. 47--52. {IEEE} (2020).
\newblock \urlprefix\url{https://doi.org/10.1109/ICTAI50040.2020.00018}

\bibitem{Ng2020-gf}
Ng, Y.K.: {CBRec}: a book recommendation system for children using the matrix
  factorisation and content-based filtering approaches.
\newblock International Journal of Business Intelligence and Data Mining
  \textbf{16}(2), 129--149 (2020).
\newblock \urlprefix\url{https://doi.org/10.1504/IJBIDM.2020.104738}

\bibitem{DBLP:conf/icadl/NishiokaHS19}
Nishioka, C., Hauke, J., Scherp, A.: {Research Paper Recommender System with
  Serendipity Using Tweets vs. Diversification}.
\newblock In: {ICADL}'19, \emph{LNCS}, vol. 11853, pp. 63--70. Springer (2019).
\newblock \urlprefix\url{https://doi.org/10.1007/978-3-030-34058-2\_7}

\bibitem{DBLP:conf/ercimdl/NishiokaHS19}
Nishioka, C., Hauke, J., Scherp, A.: {Towards Serendipitous Research Paper
  Recommender Using Tweets and Diversification}.
\newblock In: {TPDL}'19, \emph{LNCS}, vol. 11799, pp. 339--343. Springer
  (2019).
\newblock \urlprefix\url{https://doi.org/10.1007/978-3-030-30760-8\_29}

\bibitem{DBLP:journals/peerj-cs/NishiokaHS20}
Nishioka, C., Hauke, J., Scherp, A.: Influence of tweets and diversification on
  serendipitous research paper recommender systems.
\newblock PeerJ Comput. Sci. \textbf{6}, e273 (2020).
\newblock \urlprefix\url{https://doi.org/10.7717/peerj-cs.273}

\bibitem{DBLP:journals/corr/abs-2008-00202}
Ostendorff, M.: {Contextual Document Similarity for Content-based Literature
  Recommender Systems}.
\newblock CoRR \textbf{abs/2008.00202} (2020).
\newblock \urlprefix\url{https://arxiv.org/abs/2008.00202}

\bibitem{wikipedia3}
Ostendorff, M., Breitinger, C., Gipp, B.: {A Qualitative Evaluation of User
  Preference for Link-based vs. Text-based Recommendations of Wikipedia
  Articles}.
\newblock CoRR \textbf{abs/2109.07791} (2021).
\newblock \urlprefix\url{https://arxiv.org/abs/2109.07791}

\bibitem{PRISMA}
Page, M.J., McKenzie, J.E., Bossuyt, P.M., Boutron, I., Hoffmann, T.C., Mulrow,
  C.D., Shamseer, L., Tetzlaff, J.M., Akl, E.A., Brennan, S.E., Chou, R.,
  Glanville, J., Grimshaw, J.M., Hr{\'o}bjartsson, A., Lalu, M.M., Li, T.,
  Loder, E.W., Mayo-Wilson, E., McDonald, S., McGuinness, L.A., Stewart, L.A.,
  Thomas, J., Tricco, A.C., Welch, V.A., Whiting, P., Moher, D.: {The PRISMA
  2020 statement: an updated guideline for reporting systematic reviews}.
\newblock BMJ \textbf{372} (2021).
\newblock \urlprefix\url{https://www.bmj.com/content/372/bmj.n71}

\bibitem{PATRA2020103399}
Patra, B.G., Maroufy, V., Soltanalizadeh, B., Deng, N., Zheng, W.J., Roberts,
  K., Wu, H.: A content-based literature recommendation system for datasets to
  improve data reusability – a case study on gene expression omnibus (geo)
  datasets.
\newblock Journal of Biomedical Informatics \textbf{104}, 103399 (2020).
\newblock \urlprefix\url{https://doi.org/10.1016/j.jbi.2020.103399}

\bibitem{Radev&al.09}
Radev, D.R., Joseph, M.T., Gibson, B., Muthukrishnan, P.: {A} {B}ibliometric
  and {N}etwork {A}nalysis of the field of {C}omputational {L}inguistics.
\newblock Journal of the American Society for Information Science and
  Technology  (2009).
\newblock \urlprefix\url{https://doi.org/10.1002/asi.23394}

\bibitem{Radev&al.09a}
Radev, D.R., Muthukrishnan, P., Qazvinian, V.: {The ACL Anthology Network
  Corpus}.
\newblock In: NLPIR4DL'09 (2009).
\newblock \urlprefix\url{https://doi.org/10.5555/1699750.1699759}

\bibitem{radev}
Radev, D.R., Muthukrishnan, P., Qazvinian, V., Abu-Jbara, A.: {The ACL
  anthology network corpus}.
\newblock Language Resources and Evaluation pp. 1--26 (2013).
\newblock \urlprefix\url{https://doi.org/10.1007/s10579-012-9211-2}

\bibitem{DBLP:conf/iui/RahdariB19}
Rahdari, B., Brusilovsky, P.: User-controlled hybrid recommendation for
  academic papers.
\newblock In: IUI Companion'19, pp. 99--100. {ACM} (2019).
\newblock \urlprefix\url{https://doi.org/10.1145/3308557.3308717}

\bibitem{wikipedia1}
Rahdari, B., Brusilovsky, P., Thaker, K., Barria{-}Pineda, J.:
  {Knowledge-Driven Wikipedia Article Recommendation for Electronic Textbooks}.
\newblock In: {EC-TEL}'20, \emph{LNCS}, vol. 12315, pp. 363--368. Springer
  (2020).
\newblock \urlprefix\url{https://doi.org/10.1007/978-3-030-57717-9\_28}

\bibitem{renuka}
Renuka, S., Raj~Kiran, G.S.S., Rohit, P.: {An Unsupervised Content-Based
  Article Recommendation System Using Natural Language Processing}.
\newblock In: Data Intelligence and Cognitive Informatics, pp. 165--180.
  Springer Singapore (2021).
\newblock \urlprefix\url{https://doi.org/10.1007/978-981-15-8530-2_13}

\bibitem{DBLP:conf/aist/SafaryanFYKN20}
Safaryan, A., Filchenkov, P., Yan, W., Kutuzov, A., Nikishina, I.: {Semantic
  Recommendation System for Bilingual Corpus of Academic Papers}.
\newblock In: {AIST}'20, \emph{Communications in Computer and Information
  Science}, vol. 1357, pp. 22--36. Springer (2020).
\newblock \urlprefix\url{https://doi.org/10.1007/978-3-030-71214-3\_3}

\bibitem{DBLP:journals/access/SakibAABHHG21}
Sakib, N., Ahmad, R.B., Ahsan, M., Based, M.A., Haruna, K., Haider, J.,
  Gurusamy, S.: {A Hybrid Personalized Scientific Paper Recommendation Approach
  Integrating Public Contextual Metadata}.
\newblock {IEEE} Access \textbf{9}, 83080--83091 (2021).
\newblock \urlprefix\url{https://doi.org/10.1109/ACCESS.2021.3086964}

\bibitem{DBLP:journals/access/SakibAH20}
Sakib, N., Ahmad, R.B., Haruna, K.: {A Collaborative Approach Toward Scientific
  Paper Recommendation Using Citation Context}.
\newblock {IEEE} Access \textbf{8}, 51246--51255 (2020).
\newblock \urlprefix\url{https://doi.org/10.1109/ACCESS.2020.2980589}

\bibitem{citrec2}
Samad, A., Islam, M.A., Iqbal, M.A., Aleem, M.: {Centrality-Based Paper
  Citation Recommender System}.
\newblock {EAI} Endorsed Trans. Ind. Networks Intell. Syst. \textbf{6}(19), e2
  (2019).
\newblock \urlprefix\url{https://doi.org/10.4108/eai.13-6-2019.159121}

\bibitem{DBLP:conf/clef/SchaerBCWST21}
Schaer, P., Breuer, T., Castro, L.J., Wolff, B., Schaible, J.,
  Tavakolpoursaleh, N.: {Overview of LiLAS 2021 - Living Labs for Academic
  Search (Extended Overview)}.
\newblock In: {CLEF}'21, \emph{{CEUR} Workshop Proceedings}, vol. 2936, pp.
  1668--1699. CEUR-WS.org (2021).
\newblock \urlprefix\url{http://ceur-ws.org/Vol-2936/paper-143.pdf}

\bibitem{DBLP:journals/tjs/ShahidAABZYC20}
Shahid, A., Afzal, M.T., Abdar, M., Basiri, M.E., Zhou, X., Yen, N.Y., Chang,
  J.: Insights into relevant knowledge extraction techniques: a comprehensive
  review.
\newblock J. Supercomput. \textbf{76}(3), 1695--1733 (2020).
\newblock \urlprefix\url{https://doi.org/10.1007/s11227-019-03009-y}

\bibitem{DBLP:journals/peerj-cs/ShahidAAAA21}
Shahid, A., Afzal, M.T., Alharbi, A., Aljuaid, H., Al{-}Otaibi, S.: In-text
  citation's frequencies-based recommendations of relevant research papers.
\newblock PeerJ Comput. Sci. \textbf{7}, e524 (2021).
\newblock \urlprefix\url{https://doi.org/10.7717/peerj-cs.524}

\bibitem{shahid21a}
Shahid, A., Afzal, M.T., Saleem, M.Q., Idrees, M.S.E., Omer, M.K.: {Extension
  of Direct Citation Model Using In-Text Citations}.
\newblock Computers, Materials \& Continua \textbf{66}(3), 3121--3138 (2021).
\newblock \urlprefix\url{https://doi.org/10.32604/cmc.2021.013809}

\bibitem{Sharma}
Sharma, B., Willis, V.C., Huettner, C.S., Beaty, K., Snowdon, J.L., Xue, S.,
  South, B.R., Jackson, G.P., Weeraratne, D., Michelini, V.: {Predictive
  article recommendation using natural language processing and machine learning
  to support evidence updates in domain-specific knowledge graphs}.
\newblock JAMIA Open \textbf{3}(3), 332--337 (2020).
\newblock \urlprefix\url{https://doi.org/10.1093/jamiaopen/ooaa028}

\bibitem{DBLP:conf/ijcnn/ShiMZJLC20}
Shi, H., Ma, W., Zhang, X., Jiang, J., Liu, Y., Chen, S.: {A Hybrid Paper
  Recommendation Method by Using Heterogeneous Graph and Metadata}.
\newblock In: {IJCNN}'20, pp. 1--8. {IEEE} (2020).
\newblock \urlprefix\url{https://doi.org/10.1109/IJCNN48605.2020.9206733}

\bibitem{msa}
Sinha, A., Shen, Z., Song, Y., Ma, H., Eide, D., Hsu, B.P., Wang, K.: {An
  Overview of Microsoft Academic Service {(MAS)} and Applications}.
\newblock In: {WWW}'15, pp. 243--246. {ACM} (2015).
\newblock \urlprefix\url{https://doi.org/10.1145/2740908.2742839}

\bibitem{subathra}
Subathra, P., Kumar, P.N.: {Recommending Research Article Based on User Queries
  Using Latent Dirichlet Allocation}.
\newblock In: Soft Computing and Signal Processing, pp. 163--175. Springer
  Singapore (2020).
\newblock \urlprefix\url{https://doi.org/10.1007/978-981-15-2475-2_15}

\bibitem{DBLP:conf/jcdl/SugiyamaK10}
Sugiyama, K., Kan, M.: Scholarly paper recommendation via user's recent
  research interests.
\newblock In: {JCDL}'10, pp. 29--38. {ACM} (2010).
\newblock \urlprefix\url{https://doi.org/10.1145/1816123.1816129}

\bibitem{DBLP:conf/jcdl/SugiyamaK11}
Sugiyama, K., Kan, M.: Serendipitous recommendation for scholarly papers
  considering relations among researchers.
\newblock In: {JCDL}'11, pp. 307--310. {ACM} (2011).
\newblock \urlprefix\url{https://doi.org/10.1145/1998076.1998133}

\bibitem{DBLP:conf/jcdl/SugiyamaK13}
Sugiyama, K., Kan, M.: Exploiting potential citation papers in scholarly paper
  recommendation.
\newblock In: {JCDL}'13, pp. 153--162. {ACM} (2013).
\newblock \urlprefix\url{https://doi.org/10.1145/2467696.2467701}

\bibitem{DBLP:journals/jodl/SugiyamaK15}
Sugiyama, K., Kan, M.: A comprehensive evaluation of scholarly paper
  recommendation using potential citation papers.
\newblock Int. J. Digit. Libr. \textbf{16}(2), 91--109 (2015).
\newblock \urlprefix\url{https://doi.org/10.1007/s00799-014-0122-2}

\bibitem{news3}
Symeonidis, P., Kirjackaja, L., Zanker, M.: {Session-based news recommendations
  using SimRank on multi-modal graphs}.
\newblock Expert Syst. Appl. \textbf{180}, 115028 (2021).
\newblock \urlprefix\url{https://doi.org/10.1016/j.eswa.2021.115028}

\bibitem{DBLP:journals/concurrency/TangLQ21}
Tang, H., Liu, B., Qian, J.: Content-based and knowledge graph-based paper
  recommendation: Exploring user preferences with the knowledge graphs for
  scientific paper recommendation.
\newblock Concurr. Comput. Pract. Exp. \textbf{33}(13) (2021).
\newblock \urlprefix\url{https://doi.org/10.1002/cpe.6227}

\bibitem{DBLP:conf/kdd/TangZYLZS08}
Tang, J., Zhang, J., Yao, L., Li, J., Zhang, L., Su, Z.: {ArnetMiner}:
  extraction and mining of academic social networks.
\newblock In: {SIGKDD}'08, pp. 990--998. {ACM} (2008).
\newblock \urlprefix\url{https://doi.org/10.1145/1401890.1402008}

\bibitem{DBLP:conf/bigdataconf/TannerAH19}
Tanner, W., Akbas, E., Hasan, M.: {Paper Recommendation Based on Citation
  Relation}.
\newblock In: Big Data'19, pp. 3053--3059. {IEEE} (2019).
\newblock \urlprefix\url{https://doi.org/10.1109/BigData47090.2019.9006200}

\bibitem{tao}
Tao, M., Yang, X., Gu, G., Li, B.: Paper Recommend Based on LDA and PageRank,
  pp. 571--584.
\newblock Springer (2020).
\newblock \urlprefix\url{https://doi.org/10.1007/978-981-15-8101-4_51}

\bibitem{DBLP:journals/access/WaheedIRMK19}
Waheed, W., Imran, M., Raza, B., Malik, A.K., Khattak, H.A.: {A Hybrid Approach
  Toward Research Paper Recommendation Using Centrality Measures and Author
  Ranking}.
\newblock {IEEE} Access \textbf{7}, 33145--33158 (2019).
\newblock \urlprefix\url{https://doi.org/10.1109/ACCESS.2019.2900520}

\bibitem{DBLP:journals/corr/abs-2103-08819}
Wang, B., Weng, Z., Wang, Y.: {A Novel Paper Recommendation Method Empowered by
  Knowledge Graph: for Research Beginners}.
\newblock CoRR \textbf{abs/2103.08819} (2021).
\newblock \urlprefix\url{https://arxiv.org/abs/2103.08819}

\bibitem{DBLP:journals/asc/WangWYXYY21}
Wang, G., Wang, H., Yang, Y., Xu, D., Yang, J., Yue, F.: {Group article
  recommendation based on {ER} rule in Scientific Social Networks}.
\newblock Appl. Soft Comput. \textbf{110}, 107631 (2021).
\newblock \urlprefix\url{https://doi.org/10.1016/j.asoc.2021.107631}

\bibitem{wang21}
Wang, G., Zhang, X., Wang, H., Chu, Y., Shao, Z.: {Group-Oriented Paper
  Recommendation With Probabilistic Matrix Factorization and Evidential
  Reasoning in Scientific Social Network}.
\newblock IEEE Transactions on Systems, Man, and Cybernetics: Systems pp. 1--15
  (2021).
\newblock \urlprefix\url{https://doi.org/10.1109/TSMC.2021.3072426}

\bibitem{DBLP:conf/ijcai/WangCL13}
Wang, H., Chen, B., Li, W.: {Collaborative Topic Regression with Social
  Regularization for Tag Recommendation}.
\newblock In: {IJCAI}'13, pp. 2719--2725. {IJCAI/AAAI} (2013).
\newblock
  \urlprefix\url{http://www.aaai.org/ocs/index.php/IJCAI/IJCAI13/paper/view/7006}

\bibitem{DBLP:conf/service/WangXTWX20}
Wang, X., Xu, H., Tan, W., Wang, Z., Xu, X.: {Scholarly Paper Recommendation
  via Related Path Analysis in Knowledge Graph}.
\newblock In: {ICSS}'20, pp. 36--43. {IEEE} (2020).
\newblock \urlprefix\url{https://doi.org/10.1109/ICSS50103.2020.00014}

\bibitem{citeseer2}
Wu, J., Kim, K., Giles, C.L.: {CiteSeerX}: 20 years of service to scholarly big
  data.
\newblock In: {AIDR}'19, pp. 1:1--1:4. {ACM} (2019).
\newblock \urlprefix\url{https://doi.org/10.1145/3359115.3359119}

\bibitem{DBLP:conf/sigir/XieSB21}
Xie, Y., Sun, Y., Bertino, E.: {Learning Domain Semantics and Cross-Domain
  Correlations for Paper Recommendation}.
\newblock In: {SIGIR}'21, pp. 706--715. {ACM} (2021).
\newblock \urlprefix\url{https://doi.org/10.1145/3404835.3462975}

\bibitem{xie21}
Xie, Y., Wang, S., Pan, W., Tang, H., Sun, Y.: {Embedding Based Personalized
  New Paper Recommendation}.
\newblock In: ChineseCSCW'21, pp. 558--570. Springer Singapore (2021).
\newblock \urlprefix\url{https://doi.org/10.1007/978-981-16-2540-4_40}

\bibitem{DBLP:journals/www/YangLLLZZZZ19}
Yang, Q., Li, Z., Liu, A., Liu, G., Zhao, L., Zhang, X., Zhang, M., Zhou, X.: A
  novel hybrid publication recommendation system using compound information.
\newblock World Wide Web \textbf{22}(6), 2499--2517 (2019).
\newblock \urlprefix\url{https://doi.org/10.1007/s11280-019-00687-9}

\bibitem{DBLP:conf/ksem/YuHLZXLXY19}
Yu, M., Hu, Y., Li, X., Zhao, M., Xu, T., Liu, H., Xu, L., Yu, R.: {Paper
  Recommendation with Item-Level Collaborative Memory Network}.
\newblock In: {KSEM}'19, \emph{LNCS}, vol. 11775, pp. 141--152. Springer
  (2019).
\newblock \urlprefix\url{https://doi.org/10.1007/978-3-030-29551-6\_13}

\bibitem{zarvel}
Zavrel, J., Grotov, A., Mitnik, J.: {Building a Platform for Ensemble-Based
  Personalized Research Literature Recommendations for AI and Data Science at
  Zeta Alpha}, p. 536–537.
\newblock Association for Computing Machinery (2021).
\newblock \urlprefix\url{https://doi.org/10.1145/3460231.3474619}

\bibitem{DBLP:journals/ftir/ZhangC20}
Zhang, Y., Chen, X.: {Explainable Recommendation: {A} Survey and New
  Perspectives}.
\newblock Found. Trends Inf. Retr. \textbf{14}(1), 1--101 (2020).
\newblock \urlprefix\url{https://doi.org/10.1561/1500000066}

\bibitem{ZHANG2019616}
Zhang, Y., Wang, M., Gottwalt, F., Saberi, M., Chang, E.: Ranking scientific
  articles based on bibliometric networks with a weighting scheme.
\newblock Journal of Informetrics \textbf{13}(2), 616--634 (2019).
\newblock \urlprefix\url{https://doi.org/10.1016/j.joi.2019.03.013}

\bibitem{zhang20}
Zhang, Y., Wang, M., Saberi, M., Chang, E.: {Towards Expert Preference on
  Academic Article Recommendation Using Bibliometric Networks}.
\newblock In: PAKDD'20, pp. 11--19. Springer International Publishing (2020).
\newblock \urlprefix\url{https://doi.org/10.1007/978-3-030-60470-7_2}

\bibitem{DBLP:journals/access/ZhaoKFMN20}
Zhao, X., Kang, H., Feng, T., Meng, C., Nie, Z.: {A Hybrid Model Based on {LFM}
  and BiGRU Toward Research Paper Recommendation}.
\newblock {IEEE} Access \textbf{8}, 188628--188640 (2020).
\newblock \urlprefix\url{https://doi.org/10.1109/ACCESS.2020.3031281}

\bibitem{DBLP:journals/kbs/ZhuLLSQN21}
Zhu, Y., Lin, Q., Lu, H., Shi, K., Qiu, P., Niu, Z.: Recommending scientific
  paper via heterogeneous knowledge embedding based attentive recurrent neural
  networks.
\newblock Knowl. Based Syst. \textbf{215}, 106744 (2021).
\newblock \urlprefix\url{https://doi.org/10.1016/j.knosys.2021.106744}

\end{thebibliography}
\end{document}